\documentclass[a4paper,fleqn,usenatbib]{mnras}
\usepackage{newtxtext,newtxmath}

\usepackage[T1]{fontenc}
\usepackage{ae,aecompl}

\usepackage{amsmath}	
\usepackage{lscape}
\usepackage{graphicx}
\usepackage{bm}

\newcommand{\paperI}{\textcolor{blue}{Paper I}}
\newcommand{\crmax}{$\zeta_{\rm ion} = 10^{-15}\ {\rm s}^{-1}$}
\newcommand{\crfid}{$\zeta_{\rm ion} = 10^{-17}\ {\rm s}^{-1}$}

\newcommand{\crzero}{$\zeta_{\rm ion} = 0$}
\newcommand{\zsun}{{\rm Z}_\odot}
\newcommand{\msun}{{\rm M}_\odot}
\newcommand{\K}{\ {\rm K}}

\newcommand{\nH}{n_{\rm H}}
\newcommand{\pcc}{\ {\rm cm}^{-3}}
\newcommand{\ps}{\ {\rm s}^{-1}}
\newcommand{\zetaion}{\zeta_{\rm ion}}
\newcommand{\Rm}{{\rm Rm}}
\newcommand{\Bcr}{B_{\rm crit}}

\title[Ionization degree in clouds with different metallicities]{Ionization degree and magnetic diffusivity in star-forming clouds with different metallicities}

\author[D. Nakauchi, K. Omukai, and H. Susa]{
Daisuke Nakauchi$^{1}$\thanks{E-mail: nakauchi@astr.tohoku.ac.jp},
Kazuyuki Omukai$^{1}$\thanks{E-mail: omukai@astr.tohoku.ac.jp}, 
and Hajime Susa$^{2}$ \thanks{E-mail: susa@konan-u.ac.jp}
\\
$^{1}$Astronomical Institute, Graduate School of Science, Tohoku University, Aoba, Sendai 980-8578, Japan \\
$^{2}$Department of Physics, Faculty of Science, Konan University, Higashi-Nada, Kobe 658-0072, Japan
}
\date{Accepted XXX. Received YYY; in original form ZZZ}

\pubyear{2021}

\begin{document}
\label{firstpage}
\pagerange{\pageref{firstpage}--\pageref{lastpage}}
\maketitle

\begin{abstract}
Magnetic fields play such essential roles in star formation as transporting angular momentum and driving outflows from a star-forming cloud, thereby controlling the formation efficiency of a circumstellar disc and also multiple stellar systems.
The coupling of magnetic fields to the gas depends on its ionization degree.
We calculate the temperature evolution and ionization degree of a cloud for various metallicities of $Z/\zsun = 10^{-6}, 10^{-5}, 10^{-4}, 10^{-3}, 10^{-2}, 10^{-1}$, and $1$. We update the chemical network by reversing all the gas-phase processes and by considering grain-surface chemistry, including grain evaporation, thermal ionization of alkali metals, and thermionic emission from grains.
The ionization degree at $\nH \sim 10^{15}\mbox{-}10^{19}\pcc$ becomes up to eight orders of magnitude higher than that obtained in the previous model, owing to the thermionic emission and thermal ionization of K and Na, which have been neglected so far.
Although magnetic fields dissipate owing to ambipolar diffusion or Ohmic loss at $\nH < 10^{15}\pcc$, the fields recover strong coupling to the gas at $\nH \sim 10^{15}\pcc$, which is lower by a few orders of magnitude compared to the previous work.
We develop a reduced chemical network by choosing processes relevant to major coolants and charged species.
The reduced network consists of 104~(161) reactions among 28~(38) species in the absence~(presence, respectively) of ionization sources.
The reduced model includes H$_2$ and HD formation on grain surfaces as well as the depletion of O, C, OH, CO, and H$_2$O on grain surfaces.
\end{abstract}



\section{Introduction}\label{sec:intro}

Recent development of gravitational wave detectors, such as advanced LIGO, Virgo and KAGRA, revealed the presence of binary black holes~(BBHs) that merge within the Hubble time~\citep[e.g.,][]{Abbott2016,Abbott2019}.
Most of the BHs are as massive as $\sim 30~\msun$, and their origin is currently debated.
Some authors proposed that they are originated from isolated binaries formed in low-metallicity star-forming clouds, including Population III~(Pop III) stars with the primordial composition~\citep[e.g.,][]{Kinugawa2014,Kinugawa2016,Belczynski2016,Giacobbo2018}.
Others proposed that these binary systems are formed via gravitational interaction in low-metallicity dense star clusters~\citep[e.g.,][]{Rodriguez2016,Antonini2016,Kumamoto2020,Mapelli2016,Mapelli2020,Liu2020,Tanikawa2020}.
In any case, the merger rate of massive BBHs depends on the formation efficiency and nature of low-metallicity binaries.

Without metals and dust grains that work as efficient coolants, a Pop III star-forming cloud maintains a warm environment of several $100\K$ because of inefficient H$_2$ cooling, which makes Pop III stars typically massive~\citep{Omukai1998,Bromm1999,Bromm2002,Omukai_Palla2001,Omukai_Palla2003,Abel2002,Yoshida2006}.
Recent multi-dimensional~(D) radiation hydrodynamical~(RHD) simulations have shown that the mass of a Pop III star distributes in a wide range of $10\mbox{-}1000~\msun$~\citep{Hosokawa2011,Hosokawa2016,Stacy2012,Stacy2016,Hirano2014,Susa2014}, and that Pop III stars are formed in massive binary or multiple stellar systems as a result of the circumstellar disc formation and its fragmentation~(\citealt{Sugimura2020}, see also~\citealt{Machida_Omukai2008,Stacy2010,Clark2011,Greif2012,Susa2019,Chiaki2020}).
Pop III stars may also be born as rapid rotators~\citep{Stacy2011,Stacy2013,Takahashi2017}.

Massive star formation in a slightly metal-enriched cloud has not been studied extensively so far, and is still very uncertain.
From 2D RHD calculations, \cite{Fukushima2020} studied the protostar formation and evolution within a metal-enriched halo obtained from the cosmological simulation of \cite{Chiaki2016}.
They showed that despite the strong radiation feedback from a protostar, stars as massive as a few hundred solar masses can be formed.
With little mass-loss by radiation-driven winds~\citep{Kudritzki2002}, a low-metallicity massive star can be a progenitor of a massive BH.

However, the above mentioned works did not account for magnetic fields that could change those pictures.
When magnetic fields strongly couple to the gas, they slow down the cloud rotation by transferring the angular momentum from the center to outwards~(so-called magnetic braking), which suppresses the formation of a circumstellar disc and a multiple system via disc fragmentation~\citep[e.g.,][]{Gillis1974,Tomisaka2002,Tsukamoto2016,Hennebelle2019}.
Magnetic fields also drive outflows which eject a part of the cloud materials back into the interstellar space, decreasing the star-formation efficiency~\citep[e.g.,][]{Matzner2000,Nakamura2007,Wang2010,Cunningham2011,Machida_Hosokawa2013,Federrath2014a}.
3D magnetohydrodynamical~(MHD) calculations of low-metallicity star formation have shown that magnetic braking extracts too much angular momentum from the cloud center to form a multiple star system, if the field is stronger than $B \sim 10^{-12}\ {\rm G}$~(at $\sim 1\ {\rm cm}^{-3}$)~\citep{Machida2013,Peters2014,Sharda2020}.
About 10\% of the mass is removed from the cloud into the interstellar space by MHD outflows, if the field is stronger than $B \sim 10^{-11}\ {\rm G}$~(at $\sim 1\ {\rm cm}^{-3}$)~\citep{Machida2006,Machida2013,Tanaka2018,Higuchi2019}.

The magnetic field strength in the interstellar medium~(ISM) in a young galaxy is still uncertain.
While various scenarios are suggested for the generation of primordial magnetic fields~\citep[e.g.,][]{Ando2010,Widrow2012,Subramanian2016,McKee2020}, primordial fields are believed to be many orders of magnitude weaker than the Galactic ISM~\citep[$\sim 5\ \mu{\rm G}$;][]{Crutcher2010}.
If the ISM in a young galaxy is highly turbulent, weak magnetic fields can be amplified by the small-scale-dynamo action up to the equipartition $\sim 1\ \mu{\rm G}$ at $\sim 1\pcc$~\citep{Schleicher2010,Sur2010,Sur2012,Federrath2011,Schober2012a,Turk2012}.
The presence of micro-Gauss magnetic fields is indicated from the observation of high-$z$ galaxies~\citep{Bernet2008,Mao2017}.

The coupling of the magnetic fields to the gas is controlled by its ionization degree.
In the present-day star-formation, various authors have studied the ionization degree in a cloud by considering a simple equilibrium chemistry involving representative ions, electrons and charged grains~\citep{Oppenheimer1974,Umebayashi1980,Umebayashi1990,Nakano1986,Nakano2002}.
They set the abundances of neutral atoms and molecules constant and treat their depletion fractions on grain surfaces as model parameters.
Recent studies have accounted for a more elaborated gas-phase chemistry among H, He, C, and O compounds, as well as grain-surface chemistry consisting of the freeze-out of gas-phase species, the desorption of surface species, and molecule formation~\citep{Ilgner2006,Furuya2012,Dzyurkevich2017,Zhao2018}.
\cite{Marchand2016} also extended the classical model by considering grain vaporization, thermal ionization of alkali metals, and electron ejection from heated grains~(so-called thermionic emission), which elevate the ionization degree at high temperatures of $\gtrsim 500\K$.

The ionization degree in a low-metallicity cloud has also been studied by \cite{Maki2004,Maki2007}, and \cite{Susa2015}.
It was shown that in a primordial cloud, with its high temperature and lack of dust grains, Li$^+$ maintains the ionization degree high enough for magnetic fields to couple to the gas.
In a slightly metal-enriched cloud of $Z/ \zsun \gtrsim 10^{-6}$, with their large recombination cross section, dust grains capture electrons and ions, and decrease the ionization degree so low as to decouple the magnetic fields from the gas.
Magnetic fields recover the coupling after the dust grains evaporate and hydrogen ionization raises the ionization degree at $\nH \sim 10^{17}\mbox{-}10^{18}\pcc$.
However, the previous studies have some flaws in their chemical modelling:
(i) only a part of the reactions are reversed, and at a certain density, the abundances are switched artificially to the chemical equilibrium values of the H/He gas,
(ii) thermionic emission and thermal ionization of alkali metals are not properly considered, and
(iii) grain-surface chemistry is not included.
In the primordial cloud, where only the problem (i) is relevant, \cite{Nakauchi2019}~(hereafter \paperI ) calculated the ionization degree by reversing all the chemical processes.
They found that the ionization degree at $\nH \sim 10^{14}\mbox{-}10^{18}\pcc$ is enhanced by a few orders of magnitude, which couples the magnetic fields to the gas more strongly at these densities.

In this paper, we update the previous chemical network by accounting for the processes (i)-(iii), and compute the temperature evolution, ionization degree, and resistivity coefficients in a star-forming cloud for a wide range of metallicity, $Z/\zsun = 10^{-6}, 10^{-5}, 10^{-4}, 10^{-3}, 10^{-2}, 10^{-1}$, and $1$.
We find that the ionization degree at $\nH \sim 10^{15}\mbox{-}10^{19}\pcc$ becomes up to eight orders of magnitude higher than that obtained in the previous model.
This is due to the thermionic emission and thermal ionization of vaporized K and Na, which are neglected so far.
As a result, magnetic fields recover the strong coupling to the gas at much earlier stages~(by a few orders of magnitude in density), compared to the previous work.
We also develop a reduced chemical network by extracting the processes relevant to major coolants and charged species from the full network.
Among various molecule formation processes on grain surfaces, only H$_2$ and HD formation is included in the reduced model by using simple formulae.
The reduced model also includes the depletion of O, C, OH, CO, and H$_2$O on grain surfaces.

This paper consists of the following sections.
In Section \ref{sec:method}, we describe the method and chemical network used to calculate the temperature evolution and ionization degree in a star-forming cloud. 
More details about the dust-surface chemistry is summarized in Appendix.
In Section \ref{sec:result}, we show the results for both cases without and with ionization sources in Section \ref{subsec:zeta_zero} and \ref{subsec:zeta_nonzero}, respectively.
From the ionization degree obtained in Section \ref{sec:result}, we discuss the conditions of magnetic dissipation for both global and turbulent magnetic fields in Section \ref{sec:magnetic}.
After briefly summarizing the results, we discuss the uncertainties and implications of our work in Section \ref{sec:summary}.

\section{Method}\label{sec:method}

The gravitational contraction of a spherically symmetric cloud is calculated by way of a one-zone model neglecting rotation, turbulence and magnetic fields as in \cite{Omukai2000,Omukai2001,Omukai2012} and \cite{Omukai2005}.
Owing to higher density, the cloud core collapses at a shorter timescale with about the local free-fall time
\begin{equation}
t_{\rm ff} = \sqrt{\frac{ 3 \pi }{32 G \rho}},
\label{eq:t_ff}
\end{equation}
leaving behind a lower density envelope~(so-called runaway collapse;~\citealt{Larson1969,Penston1969}).
This runaway collapse proceeds in a self-similar way such that the density in the cloud core distributes uniformly across the Jeans length scale
\begin{equation}
\lambda_{\rm J} = \sqrt{\frac{\pi k_{\rm B} T}{G \mu m_{\rm H} \rho}},
\label{eq:lambda_jeans}
\end{equation}
and the density in the envelope declines with radius following a power law.
 
In our model, the physical quantities in the cloud core are calculated.
The density increases in the free-fall time as
\begin{equation}
\frac{d \rho}{dt} = \frac{\rho}{t_{\rm ff}},
\label{eq:free_fall}
\end{equation}
and the temperature $T$ is determined by the energy equation:
\begin{equation}
\frac{de}{dt} = - P \frac{d}{dt}\left(\frac{1}{\rho}\right) - \Lambda_{\rm net},
\label{eq:EoE}
\end{equation}
where $P$ is the pressure, $e$ the internal energy per unit mass, and $\Lambda_{\rm net}$ the net cooling rate per unit mass.
The following five processes contribute to the net cooling rate:
\begin{equation}
\Lambda_{\rm net} = \Lambda_{\rm line} + \Lambda_{\rm chem}  + \Lambda_{\rm grain} + \Lambda_{\rm cont} - \Gamma_{\rm ion},
\label{eq:cooling_rate}
\end{equation}
where $\Lambda_{\rm line}$ includes cooling by H Ly$\alpha$ emission, molecular line emissions of H$_2$, HD, CO, OH, and H$_2$O, and fine-structure line emissions of CII, CI, and OI, $\Lambda_{\rm cont}$ the cooling by thermal emissions from gas~(e.g, H$_2$ collision-induced emission; CIE) and dust grains, $\Lambda_{\rm chem}$ cooling/heating associated with the chemical reactions, and $\Gamma_{\rm ion}$ the ionization heating by cosmic-ray~(CR) injection and decay of radioactive elements~(RE).
The formulations of H Ly$\alpha$, H$_2$, and HD cooling are referred from \paperI, those of CO, OH, and H$_2$O cooling from \cite{Omukai2010}, and those of fine-structure line cooling, $\Lambda_{\rm cont}$ and $\Lambda_{\rm chem}$ from \cite{Omukai2000}~(with some updates for CI and OI cooling by \citealt{Nakauchi2018}).
When the cloud becomes opaque, the radiative cooling rates are damped depending on the hydrogen column density of the cloud: $N_{\rm H} = n_{\rm H} \lambda_{\rm J}$.
The ionization heating rate $\Gamma_{\rm ion}$ is estimated by assuming that the gas obtains 3.4 eV of heat per ionization~\citep{Spitzer1969}.

In the gas phase, 1184 chemical reactions are considered among the following 63 species:
H, H$_2$, e$^-$, H$^+$, H$_2^+$, H$_3^+$, H$^-$, He, He$^+$, He$^{2+}$, HeH$^+$,
D, HD, D$^+$, HD$^+$, D$^-$, C, C$_2$, CH, CH$_2$, CH$_3$, CH$_4$,
C$^+$, C$_2^+$, CH$^+$, CH$_2^+$, CH$_3^+$, CH$_4^+$, CH$_5^+$,
O, O$_2$, OH, CO, H$_2$O, HCO, O$_2$H, CO$_2$, H$_2$CO, H$_2$O$_2$,
O$^+$, O$_2^+$, OH$^+$, CO$^+$, H$_2$O$^+$, HCO$^+$, O$_2$H$^+$, H$_3$O$^+$,
H$_2$CO$^+$, HCO$_2^+$, H$_3$CO$^+$,
Li, LiH, Li$^+$, Li$^-$, LiH$^+$, Li$^{2+}$, Li$^{3+}$,
K, K$^{+}$, Na, Na$^{+}$, Mg, Mg$^{+}$.
The primordial-gas chemistry consists of 214 reactions, i.e., 107 forward and reverse pairs, which are listed with their references in Table 1 of \paperI.
The ionization processes of Li, Na, and K via H$_2$ collision and their inverses:
\begin{align}
&{\rm H}_2 + {\rm Li} \rightleftharpoons {\rm H}_2 + {\rm Li}^+ + e \nonumber    \\
&{\rm H}_2 + {\rm K} \rightleftharpoons {\rm H}_2 + {\rm K}^+ + e \nonumber    \\
&{\rm H}_2 + {\rm Na} \rightleftharpoons {\rm H}_2 + {\rm Na}^+ + e
\label{eq:therm_ionization}
\end{align}
are referred from \cite{Ashton1973}.
The other 964 reactions~(464 forward-reverse pairs and 36 CR-induced processes) are referred from the UMIST database~\citep{McElroy2013}.
From the large number of reactions listed in the database, we choose those reactions which contain the following species as reactants or products: H, H$_2$, e$^-$, H$^+$, H$_2^+$, H$_3^+$, H$^-$, He, He$^+$, HeH$^+$, C, C$_2$, CH, CH$_2$, CH$_3$, CH$_4$, C$^+$, C$_2^+$, CH$^+$, CH$_2^+$, CH$_3^+$, CH$_4^+$, CH$_5^+$, O, O$_2$, OH, CO, H$_2$O, HCO, O$_2$H, CO$_2$, H$_2$CO, H$_2$O$_2$, O$^+$, O$_2^+$, OH$^+$, CO$^+$, H$_2$O$^+$, HCO$^+$, O$_2$H$^+$, H$_3$O$^+$, H$_2$CO$^+$, HCO$_2^+$, H$_3$CO$^+$, K, K$^{+}$, Na, Na$^{+}$, Mg, Mg$^{+}$, but we remove those reactions overlapping with the primordial-gas chemistry.
In case a forward-reverse pair is found in the database, we regard the reaction with a positive value for the heat of reaction $\Delta E$ as forward and calculate its reverse rate coefficient by the method explained below.

All the gas-phase reactions are reversed so that the fractional abundance is calculated correctly both in the non-equilibrium and equilibrium cases.
The rate coefficients for the forward and reverse reactions are related to each other through the detailed balance principle~\citep[e.g.,][]{Draine2011}:
\begin{equation}
k_{\rm rev} = k_{\rm fwd} K_{\rm eq}(T),
\label{eq:detailed_balance}
\end{equation}
where $K_{\rm eq}(T)$ is the equilibrium constant.
For a reaction where $M$ reactants ${\rm R}_1, {\rm R}_2, ..., {\rm R}_M$ change into $N$ products ${\rm P}_1, {\rm P}_2, ..., {\rm P}_N$, $K_{\rm eq}(T)$ is calculated from
\begin{equation}
\begin{split}
K_{\rm eq}(T) &= \left(\frac{2 \pi k_{\rm B} T}{h_{\rm P}^2}\right)^{\frac{3}{2}(M-N)} 
\left(\frac{m_{{\rm R}_1}...m_{{\rm R}_M}}{m_{{\rm P}_1}...m_{{\rm P}_N}}\right)^{\frac{3}{2}} \\
&\times \left(\frac{z({{\rm R}_1})...z({{\rm R}_M})}{z({{\rm P}_1})...z({{\rm P}_N})}\right) e^{-\Delta E/k_{\rm B} T},
\end{split}
\label{eq:eqb_const}
\end{equation}
where $m(i)$ and $z(i)$ are the mass and partition function of each atom or molecule, and $\Delta E$ the heat of reaction, whose values are adopted from the references summarized in Appendix \ref{sec:part_fnc}.
In a sufficiently opaque cloud, atoms~(or molecules) are ionized~(or dissociated) by thermal radiation trapped in the cloud.
In this case, the radiation field has the black-body spectrum~($J_\nu = B_\nu(T)$), and the rate coefficient for a radiative-dissociation reaction $k_{\rm dissoc}$ is calculated by that of its reverse reaction, i.e., radiative association $k_{\rm assoc}$ through Eq. \eqref{eq:eqb_const}.
When the cloud is still optically thin, since $k_{\rm dissoc}$ scales linearly with the radiation intensity $J_\nu = (1-e^{-\tau_{\rm cont}})B_\nu(T)$~($\tau_{\rm cont}$ is the continuum optical depth), $k_{\rm dissoc}$ is calculated from~(\paperI)
\begin{equation}
k_{\rm dissoc} = (1-e^{-\tau_{\rm cont}}) k_{\rm assoc} K_{\rm eq}(T).
\label{eq:detailed_balance_photo}
\end{equation}

We adopt the dust model of \cite{Pollack1994}, where the dust grains are assumed to be composed of water ice, organics, troilite, metallic iron, and silicate~(olivine and orthopyroxene).
Below the vaporization temperature of water ice~(100-200 K), the dust-to-gas mass ratio is $9.4 \times 10^{-3}$.
When the grain temperature $T_{\rm gr}$ exceeds the vaporization temperature of each constituent, the dust-to-gas mass ratio is decreased by the abundance of each constituent.
The grain temperature $T_{\rm gr}$ is determined by the energy-balance equation of the dust grains:
\begin{equation}
\begin{split}
4 \sigma_{\rm SB} \kappa_{\rm gr} T_{\rm gr}^4 \beta_{\rm esc} &= \Gamma_{{\rm gas}-{\rm dust}} \\
&+ 4 \sigma_{\rm SB} \kappa_{\rm gr} \left(T^4 (1-e^{-\tau_{\rm cont}}) + T_{\rm CMB}^4 e^{-\tau_{\rm cont}}\right),
\end{split}
\label{eq:grain_temp}
\end{equation}
where $\kappa_{\rm gr}$ is the Planck mean opacity of dust grains calculated by \cite{Semenov2003}, $\beta_{\rm esc} = {\rm min}(1, \tau_{\rm cont}^{-2})$ a factor representing the radiative diffusion effect~\citep{Masunaga1998}, $\Gamma_{{\rm gas}-{\rm dust}}$ the energy transfer rate via gas-grain collision~\citep{Hollenbach1979}, and $T_{\rm CMB} = 2.725\K$ the CMB temperature.
The grain size distribution is assumed to follow the Mathis, Rumpl \& Nordsieck~(MRN)-type power law~\citep{Mathis1977,Pollack1985}:
\begin{equation}
y_{\rm gr}(a_{\rm gr})
= C_{\rm gr} \begin{cases}
\left(a_{\rm gr}/a_{\rm mid}\right)^{-\lambda_1} & \text{$a_{\rm min} \le a_{\rm gr} \le a_{\rm mid}$}, \\
\left(a_{\rm gr}/a_{\rm mid}\right)^{-\lambda_2} & \text{$a_{\rm mid} \le a_{\rm gr} \le a_{\rm max}$},
\end{cases}
\end{equation}
where $\lambda_1 = 3.5, \lambda_2 = 5.5, a_{\rm min} = 0.005\ \mu{\rm m}, a_{\rm mid} = 1\ \mu{\rm m}$, and $a_{\rm max} = 5\ \mu{\rm m}$.
When we take an integration or average of a physical quantity over the grain size distribution, the distribution is divided equally in $\log a_{\rm gr}$ into 15 bins for $a_{\rm min} \le a_{\rm gr} \le a_{\rm mid}$ and 5 bins for $a_{\rm mid} \le a_{\rm gr} \le a_{\rm max}$.
For simplicity, we neglect the grain growth by the accretion of gas-phase species and by grain-grain collisions.

In addition to the gas-phase chemistry, we consider the grain-surface chemistry, the details of which are described in Appendix \ref{sec:dust_chem}.
The grain-surface chemistry is divided into three categories: (i) the adsorption of a gas-phase species onto the grain surface, (ii) the desorption of a grain-surface species into the gas-phase, and (iii) molecule formation~\citep[e.g.,][]{Hasegawa1992}.
Dust grains also obtain electric charges when gas-phase ions or electrons recombine with grain-surface species.
These charge transfer reactions from gas to grain and grain to grain are taken into account following \cite{Draine1987}.
Dust grains are assumed to hold up to two charges, i.e., gr$^0$, gr$^{\pm}$, and gr$^{2\pm}$, since the abundance of more than triply charged grains is negligibly small~\citep[e.g.,][]{Nakano2002}.
In a sufficiently warm cloud of $\gtrsim 500\K$, the ejection of thermal electrons from grain surfaces~(so-called thermionic emission) is also taken into account following \cite{Desch2015}.

The fractional abundance of He, D, and Li nuclei relative to H nuclei is set to $y_{\rm He} = 8.3 \times 10^{-2}, y_{\rm D} = 2.6 \times 10^{-5}$, and $y_{\rm Li} = 4.7 \times 10^{-10}$, which are derived from the standard Big Bang nucleosynthesis~(BBN) theory by using the baryon-to-photon ratio of the Planck observation~\citep{Cyburt2016}.
In the solar metallicity case~($Z={\rm Z}_\odot$), the abundance of C, O, Na, Mg, and K nuclei is adopted from the photospheric values of the Sun: $y_{{\rm C},\odot} = 2.7 \times 10^{-4}, y_{{\rm O},\odot} = 4.9 \times 10^{-4}$, $y_{{\rm Na},\odot} = 1.7 \times 10^{-6}$, $y_{{\rm Mg},\odot} = 4.0 \times 10^{-5}$, and $y_{{\rm K},\odot} = 1.1 \times 10^{-7}$~\citep{Asplund2009}.
In a metal-poor cloud, the abundance of a heavy element is decreased in proportion to the metallicity.
Below the water ice vaporization temperature, 72\% of C, 46\% of O, 98\% of Mg, and, 100\% of Na and K are depleted into dust grains.
Above the vaporization temperature of the silicate dust~($1000\mbox{-}1500\K$ for olivine and orthopyroxene), the depleted metals are released into the gas phase, and their  gas-phase abundance is increased in proportion to the decreased fraction of the silicate dust~\citep{Finocchi1997}.

The calculations are started from the density of $n_{\rm H,0} = 1\pcc$ and the temperature of $T_0 = 100\K$.
For the light elements, H, D, He, and Li, the initial abundances are set to be same as the intergalactic values in the post-recombination era~\citep{Galli2013}: $y({{\rm H}^+}) = 10^{-4}$, $y({{\rm H}_2}) = 6 \times 10^{-7}$, $y({\rm HD}) = 4 \times 10^{-10}$, and $y({\rm Li}^+) = y_{\rm Li}$, and the remaining H, D, and He nuclei are in the neutral atomic state.
For heavy elements, with their high ionization potentials of 11.3 eV for carbon and 13.6 eV for oxygen, all the C and O nuclei are assumed to exist as neutral atoms, while with its low ionization energy of 7.6 eV, all Mg is assumed to present as Mg$^+$.

The injection of CR particles and the decay of RE are important ionization sources.
The CR ionization rate is calculated with the shielding effect as~\citep{Nakano1986}:
\begin{equation}
\zeta_{\rm CR} = \zetaion \exp\left(-\frac{N_{\rm H}}{4.3 \times 10^{25}\ {\rm cm}^{-2}}\right)\ {\rm s}^{-1}. 
\label{eq:CR}
\end{equation}
The CR intensity in the Galactic ISM was once estimated as $\zetaion \sim 10^{-17}\ {\rm s}^{-1}$~\citep{Spitzer1969}, but the recent observations suggest a much larger value of $2.3 \times 10^{-16}\ {\rm s}^{-1}$~\citep{Neufeld2017}.
The CR intensity in the ISM in the first galaxies is even more uncertain.
Since CR particles are generated by the shock-acceleration in a supernova~(SN) remnant, the CR intensity would be stronger in a cloud closer to a star-forming galaxy~\citep{Stacy2007,Nakauchi2014}.
This motivates us to consider a wide range for the CR intensity encompassing the Galactic ISM value: $\zetaion = 0, 10^{-19}, 10^{-17}$, and $10^{-15}\ps$.
On the other hand, REs are synthesized in a SN explosion.  
They are classified into two types depending on the decay time~\citep{Umebayashi2009, Susa2015}.
Long-lived REs represented by $^{40}$K have a decay time of $> 1\ {\rm Gyr}$, which is longer than the age of Universe at $z > 6$, and these REs accumulate in the ISM in a galaxy like ordinary metals.
We assume that the ionization rate is given by the Galactic value in the solar metallicity case and the rate scales linearly with the metallicity:
\begin{equation}
\zeta_{\rm RE}^{\rm long} = 1.4 \times 10^{-22} \frac{Z}{{\rm Z}_\odot}\ {\rm s}^{-1}.
\label{eq:RE_long}
\end{equation}
On the other hand, short-lived REs represented by $^{26}$Al have a decay time of $\sim 1\ {\rm Myr}$, which is too short for these elements to be stored in the ISM during galaxy evolution.
Considering the fact that the presence of short-lived REs reflects the most recent star-forming activity like CR-particle generation, we assume that the ionization rate is given by the Galactic value when \crfid, and the rate scales linearly with the CR intensity:
\begin{equation}
\zeta_{\rm RE}^{\rm short} = 7.6 \times 10^{-19} \left(\frac{\zetaion}{10^{-17}\ {\rm s}^{-1}}\right)\ {\rm s}^{-1}.
\label{eq:RE_short}
\end{equation}
The total ionization rate is given by the sum of the contribution from the three sources:
\begin{equation}
\zeta_{\rm total} = \zeta_{\rm CR} + \zeta_{\rm RE}^{\rm long} + \zeta_{\rm RE}^{\rm short}.
\label{eq:zeta_total}
\end{equation}

\vspace{10mm}

\section{Temperature Evolution and Ionization Degree}
\label{sec:result}

\subsection{Cases without ionization sources}
\label{subsec:zeta_zero}

In this subsection, we describe the results without ionization sources~(\crzero).

\subsubsection{Temperature Evolution}

\begin{figure}
\begin{center}
\includegraphics[scale=1.0]{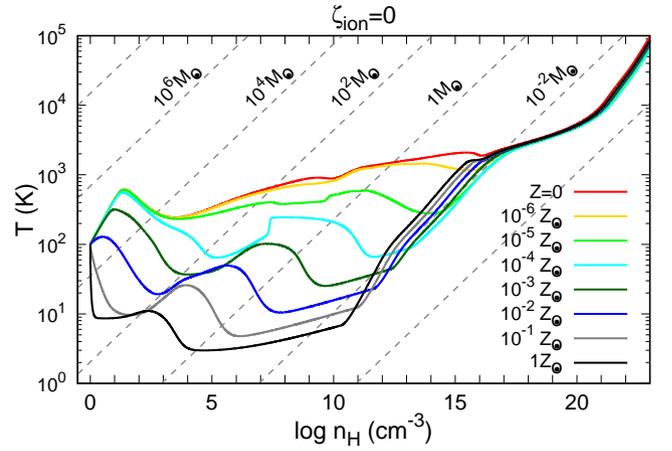}
\caption{Temperature evolution of star-forming clouds without ionization sources. Individual colored curves correspond to the results with various metallicities indicated in the legend. The oblique dashed lines indicate the loci of constant Jeans mass.}
\label{fig:nT_wo_cr}
\end{center}
\end{figure}

\begin{figure*}
\begin{center}
\begin{tabular}{cc}
{\includegraphics[scale=0.8]{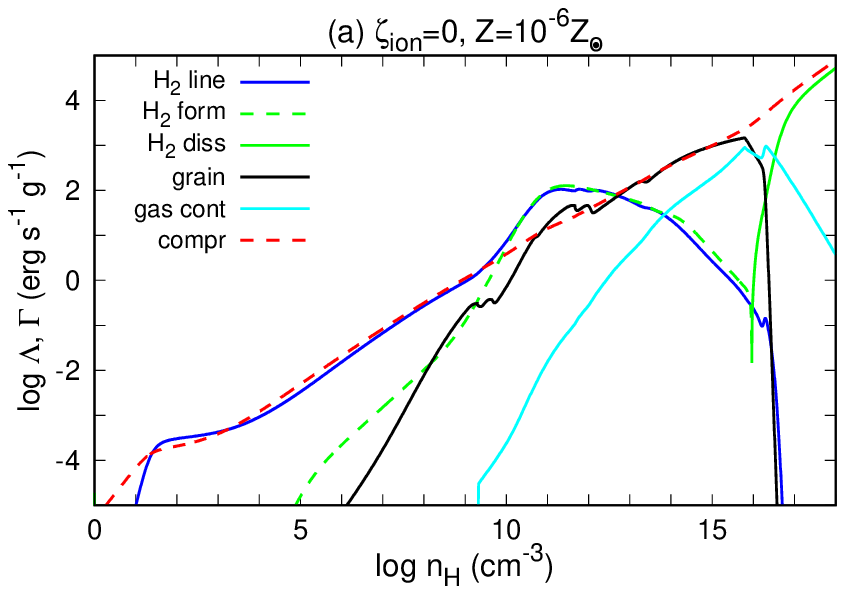}}
{\includegraphics[scale=0.8]{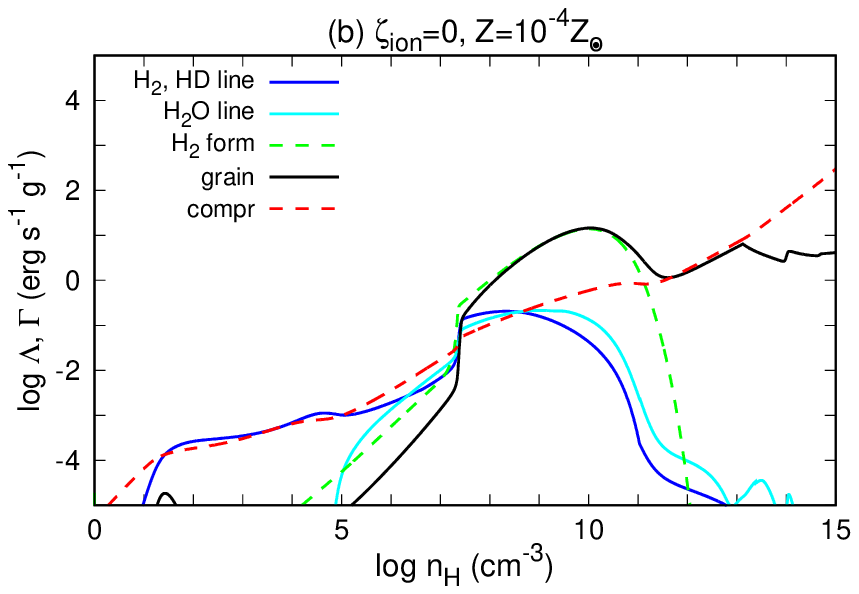}}\\
{\includegraphics[scale=0.8]{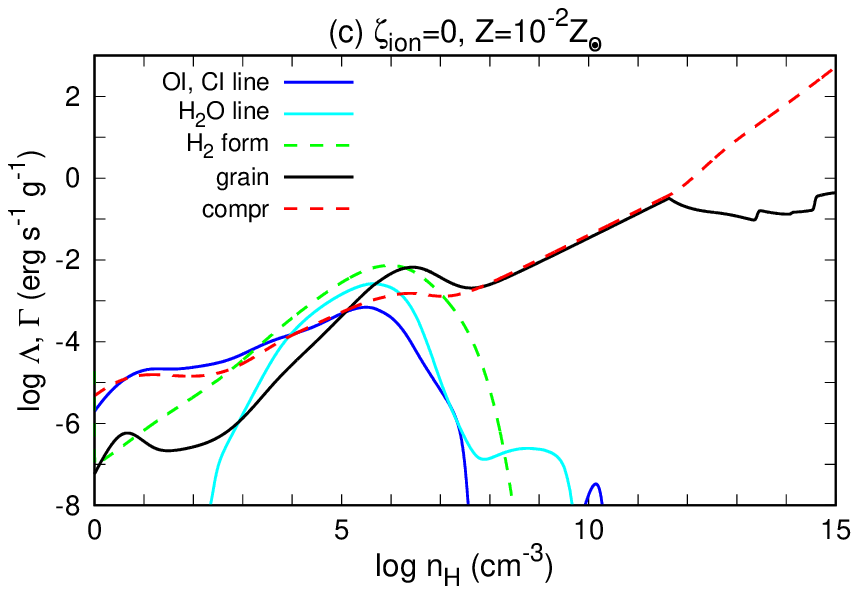}}
{\includegraphics[scale=0.8]{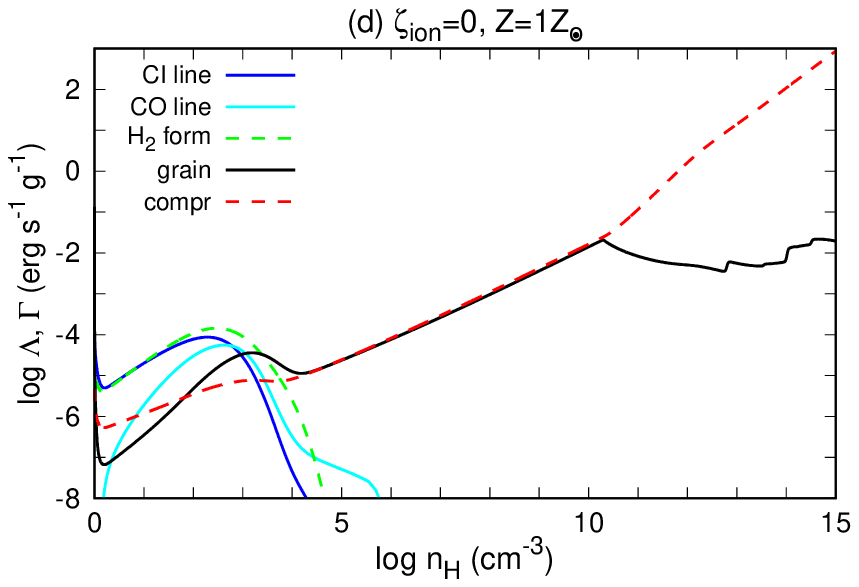}}\\
\end{tabular}
\caption{Contribution to the cooling and heating rates by individual processes for the clouds without ionization sources. 
Different panels correspond to the results with $Z/\zsun=$ (a) $10^{-6}$, (b) $10^{-4}$, (c) $10^{-2}$, and (d) $1$, respectively.}
\label{fig:cool_wo_cr}
\end{center}
\end{figure*}

\begin{figure*}
\begin{center}
\begin{tabular}{cc}
{\includegraphics[scale=0.8]{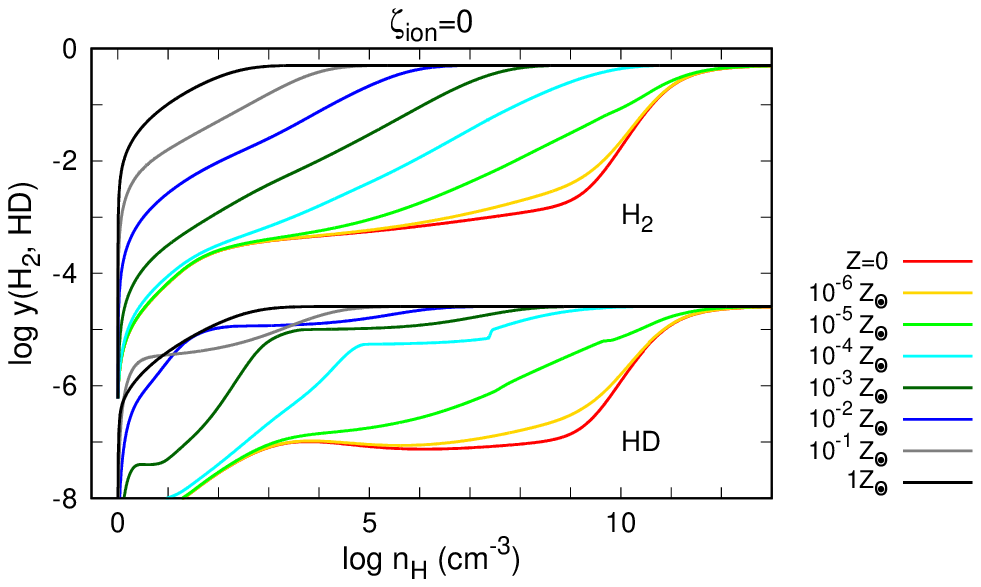}}
{\includegraphics[scale=0.8]{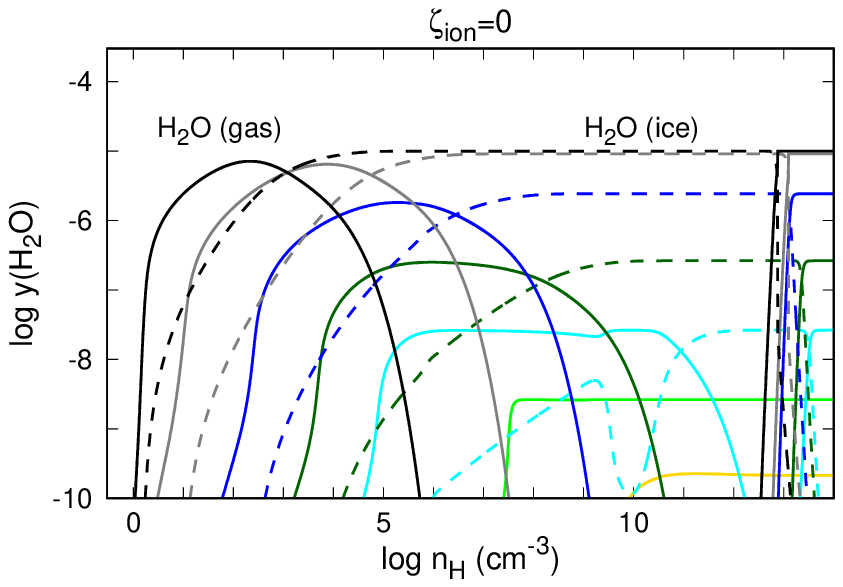}}\\
\end{tabular}
\caption{Fractional abundances of H$_2$, HD~(left panel), and H$_2$O~(right panel) for the cases shown in Figure 1.
In the right panel, the solid and dashed curves indicate the H$_2$O abundances in the gas and ice phases, respectively.}
\label{fig:chem_wo_cr}
\end{center}
\end{figure*}


Figure \ref{fig:nT_wo_cr} shows the temperature evolution of star-forming clouds with various metallicities indicated in the legend.
The oblique dashed lines indicate the loci of constant Jeans mass.
Figure \ref{fig:cool_wo_cr} shows the cooling and heating processes in the clouds with $Z/\zsun=$ (a) $10^{-6}$, (b) $10^{-4}$, (c) $10^{-2}$, and (d) $1$.
In Figure \ref{fig:chem_wo_cr}, we show the fractional abundances of major coolants, i.e., H$_2$, HD~(left panel), and H$_2$O~(right panel).
In the right panel, solid and dashed curves indicate the H$_2$O abundances in the gas and ice phases, respectively.

In almost primordial cases~($Z/\zsun \lesssim 10^{-6}$; Figure \ref{fig:cool_wo_cr}a), after the short adiabatic contraction phase without enough coolants, H$_2$ formation proceeds up to $y({\rm H}_2) \sim 10^{-3}$ via the H$^-$ channel~\citep{Peebles1968,Hirasawa_Aizu1969}:
\begin{align}
&{\rm H} + e \rightarrow {\rm H}^- + \gamma \nonumber\\
&{\rm H} + {\rm H}^- \rightarrow {\rm H}_2 + e,
\label{eq:H-channel}
\end{align}
and H$_2$ cooling lowers the temperature down to the local minimum of $\sim 200 \K$.
At $\nH \gtrsim 10^4 \pcc$, H$_2$ rotational levels reach the local thermodynamic equilibrium~(LTE), which decreases the cooling efficiency and raises the temperature again gradually.
At $\nH \gtrsim 10^8 \pcc$, the hydrogen becomes fully molecular via the three-body reactions~\citep{Palla1983}
\begin{align}
&{\rm H} + {\rm H} + {\rm H} \rightarrow {\rm H} + {\rm H}_2 \nonumber\\
&{\rm H} + {\rm H} + {\rm H}_2 \rightarrow 2{\rm H}_2.
\label{eq:three_body}
\end{align}
The elevated H$_2$ fraction by these reactions raises the cooling efficiency, but interrupts the temperature increase only temporarily at $\nH \sim 10^{10} \pcc$, because the H$_2$ lines become optically thick and H$_2$ formation also contributes to gas heating.
The temperature dip at $\nH \sim 10^{16} \pcc$ is caused by the H$_2$ collision-induced emission~\citep[CIE;][]{Omukai1998}.
The cloud soon becomes optically thick to the collision-induced absorption. 
The very brief adiabatic temperature increase at this moment is followed by gradual temperature increase due to the H$_2$ dissociation cooling.

In extremely metal-poor cases~($Z/\zsun = 10^{-5}\mbox{-}10^{-3}$; Figure \ref{fig:cool_wo_cr}b), H$_2$ formation proceeds on grain surfaces, in addition to the H$^-$ channel, so that the cloud contracts with lower temperatures via the enhanced H$_2$ cooling.
Once the temperature decreases below $\sim 150\K$, HD formation proceeds via the exothermic reactions
\begin{align}
&{\rm H}^+ + {\rm D} \rightleftharpoons {\rm H} + {\rm D}^+ \nonumber\\
&{\rm H}_2 + {\rm D}^+ \rightleftharpoons {\rm H}^+ + {\rm HD},
\label{eq:HD_form}
\end{align}
and the HD cooling lowers the temperature further until the HD rotational levels reach the LTE and its cooling efficiency is reduced.
At $\nH \sim 10^7, 10^5$, and $10^4\pcc$ for $Z/\zsun = 10^{-5}, 10^{-4}$, and $10^{-3}$, respectively, OH and H$_2$O are formed via the following reactions~(Figure \ref{fig:chem_wo_cr} right)
\begin{align}
&{\rm H} + {\rm O} \rightarrow {\rm OH} + \gamma \nonumber\\
&{\rm H}_2 + {\rm OH} \rightarrow {\rm H} + {\rm H}_2{\rm O} \nonumber\\
&{\rm H} + {\rm OH} \rightarrow {\rm H}_2{\rm O} + \gamma. 
\label{eq:H2O_form}
\end{align}
While H$_2$O cooling becomes dominant, it is counteracted by the H$_2$ formation heating on grain surfaces, and the temperature keeps increasing gradually.
Note that H$_2$O formation proceeds mainly via the gas-phase reactions~(Eq. \ref{eq:H2O_form}), and the pathway via grain-surface reactions has only a minor effect.
After H$_2$ formation is over at $\nH \sim 10^{12}, 10^{10}$, and $10^8\pcc$~(for $Z/\zsun = 10^{-5}, 10^{-4}$, and $10^{-3}$), the cloud cools efficiently via dust thermal emission up to the second local minimum.
Once the cloud becomes optically thick to the absorption of the dust thermal radiation at $\nH \sim 10^{13}\mbox{-}10^{15}\pcc$, the temperature begins to increase adiabatically with contraction.

In the metal-enriched cases of $Z/\zsun \gtrsim 10^{-2}$~(Figure \ref{fig:cool_wo_cr}c, d), OI, CI, and CO line cooling becomes effective, and the cloud cools down to lower temperatures than in more metal-poor clouds.
As H$_2$ formation proceeds via the grain-surface reaction and its heating rate rises, the temperature starts to increase.
After the H$_2$ formation is over, the temperature remains low~($\lesssim 10\K$) via dust cooling until the cloud becomes opaque to the thermal radiation.
In the case of $Z/\zsun \gtrsim 10^{-2}$, HD formation proceeds via the grain-surface processes rather than via the gas-phase reactions~(Eq. \ref{eq:HD_form}).
However, HD cooling remains a minor process throughout the evolution in these metallicities.
Therefore, among the molecule formation processes on grain surfaces, only H$_2$ formation plays an important role in the temperature evolution.

Once the cloud center becomes opaque to the thermal radiation, the temperature increases adiabatically as $T \propto \nH^{\gamma_{\rm eff}-1}$, with the effective adiabatic index~($\gamma_{\rm eff}$) larger than the critical value $\gamma_{\rm crit}=4/3$ for gravitational contraction.
Then the cloud center is supported by thermal pressure, slowing down dynamical contraction.
This occurs at lower densities with increasing metallicity.
When $Z/\zsun \gtrsim 10^{-3}$, the adiabatic phase lasts long enough time to stop dynamical contraction, forming a first hydrostatic core~\citep[e.g.,][]{Larson1969,Masunaga1998,Omukai2010}.
However, the cloud center contracts in a quasi-static way by accreting the surrounding gas and increasing its mass.
As the temperature approaches $\sim 2000\K$, collisional dissociation of H$_2$ occurs and works as the effective cooling.
This makes the temperature increase shallower with $\gamma_{\rm eff} < 4/3$, enabling the dynamical contraction of the cloud again.
After almost all the H$_2$ is dissociated into hydrogen atoms, $\gamma_{\rm eff}$ is again elevated to $\gamma_{\rm eff} = 5/3$, and a second hydrostatic core, or a protostar, is formed.

\subsubsection{Ionization degree}
\begin{figure*}
\begin{center}
\begin{tabular}{cc}
{\includegraphics[scale=0.8]{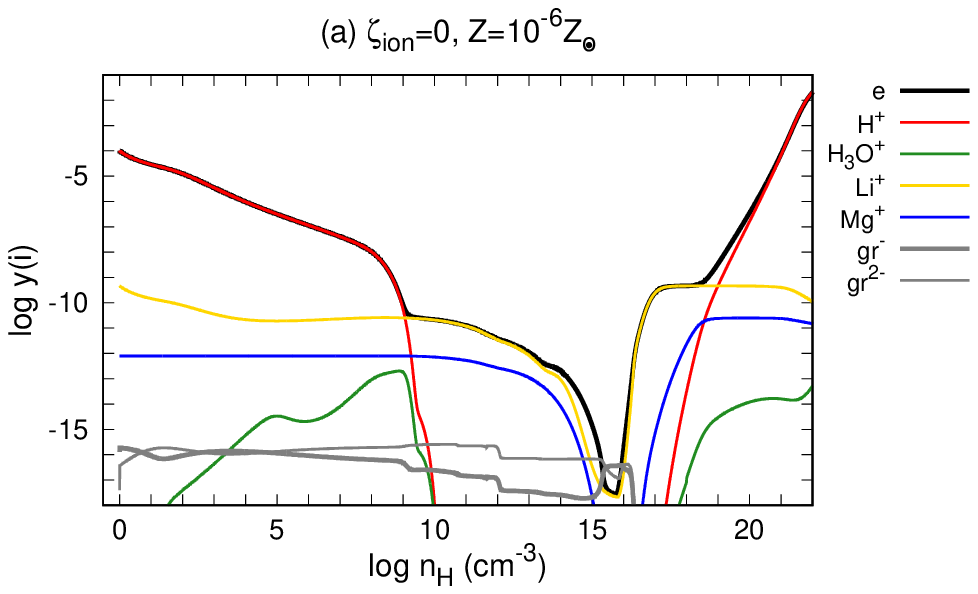}}
{\includegraphics[scale=0.8]{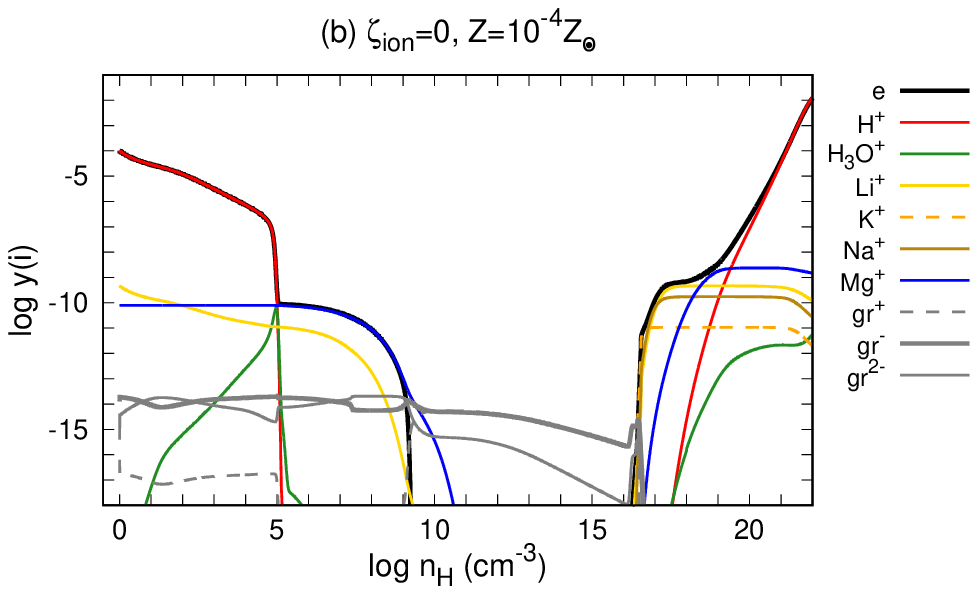}}\\
{\includegraphics[scale=0.8]{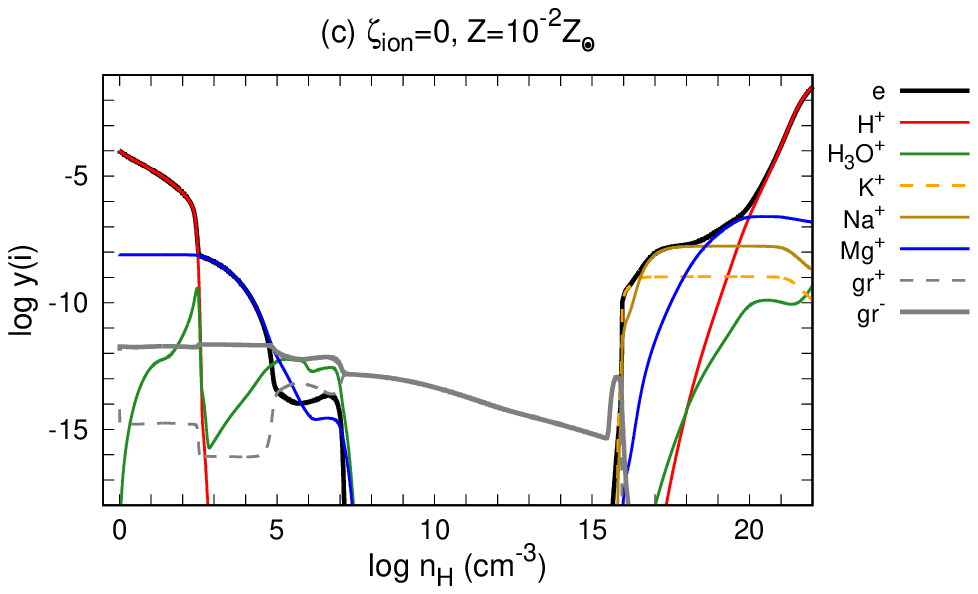}}
{\includegraphics[scale=0.8]{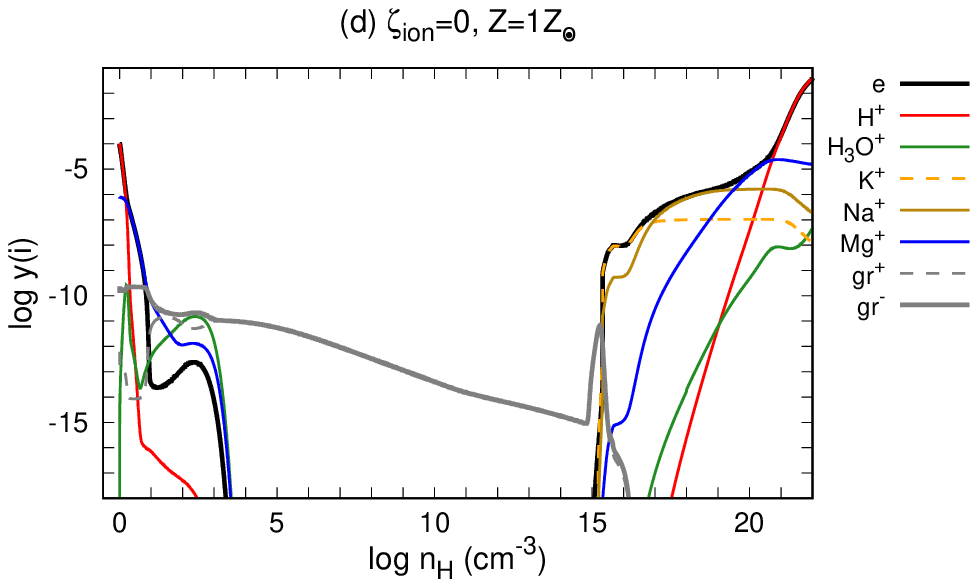}}\\
\end{tabular}
\caption{Fractional abundances of the major charged species as a function of density, for the same cases with Figure \ref{fig:cool_wo_cr}.}
\label{fig:charge_CR00}
\end{center}
\end{figure*}

In Figure \ref{fig:charge_CR00}, we show the fractional abundances of the major charged species as a function of density, for the same cases with Figure \ref{fig:cool_wo_cr}.

In the beginning of the collapse, $e$ and H$^+$ are the major charged species, and the ionization degree decreases slowly via H$^+$ radiative recombination.
At $\nH \sim 10^8, 10^5, 10^3$, and $1\pcc$~(for $Z/\zsun = 10^{-6}, 10^{-4}, 10^{-2}$, and $1$), along with the H$_2$O formation via Eq. \eqref{eq:H2O_form}~(Figure \ref{fig:chem_wo_cr} right), molecular ions such as H$_2$O$^+$ and H$_3$O$^+$ are produced via
\begin{align}
&{\rm H}^+  +   {\rm H}_2{\rm O}        \rightarrow      {\rm H}   +   {\rm H}_2{\rm O}^+    \nonumber \\
&{\rm H}_2  +   {\rm H}_2{\rm O}^+    \rightarrow      {\rm H}   +   {\rm H}_3{\rm O}^+ . 
\label{eq:H3O_form}
\end{align}
They are disrupted immediately via the dissociative recombination:
\begin{align}
&e   +   {\rm H}_3{\rm O}^+                \rightarrow      {\rm H}   +    {\rm H}_2{\rm O} \nonumber    \\
&e   +    {\rm H}_3{\rm O}^+               \rightarrow      2{\rm H} +      {\rm OH}      \nonumber\\
&e   +   {\rm H}_3{\rm O}^+                 \rightarrow      {\rm H}_2   +     {\rm OH}, 
\label{eq:H3O_recomb}
\end{align}
and the ionization degree drops rapidly.
Afterwards, the evolution of ionization degree differs between the cases with different metallicities.

In the case of $Z/\zsun=10^{-6}$~(Figure \ref{fig:charge_CR00}a), Li$^+$ takes over the major cation species at $\nH \sim 10^{10}\pcc$.
The ionization degree continues decreasing as Li$^+$ and $e$ recombine on grain surfaces:
\begin{align}
&{\rm Li}^+   +   {\rm gr}^{--}  \rightarrow   {\rm Li}  +   {\rm gr}^{-} \nonumber    \\
&e   +   {\rm gr}^{-}              \rightarrow      {\rm gr}^{--}.
\label{eq:Li_recomb}
\end{align}
The grain-surface recombination proceeds more rapidly with increasing density, and at $\nH \sim 10^{15}\pcc$, charged dust grains gr$^\pm$ become the dominant charge carriers, instead of Li$^+$ and $e$.
However, this occurs only temporarily, since the temperature soon becomes high enough~($T \sim 1500\K$) to evaporate dust grains at $\nH \sim 10^{16}\pcc$.
Along with grain vaporization, the ionization degree rises rapidly until Li is ionized completely by the thermal radiation trapped in the opaque cloud and by collision with H$_2$:
\begin{align}
&{\rm Li}    +    \gamma \rightarrow e    +    {\rm Li}^+  \nonumber   \\
&{\rm H}_2 +  {\rm Li}    \rightarrow {\rm H}_2    +     e    +    {\rm Li}^+.   
\label{eq:Li_ionize}
\end{align}
After Li ionization is completed, hydrogen ionization begins successively and the ionization degree increases monotonically with increasing temperature.

In the cases of $Z/\zsun \gtrsim 10^{-5}$~(Figure \ref{fig:charge_CR00}b-d), after the ionization degree drops rapidly via the H$_3$O$^+$ dissociative recombination~(Eq. \ref{eq:H3O_recomb}), Mg$^+$ takes over the major cation species.
The abundances of Mg$^+$ and $e$ are lowered via the grain-surface recombination:
\begin{align}
&{\rm Mg}^+   +   {\rm gr}^{-}  \rightarrow   {\rm Mg}  +   {\rm gr} \nonumber    \\
&e   +   {\rm gr}              \rightarrow      {\rm gr}^{-},
\label{eq:Mg_recomb}
\end{align}
and finally Mg$^+$ and $e$ are superseded by charged dust grains gr$^\pm$ at $\nH \sim 10^{10}, 10^6$, and $10^2\pcc$~(for $Z/\zsun = 10^{-4}, 10^{-2}$, and $1$).
The ionization degree decreases further as charged dust grains neutralize each other:
\begin{equation}
{\rm gr}^+ + {\rm gr}^- \rightarrow 2 {\rm gr}.
\label{eq:gr_recomb}
\end{equation}
When the temperature reaches $\sim 1000\K$ at $\nH \sim 10^{15}\pcc$, thermionic emission from neutral grain surfaces becomes important. 
The ejected thermal electrons are soon absorbed by other neutral grains:
\begin{align}
&{\rm gr} \rightarrow {\rm gr}^+ + e \nonumber    \\
&e   +   {\rm gr}              \rightarrow      {\rm gr}^{-}.
\label{eq:thermionic_emission}
\end{align}
Through these two processes, the abundance of charged grains gr$^\pm$ increases rapidly.
When the temperature exceeds $\sim 1400\K$, depleted heavy elements such as K and Na are released into the gas phase, along with the vaporization of silicate dust.
With high cloud temperature of a few thousand K and their low ionization potentials, these alkali metals are immediately ionized, leading to jump up of the ionization degree.
After alkali metals are ionized completely, the ionization degree is raised dominantly by H ionization until the end of the calculation.

\begin{figure*}
\begin{center}
\begin{tabular}{cc}
{\includegraphics[scale=0.8]{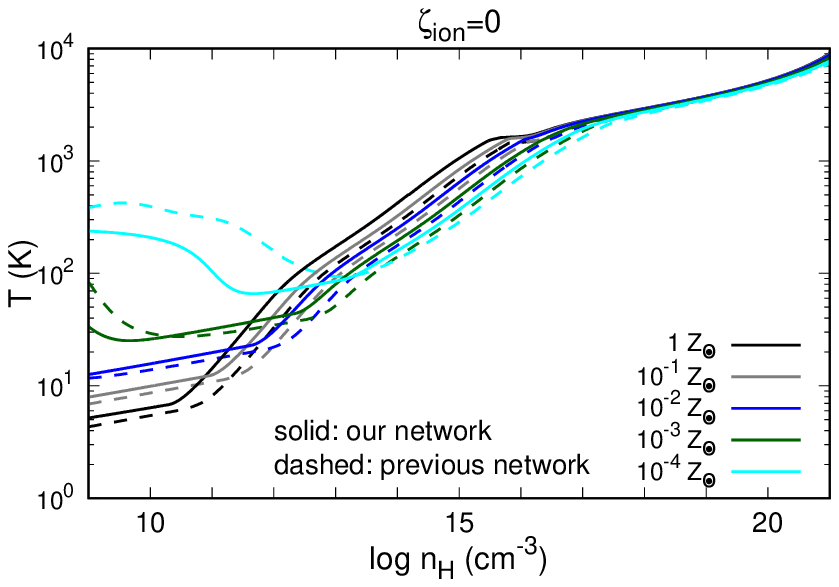}}
{\includegraphics[scale=0.8]{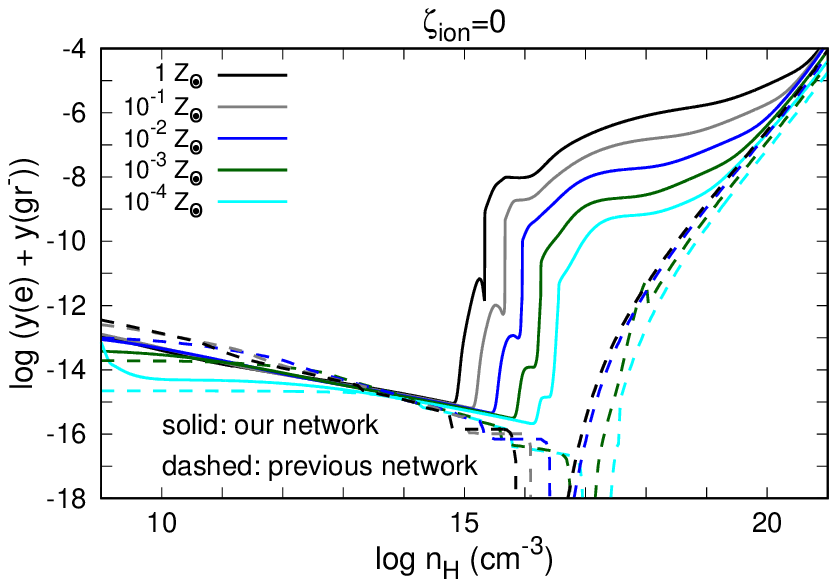}}\\
\end{tabular}
\caption{Comparison for the temperature evolution~(left panel) and the ionization degree~(right panel) calculated by our~(solid) and previous~(dashed curves) chemical networks.}
\label{fig:CR00_ion_comp}
\end{center}
\end{figure*}

Finally, we discuss the differences in the evolution of the temperature and ionization degree obtained by our and previous models~\citep{Susa2015}.
As mentioned in Section \ref{sec:intro}, previous chemical model has the following problems: (i) thermionic emission and thermal ionization of vaporized alkali metals are neglected, (ii) only a part of the reactions are reversed, and the abundances are switched artificially to the equilibrium values of the gas composed of H, H$_2$, e, H$^+$, He, He$^+$, and He$^{++}$.
In Figure \ref{fig:CR00_ion_comp}, we compare the temperature evolution~(left panel) and the ionization degree~(right panel) calculated by our~(solid) and previous~(dashed curves) chemical networks, for the models with $Z/\zsun = 10^{-4}, 10^{-3}, 10^{-2}, 10^{-1}$, and $1$.
In the left panel, the temperature evolves along the qualitatively similar track in both models, although in our model the dust cooling works from the lower densities and the cloud evolves with lower temperatures at $\nH < 10^{13}\pcc$~(or $10^{10}\pcc$) for $Z/\zsun = 10^{-4}$~(or $10^{-3}$, respectively).
In the right panel, the ionization degree in both models decreases slowly following almost the same track up to $\nH \sim 10^{15}\pcc$.
At $\nH \sim 10^{15}\pcc$, the ionization degree in our model jumps up via thermionic emission and thermal ionization of vaporized K and Na.
At higher densities, the ionization degree increases slowly, following the chemical equilibrium abundances between all the 63 gas-phase species.
On the other hand, due to the lack of these processes, the ionization degree in the previous model keeps decreasing until the dust grains evaporate completely.
After that, the ionization degree increases following the equilibrium value of the H/He gas.
From Figure \ref{fig:CR00_ion_comp}, we find the ionization degree at $\nH \sim 10^{15}\mbox{-}10^{19}\pcc$ higher by up to eight orders of magnitude than that predicted in the previous model.

\subsection{Cases with ionization sources}
\label{subsec:zeta_nonzero}

Next, we describe the effect of ionization sources on the temperature evolution and ionization degree.
CR particles propagating in a cloud lose their energy via the gas ionization, and are thus attenuated when the density becomes higher than $\nH \sim 10^{11}\pcc$.
Therefore, CR ionization dominates only at low densities and is superseded by radioactive ionization at $\nH \sim 10^{11}\pcc$.

\subsubsection{Temperature Evolution}
\begin{figure*}
\begin{center}
\begin{tabular}{cc}
{\includegraphics[scale=0.8]{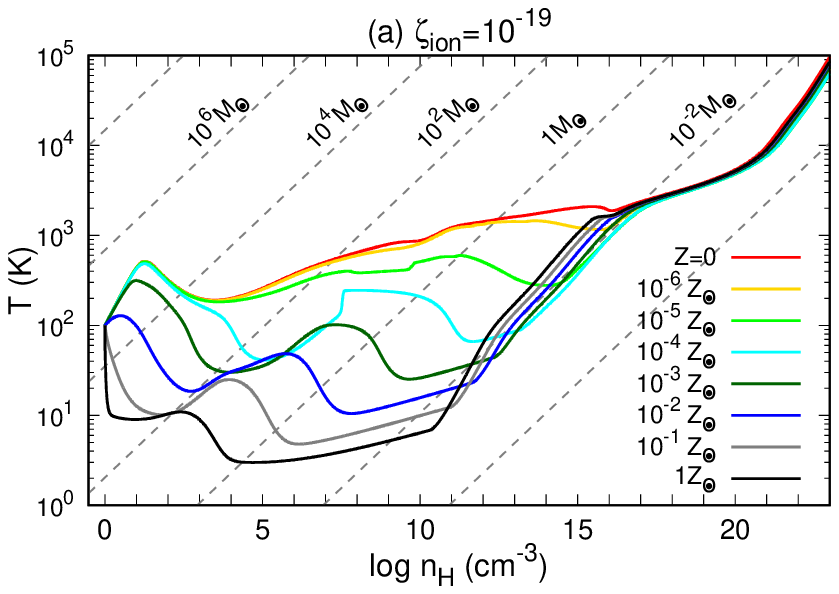}}
{\includegraphics[scale=0.8]{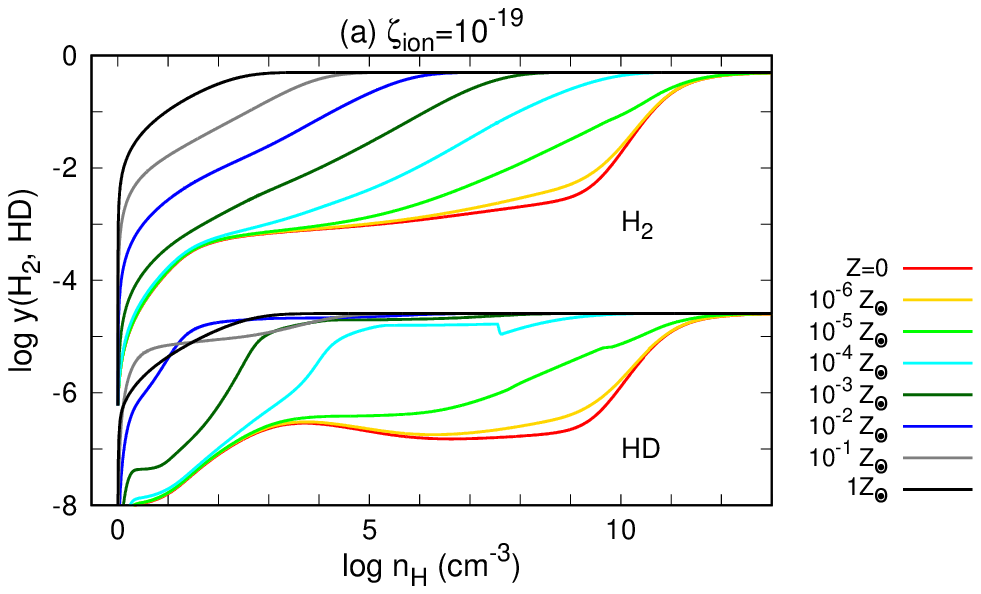}}\\
{\includegraphics[scale=0.8]{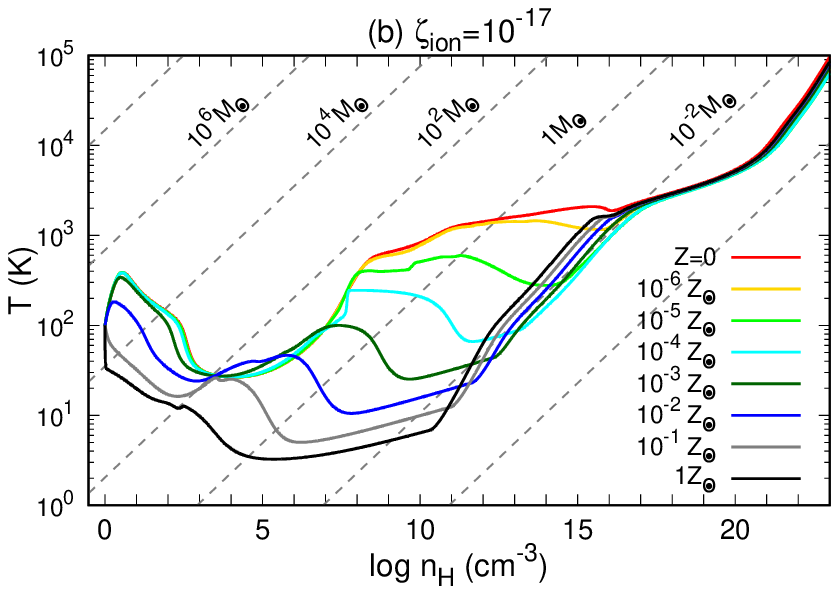}}
{\includegraphics[scale=0.8]{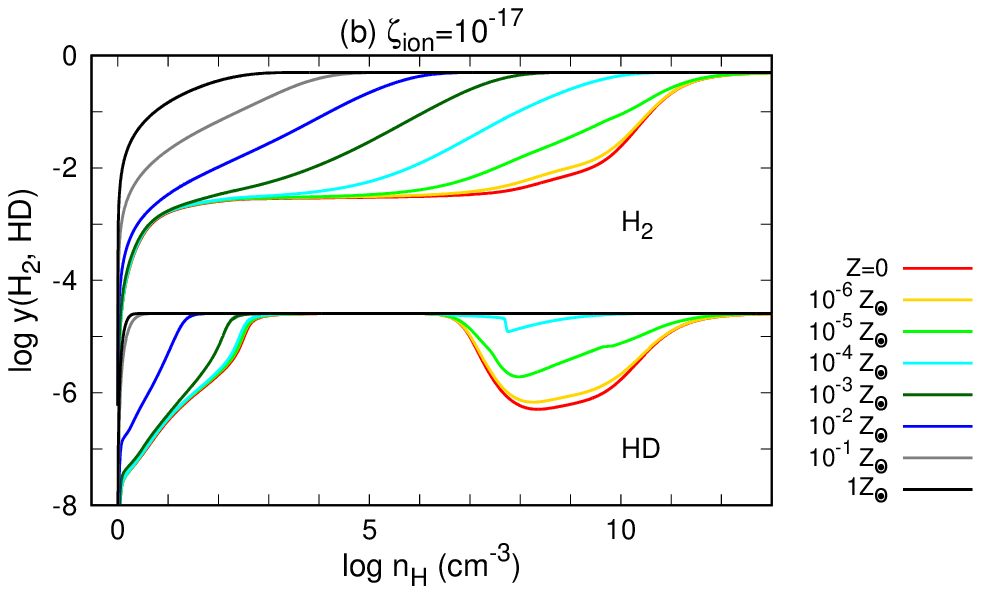}}\\
{\includegraphics[scale=0.8]{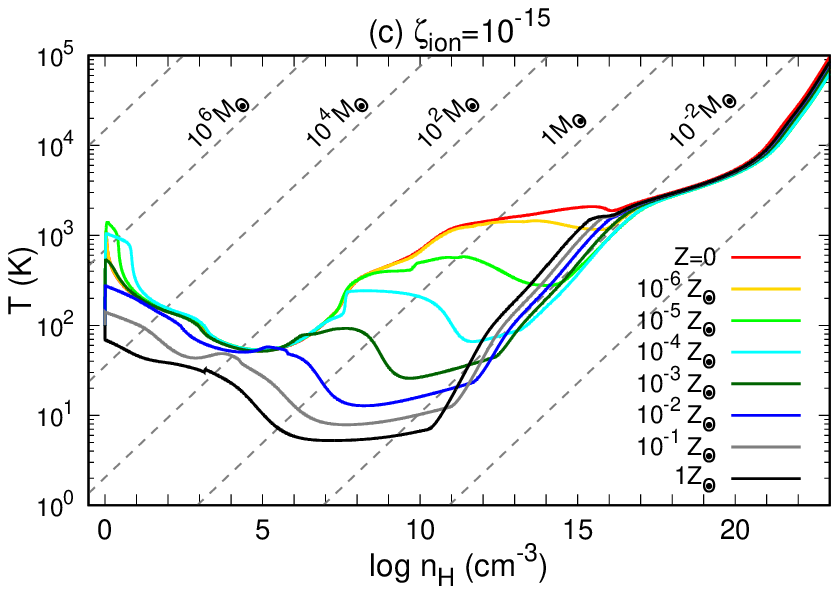}}
{\includegraphics[scale=0.8]{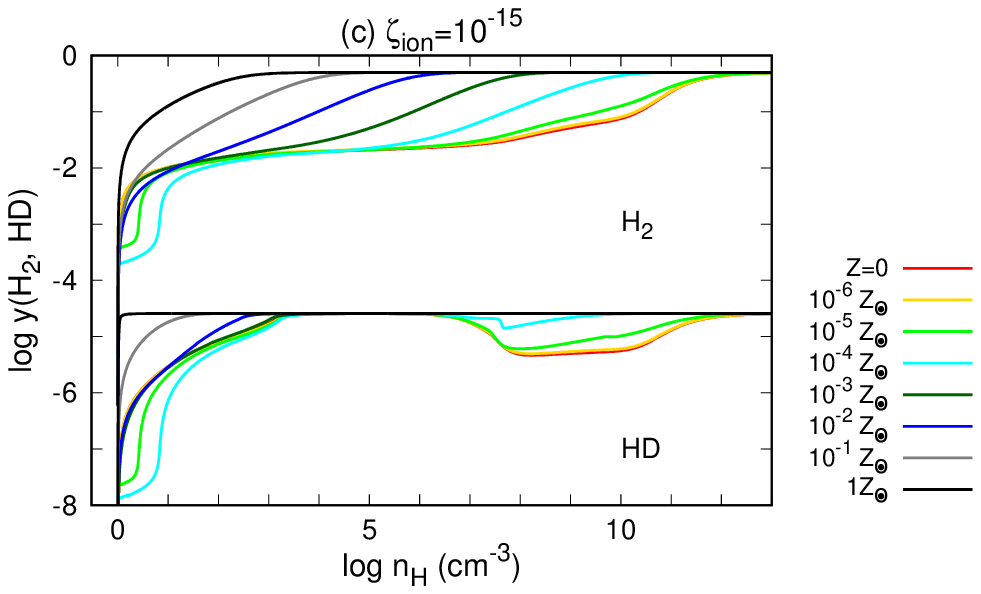}}
\end{tabular}
\caption{Effects of ionization sources on the temperature evolution~(left column) and H$_2$ and HD abundances~(right column) in the clouds with different metallicities.
Individual panels correspond to the results with $\zeta_{\rm ion}=$ (a) $10^{-19}$, (b) $10^{-17}$, and (c) $10^{-15}\ps$, respectively.}
\label{fig:nT_w_cr}
\end{center}
\end{figure*}


In Figure \ref{fig:nT_w_cr}, we show the effects of ionization sources on the temperature evolution~(left column) and the H$_2$ and HD abundances~(right column) for the cases with $\zeta_{\rm ion}=$ (a) $10^{-19}$, (b) $10^{-17}$, and (c) $10^{-15}\ps$, respectively.
Ionization sources not only heat the gas directly, but also enhance the H$_2$-cooling efficiency by activating H$_2$ formation via the electron-catalyzed H$^-$ channel~(Eq. \ref{eq:H-channel}).
Both effects are observed at low densities of $\nH \lesssim 10^{11}\pcc$, where the CR ionization dominates over radioactive ionization.

The direct heating by CR ionization is observed in all the metallicity cases.
As the CR intensity becomes stronger, the cloud evolves with higher temperatures at $\nH \sim 1\mbox{-}10^3\pcc$.
On the other hand, the effect of enhanced H$_2$ cooling appears only in the low-metallicity cases of $Z/\zsun \lesssim 10^{-3}$.
In these cases, the H$_2$ abundance at each density becomes higher with increasing $\zetaion$~(Figure \ref{fig:nT_w_cr} right).
Once the cloud cools via the enhanced H$_2$ cooling up to $T \lesssim 150\K$, HD formation kicks in.
Due to the efficient HD cooling, these low-metallicity clouds~($Z/\zsun \lesssim 10^{-3}$) evolve along a similar track at $\nH \lesssim 10^8\pcc$.
At higher densities, the temperature evolution is controlled by H$_2$ formation heating and dust cooling, both of which are independent of the ionization degree.
Therefore, the temperature evolution afterwards does not depend on the presence of ionization sources.

\subsubsection{Ionization degree}
\begin{figure*}
\begin{center}
\begin{tabular}{cc}
{\includegraphics[scale=0.8]{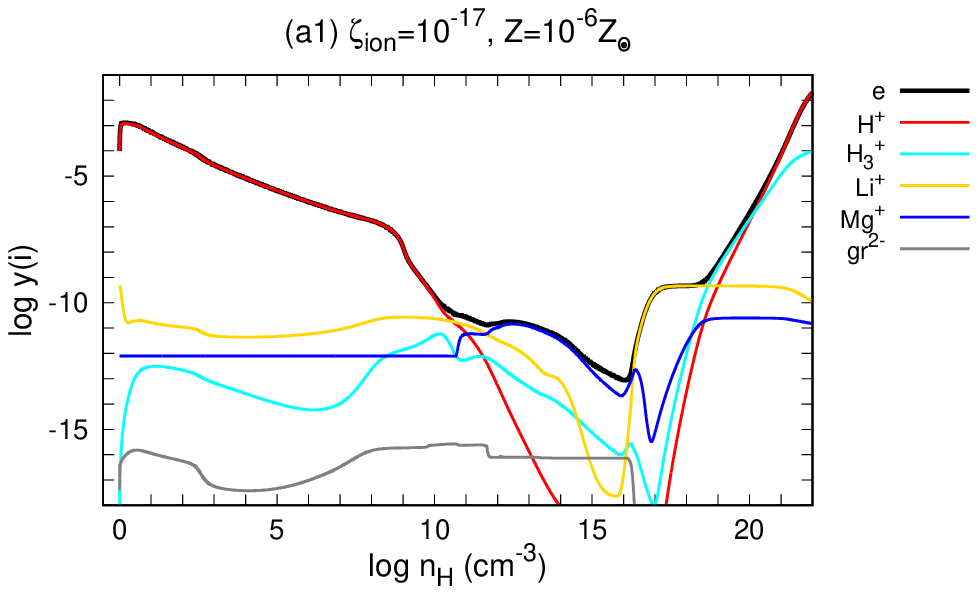}}
{\includegraphics[scale=0.8]{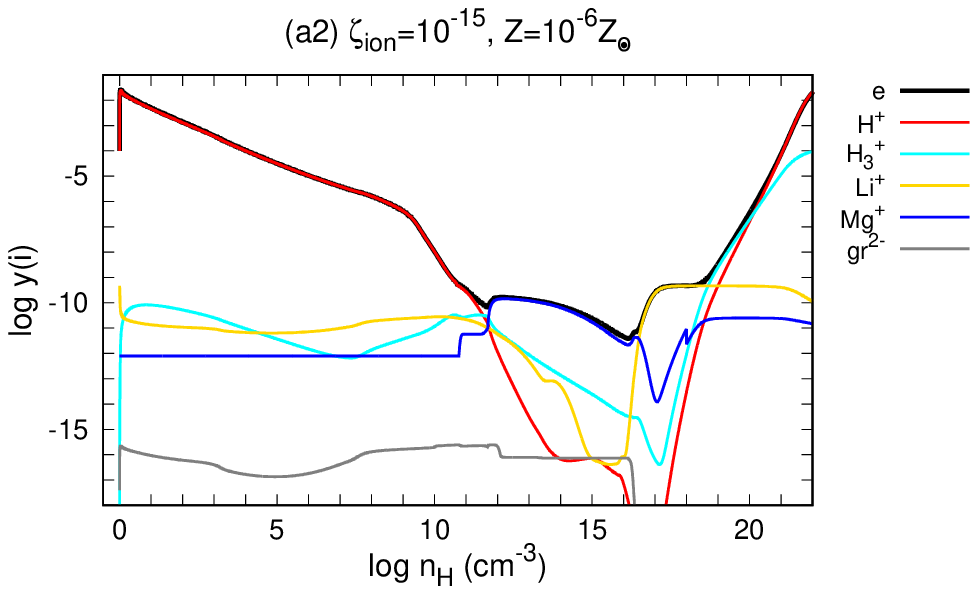}}\\
{\includegraphics[scale=0.8]{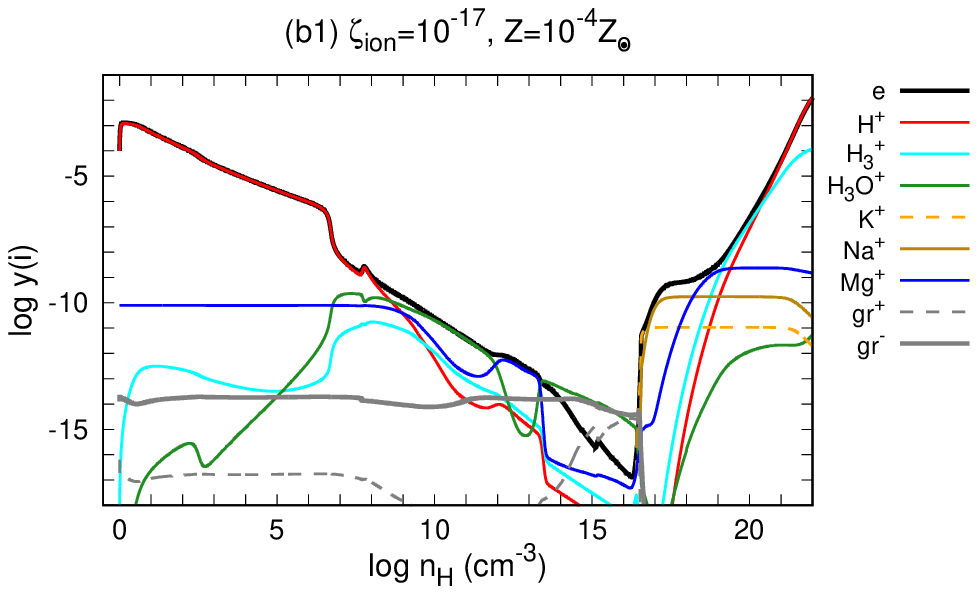}}
{\includegraphics[scale=0.8]{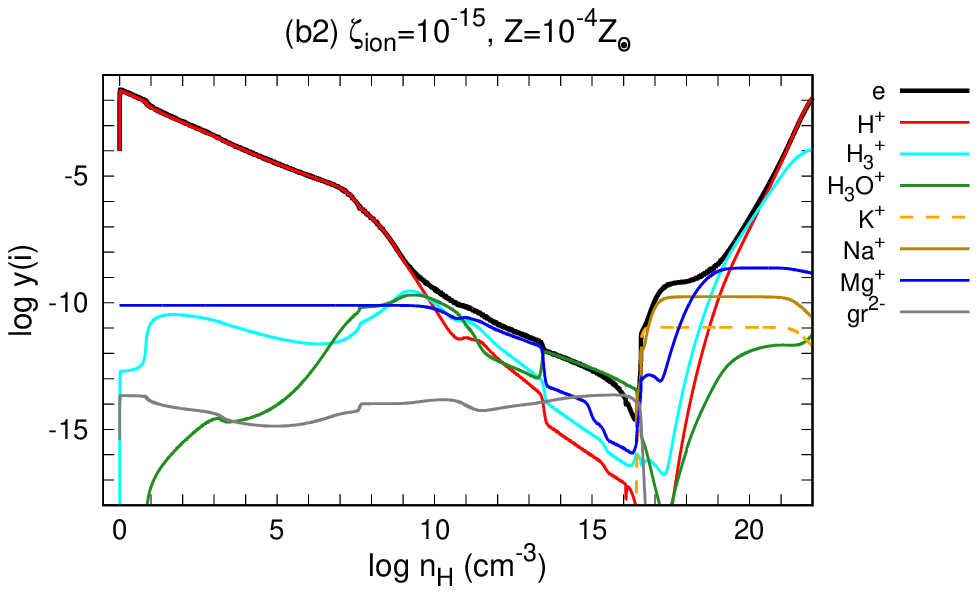}}\\
{\includegraphics[scale=0.8]{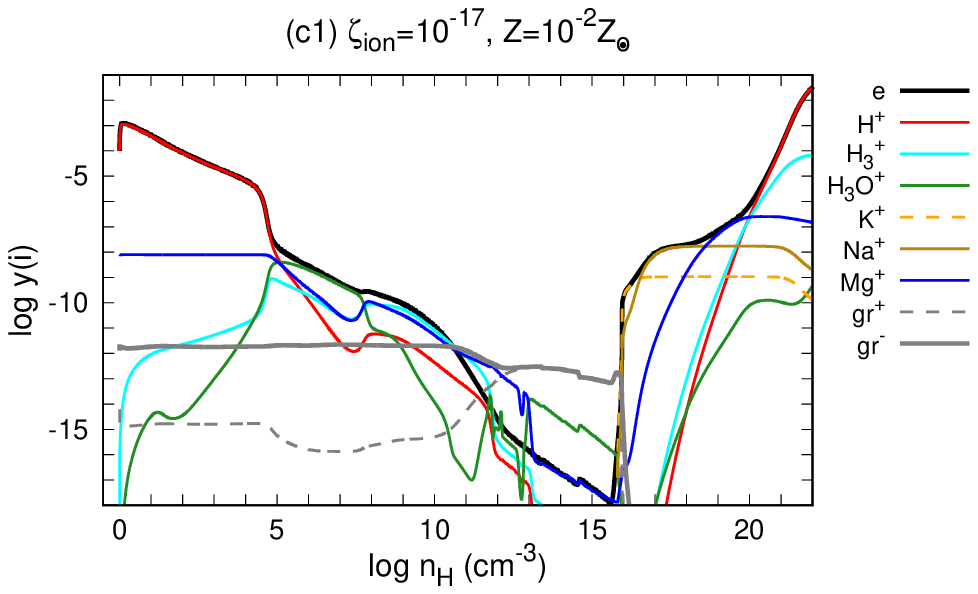}}
{\includegraphics[scale=0.8]{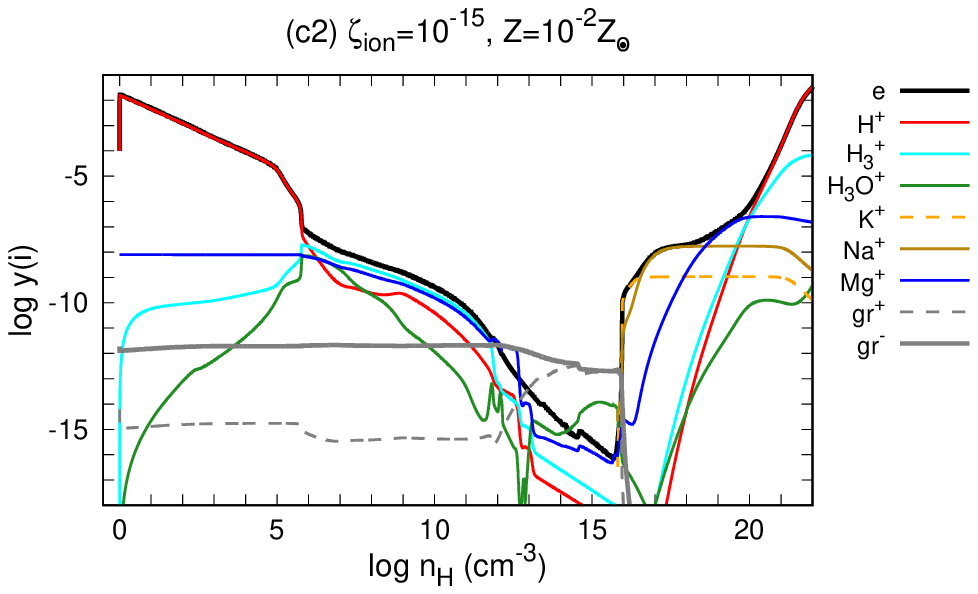}}\\
{\includegraphics[scale=0.8]{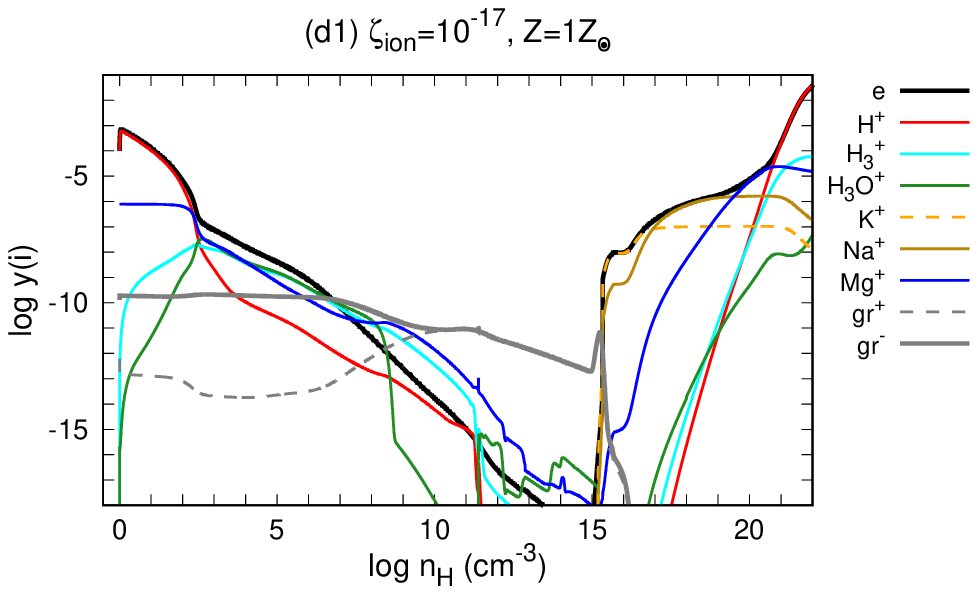}}
{\includegraphics[scale=0.8]{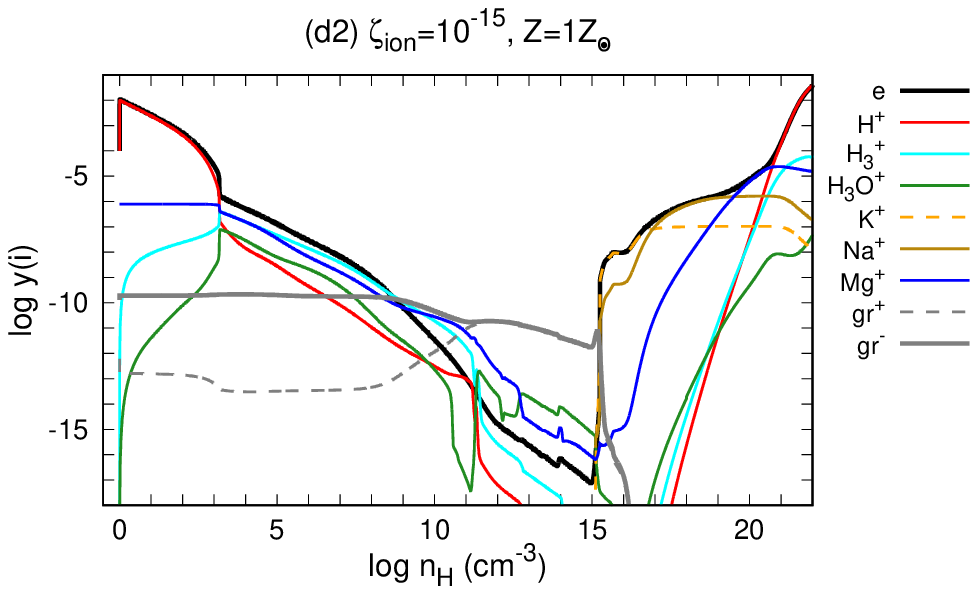}}\\
\end{tabular}
\caption{Same as Figure \ref{fig:charge_CR00}, but for the models with \crfid~(left column) and \crmax~(right column).}
\label{fig:charge_w_cr}
\end{center}
\end{figure*}

\begin{figure}
\begin{center}
\includegraphics[scale=0.3,angle=-90]{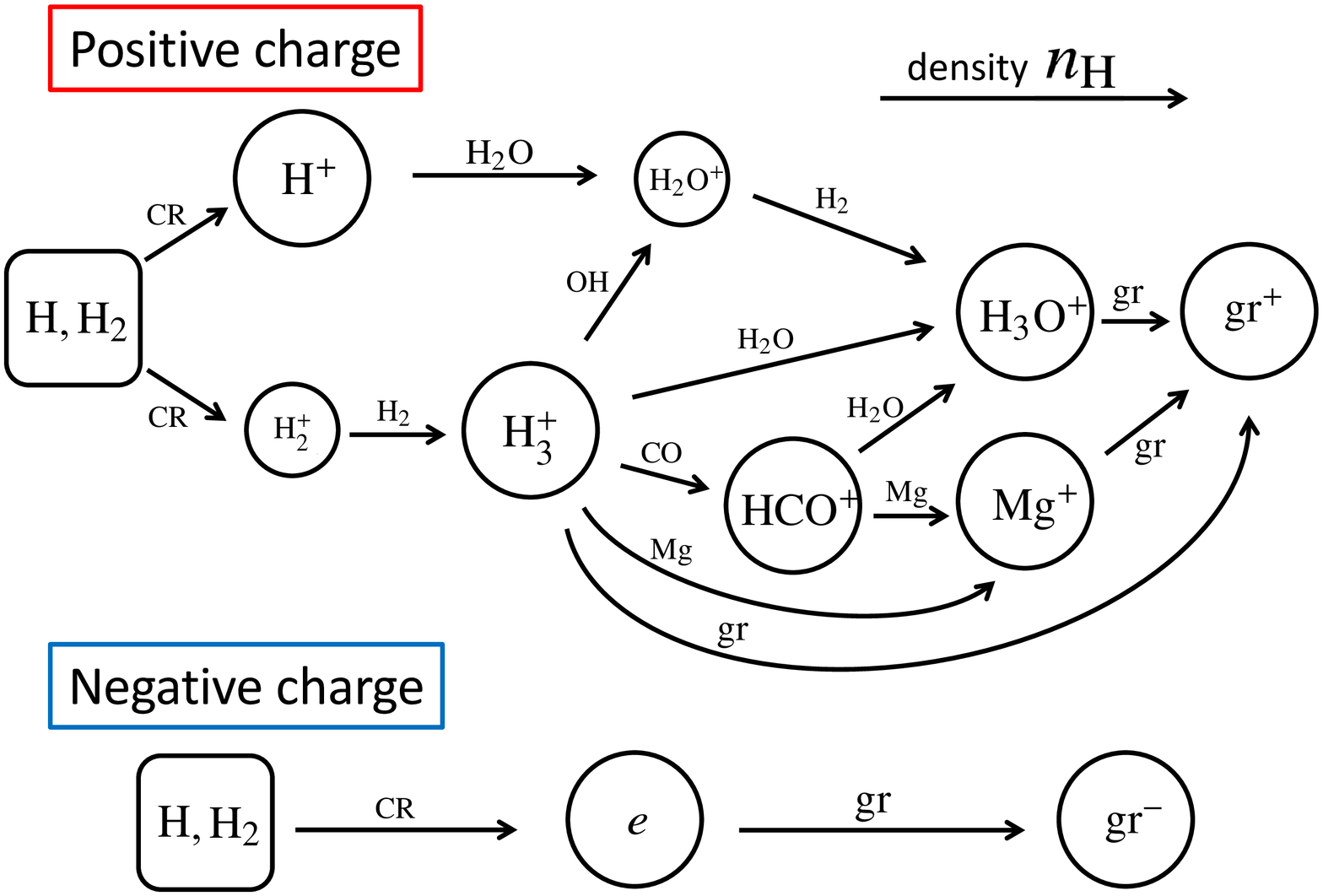}
\caption{Schematic diagram representing the transition of the major charged species~(species enclosed by large circles) with increasing density. Species enclosed by small circles indicate the intermediate products of the major charged species.}
\label{fig:flow_chart}
\end{center}
\end{figure}

\begin{figure*}
\begin{center}
\begin{tabular}{cc}
{\includegraphics[scale=0.8]{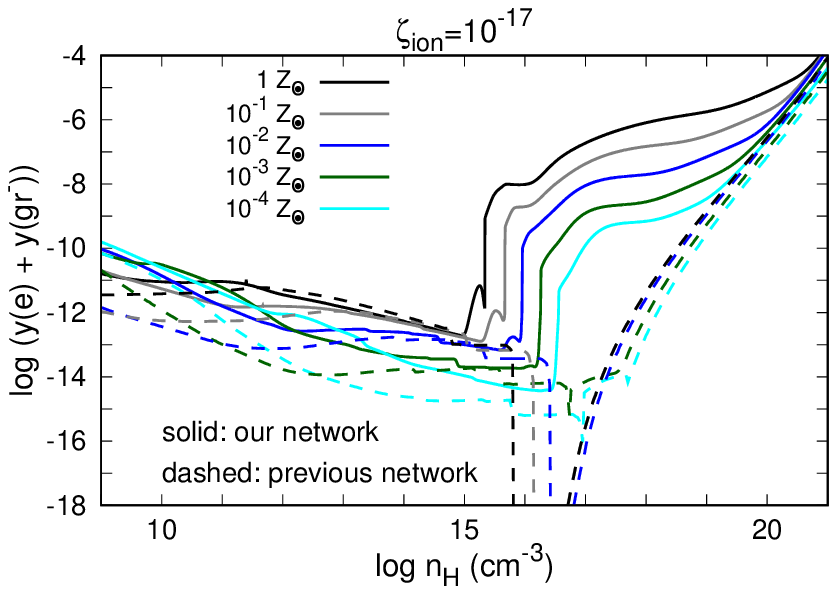}}
{\includegraphics[scale=0.8]{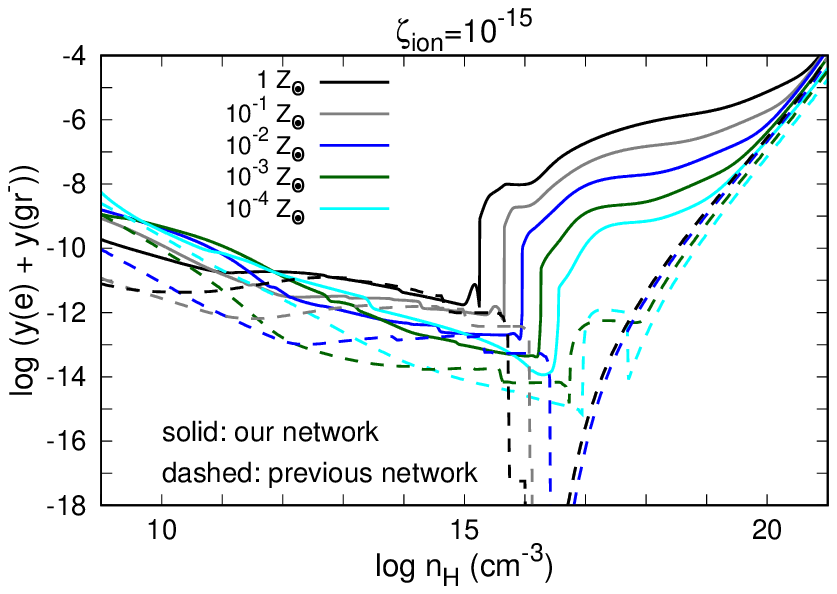}}\\
\end{tabular}
\caption{Comparison of the fractional ionization calculated by our~(solid) and previous~(dashed curves) chemical networks for the cases with \crfid~(left panel) and with \crmax~(right panel). In both models, the temperature evolution is completely same as the case with \crzero~for $\nH > 10^9\pcc$~(Figure \ref{fig:CR00_ion_comp} left panel) and is not shown.}
\label{fig:CR_ion_comp}
\end{center}
\end{figure*}


In Figure \ref{fig:charge_w_cr}, the evolution of the major charged species is plotted for the same metallicity cases with Figure \ref{fig:charge_CR00}, but for the different ionization rates of \crfid~(left column) and \crmax~(right column), respectively.
According to Figure \ref{fig:charge_w_cr}, the ionization degree takes a higher value at each density with increasing ionization rate.
Gas-phase ions and electrons remain as major charge carriers over a wider range of densities compared to the case without ionization.
Charged dust grains never become dominant in the very low-metallicity cases of $Z/\zsun \lesssim 10^{-5}$~(or $10^{-4}$) for \crfid~(or \crmax, respectively).
With ionization sources, a positive charge is carried successively by various species H$^+$, H$_3^+$, H$_3$O$^+$, HCO$^+$, Mg$^+$, and gr$^+$, whereas a negative charge is carried only by $e$ and gr$^-$.
The transition in the major charged species is represented by a schematic diagram in Figure \ref{fig:flow_chart}, and is described extensively below.

In the beginning of the collapse, electrons are produced via the CR ionization of hydrogen atoms and removed via the radiative recombination of H$^+$.
As a result of the balance between these processes, the ionization degree decreases slowly as
\begin{equation}
y(e) \simeq y({\rm H}^+) \simeq \left(\frac{\zetaion}{k_{\rm H, rec} \nH}\right)^{1/2}.
\label{eq:ye_H}
\end{equation}
CR particles also ionize H$_2$, produced via the H$^-$ channel~(Eq. \ref{eq:H-channel}) or grain-surface reactions:
\begin{equation}
{\rm H}_2    +    {\rm CR}              \rightarrow                  e   +    {\rm H}_2^+.
\label{eq:H2_CR}
\end{equation}
The resultant H$_2^+$ immediately reacts with H$_2$ and produces H$_3^+$ via~\citep{Oppenheimer1974}:
\begin{equation}
{\rm H}_2  +     {\rm H}_2^+          \rightarrow          {\rm H}  +     {\rm H}_3^+.    
\label{eq:H3p_form}
\end{equation}
As the formation of H$_2$O and CO proceeds, positive charges are passed from H$_3^+$ to other molecular ions, H$_3$O$^+$ and HCO$^+$, via the following ion-molecule reactions:
\begin{align}
&{\rm H}_3^+    +    {\rm H}_2{\rm O}  \rightarrow                 {\rm H}_2   +   {\rm H}_3{\rm O}^+,    \nonumber \\
&{\rm H}_3^+    +     {\rm CO}              \rightarrow                {\rm H}_2   +   {\rm HCO}^+,  \nonumber \\
&{\rm H}_2{\rm O} + {\rm HCO}^+ \rightarrow     {\rm CO}   +    {\rm H}_3{\rm O}^+.      
\label{eq:molecular_ion_form}
\end{align}
Positive charges are also transferred to metallic ion Mg$^+$ via the following reactions:
\begin{align}
&{\rm H}_3^+    +    {\rm Mg}  \rightarrow    {\rm H} + {\rm H}_2 + {\rm Mg}^+,    \nonumber \\
&{\rm HCO}^+   +    {\rm Mg}  \rightarrow    {\rm HCO} + {\rm Mg}^+.      
\label{eq:metal_ion_form}
\end{align}
Through these processes, positive charges are carried by molecular ions m$^+$~(=~H$_3^+$, H$_3$O$^+$, HCO$^+$) and metallic ion Mg$^+$, instead of H$^+$.

At the densities where molecular ions are dominant cation species, electrons are removed by the dissociative recombination of m$^+$~(e.g., Eq. \ref{eq:H3O_recomb}), whereas they are produced by CR or radioactive ionization of H$_2$.
Through the balance between these processes, the ionization degree decreases with contraction as
\begin{equation}
y(e) \simeq y({\rm m}^+) \simeq \left(\frac{\zetaion}{k_{\rm m, rec} \nH}\right)^{1/2}.
\label{eq:ye_m}
\end{equation}

Dust grains obtain a negative charge by capturing electrons provided via CR or radioactive ionization of H$_2$.
Grains also obtain a positive charge by colliding with molecular ions or metallic ions.
Grains can become dominant charge carriers, if the ionization degree determined by Eq. \eqref{eq:ye_m} decreases below the total grain fraction $y_{\rm gr} = 2.6 \times 10^{-10} Z/\zsun$ before the grain evaporation at $\nH \sim 10^{17}\pcc$.
This condition can be expressed with respect to metallicity as
\begin{equation}
Z/\zsun \simeq 10^{-5} \left(\frac{\zetaion}{10^{-17}\ps}\right)^{1/2},
\label{eq:z_crit}
\end{equation}
where $k_{\rm m, rec} = 3 \times 10^{-7}\ {\rm cm}^3\ {\rm s}^{-1}$ is adopted.
This result is consistent with the numerical calculation.
When the metallicity is higher than Eq. \eqref{eq:z_crit}, dust grains become the dominant population of charged species~(Figures \ref{fig:charge_w_cr} c1-d2).

When charged grains are prevalent, negative grains are produced via electron capture of neutral grains and are destroyed via mutual neutralization:
\begin{align}
&e + {\rm gr} \rightarrow {\rm gr}^- \nonumber\\
&{\rm gr}^+ + {\rm gr}^- \rightarrow 2 {\rm gr}.
\label{eq:e_capture}
\end{align}
These processes balance each other, controlling their abundances as
\begin{equation}
y_{{\rm gr}^-} \simeq y_{{\rm gr}^+} \simeq \left(\frac{k_{e-{\rm gr}}}{k_{{\rm gr}^+-{\rm gr}^-}} y({\rm gr}) y(e)\right)^{1/2}.
\label{eq:ygrp_1}
\end{equation}
Since the electron fraction is determined by the balance between the radioactive ionization of H$_2$ and electron capture of neutral grains, Eq. \eqref{eq:ygrp_1} is rewritten as
\begin{equation}
y_{{\rm gr}^-} \simeq y_{{\rm gr}^+} \simeq \left(\frac{\zetaion}{k_{{\rm gr}^+-{\rm gr}^-} \nH}\right)^{1/2}.
\label{eq:ym_cr}
\end{equation}
Charged grain abundances decrease with increasing density, following Eq. \eqref{eq:ym_cr}.

Once the temperature exceeds $T \simeq 1500\K$ at $\nH \sim 10^{15}\mbox{-}10^{17}\pcc$, dust grains evaporate, and the ionization degree increases rapidly via thermal ionization of vaporized alkali metals.
The thermal ionization proceeds much more rapidly than radioactive ionization.
Therefore, the presence of the ionization sources has no effect on the ionization degree afterwards.

Finally, we discuss the differences of the above results from those of the previous models~\citep{Susa2015}.
In Figure \ref{fig:CR_ion_comp}, we compare the fractional ionization calculated by our~(solid) and previous~(dashed curves) chemical networks for the cases with \crfid~(left panel) and with \crmax~(right panel).
For $\nH > 10^9\pcc$, CR or radioactive ionization works as minor heating process, so that the temperature evolution in both models is the same as in the case with \crzero~(Figure \ref{fig:CR00_ion_comp} left panel).
Before the grain evaporation at $\nH \sim 10^{15}\pcc$, the fractional ionization in both models decreases slowly in a qualitatively similar way.
At the densities where ions and electrons are the dominant charge carriers, the ionization degree in our model takes higher values, owing to the updated rate coefficients for the gas-phase process.
Along with the grain evaporation at $\nH \sim 10^{15}\pcc$, the ionization degree in our model jumps up via thermal ionization of K and Na, while that in the previous model drops immediately below $<10^{-18}$.
Afterwards, the ionization degree does not depend on the radioactive ionization rate and evolves along the same track as in the case with \crzero~(Figure \ref{fig:CR00_ion_comp} right panel).
As a result, the ionization degree in our model is up to eight orders of magnitude higher than that in the previous model, irrespective of the CR or radioactive ionization rate.

\subsection{Reduced chemical networks}
\begin{table*}
\caption{Reduced chemical network.}
\begin{center}
{\begin{tabular}{llllll}
\hline
Number & Reaction & Number & Reaction & Number & Reaction \\
\hline

         R1     & $   e   +      {\rm H}^+           \rightleftharpoons              {\rm H}     +   \gamma       $                    &
         R37$^\ast$     & $  {\rm H}    +    {\rm H}^+     \rightleftharpoons          {\rm H}_2^+   +      \gamma $         &
       R67    & $                e    +     {\rm gr}                             \rightleftharpoons                          {\rm gr}^- $     \\         

         R2     & $  {\rm H}    +     e                \rightleftharpoons              {\rm H}^-    +     \gamma    $                   &
        R38$^\ast$    & $   {\rm H}   +    {\rm H}_2^+  \rightleftharpoons                 {\rm H}_2     +   {\rm H}^+ $    &
       R68    & $                e    +    {\rm gr}^+                           \rightleftharpoons           {\rm gr} $                       \\

         R3     & $  {\rm H}    +    {\rm H}^-     \rightleftharpoons                 {\rm H}_2     +    e $                             & 
        R39$^\ast$    & $ {\rm H}_2  +     {\rm H}_2^+  \rightleftharpoons               {\rm H}  +     {\rm H}_3^+ $     &
       R69    & $                e  +      {\rm gr}^-                             \rightleftharpoons                         {\rm gr}^{--} $  \\

        R4    & $  3{\rm H}                             \rightleftharpoons                  {\rm H}     +   {\rm H}_2 $                  &
        R40$^\ast$    & $  e     +   {\rm H}_3^+                \rightleftharpoons                  {\rm H}  +      {\rm H}_2 $     &
       R70   & $                {\rm H}^+    +    {\rm gr}          \rightarrow                  {\rm H}   +    {\rm gr}^+ $     \\

        R5    & $   2{\rm H}_2                        \rightleftharpoons        2{\rm H}    +    {\rm H}_2 $     &
        R41$^\ast$    & $  e     +   {\rm H}_3^+                   \rightleftharpoons        3{\rm H} $     &
       R71$^\ast$   & $               {\rm H}_3^+    +     {\rm gr}        \rightarrow        {\rm H}    +    {\rm H}_2     +  {\rm gr}^+ $     \\

        R6    & $   2{\rm H} + {\rm grain}         \rightarrow                           {\rm H}_2 $     &
       R42$^\ast$  & $   {\rm H}_3^+   +       {\rm O}              \rightleftharpoons                 {\rm H}_2    +   {\rm OH}^+ $     &
        R72$^\ast$   & $                {\rm C}^+    +     {\rm gr}            \rightarrow             {\rm C}    +    {\rm gr}^+ $     \\

        R7    & $ {\rm H}   +     {\rm HD}                \rightleftharpoons        2{\rm H} +       {\rm D} $     &
       R43$^\ast$  & $   {\rm H}_3^+    +     {\rm OH}              \rightleftharpoons                 {\rm H}_2    +  {\rm H}_2{\rm O}^+ $     &
         R73   & $              {\rm H}_3{\rm O}^+    +     {\rm gr} \rightarrow       {\rm H}_2     +   {\rm OH}  +     {\rm gr}^+ $     \\
 
        R8    & $  {\rm H}^+   +      {\rm D}            \rightleftharpoons                  {\rm H}     +   {\rm D}^+ $     &
       R44$^\ast$  & $     {\rm H}_3^+    +    {\rm H}_2{\rm O}  \rightleftharpoons                 {\rm H}_2   +   {\rm H}_3{\rm O}^+ $    &
       R74    & $              {\rm Li}^+   +    {\rm gr}                \rightarrow                 {\rm Li}   +    {\rm gr}^+ $     \\
 
        R9   & $  {\rm H}_2    +     {\rm D}             \rightleftharpoons                  {\rm H}     +   {\rm HD} $     &
       R45$^\ast$    & $    {\rm H}_3^+    +     {\rm CO}             \rightleftharpoons                 {\rm H}_2   +   {\rm HCO}^+ $     &
       R75    & $              {\rm Mg}^+    +     {\rm gr}              \rightarrow                 {\rm Mg}  +     {\rm gr}^+ $     \\
 
        R10    & $  {\rm H}_2    +    {\rm D}^+           \rightleftharpoons                 {\rm H}^+     +   {\rm HD} $     &
       R46$^\ast$  & $        {\rm H}_3^+      +    {\rm O}               \rightleftharpoons                  {\rm H}   +   {\rm H}_2{\rm O}^+ $    &
       R76   & $               {\rm H}^+    +   {\rm gr}^-                \rightarrow                  {\rm H}    +    {\rm gr} $     \\
 
       R11  & $    {\rm H}    +     {\rm D}    + {\rm grain}           \rightarrow                           {\rm HD} $     &
         R47$^\ast$     & $   e   +     {\rm He}^+          \rightleftharpoons              {\rm He}   +     \gamma     $        &
       R77$^\ast$    & $              {\rm H}_3^+   +     {\rm gr}^-            \rightarrow        {\rm H}     +   {\rm H}_2    +    {\rm gr} $     \\
 
       R12  & $   {\rm H}   +  {\rm OH}                   \rightleftharpoons {\rm O}   + {\rm H}_2 $       & 
        R48$^\ast$    & $  {\rm H}_2  +     {\rm He}^+  \rightleftharpoons        {\rm H}   +     {\rm H}^+    +    {\rm He}$     &
       R78$^\ast$    & $              {\rm He}^+    +    {\rm gr}^-              \rightarrow                 {\rm He}   +     {\rm gr} $     \\
 
       R13  & $  {\rm H}_2     +   {\rm OH}              \rightleftharpoons                  {\rm H}   +    {\rm H}_2{\rm O} $     &
        R49$^\ast$    & $ {\rm H}    +   {\rm He}^+           \rightleftharpoons                 {\rm H}^+    +    {\rm He} $     &
       R79$^\ast$    & $               {\rm C}^+     +   {\rm gr}^-                 \rightarrow                  {\rm C}      +   {\rm gr} $     \\
 
       R14  & $    {\rm C}     +    {\rm OH}              \rightleftharpoons                  {\rm H}    +    {\rm CO} $     &
        R50$^\ast$    & $ {\rm H}_2   +    {\rm He}^+        \rightleftharpoons                {\rm H}_2^+    +     {\rm He} $    &
       R80    & $             {\rm H}_3{\rm O}^+     +   {\rm gr}^-  \rightarrow       {\rm H}_2    +    {\rm OH}    +    {\rm gr} $     \\
 
       R15  & $   {\rm H}   +      {\rm O}                   \rightleftharpoons                 {\rm OH}    +    \gamma $     &
       R51$^\ast$   & $    {\rm He}^+    +    {\rm H}_2{\rm O}  \rightleftharpoons        {\rm H}    +    {\rm He}  +     {\rm OH}^+ $     &
       R81    & $              {\rm Li}^+   +     {\rm gr}^-                  \rightarrow                 {\rm Li}     +   {\rm gr} $     \\
 
       R16   & $    {\rm H}     +   {\rm OH}                        \rightleftharpoons                {\rm H}_2{\rm O}    +     \gamma $    &
       R52$^\ast$   & $    {\rm He}^+    +     {\rm CO}               \rightleftharpoons       {\rm He}  +      {\rm C}^+      +    {\rm O} $     &
       R82    & $              {\rm Mg}^+    +    {\rm gr}^-                \rightarrow                 {\rm Mg}    +    {\rm gr} $     \\
 
       R17  & $   {\rm H}^+   +     {\rm OH}              \rightleftharpoons                  {\rm H}    +   {\rm OH}^+ $     &
       R53$^\ast$  & $      e    +     {\rm C}^+                                  \rightleftharpoons                  {\rm C}     +    \gamma $    &
       R83    & $               {\rm H}^+    +  {\rm gr}^{--}                 \rightarrow                  {\rm H}    +   {\rm gr}^- $     \\
 
       R18  & $   {\rm H}^+    +   {\rm H}_2{\rm O}  \rightleftharpoons                  {\rm H}   +   {\rm H}_2{\rm O}^+ $     &
       R54$^\ast$  & $    {\rm C}     +    {\rm O}_2              \rightleftharpoons                  {\rm O}    +     {\rm CO} $     &
       R84$^\ast$    & $              {\rm H}_3^+    +   {\rm gr}^{--}                \rightarrow        {\rm H}     +   {\rm H}_2   +    {\rm gr}^- $     \\
 
       R19  & $     {\rm H}     +   {\rm O}^+                   \rightleftharpoons                 {\rm H}^+   +      {\rm O} $     &
       R55$^\ast$  & $        {\rm O}      +   {\rm OH}                 \rightleftharpoons                  {\rm H}    +    {\rm O}_2 $     &
       R85$^\ast$    & $               {\rm C}^+    +   {\rm gr}^{--}                    \rightarrow                  {\rm C}    +    {\rm gr}^- $     \\
 
       R20  & $     {\rm H}_2    +    {\rm O}^+                \rightleftharpoons                  {\rm H}    +   {\rm OH}^+ $    &
       R56$^\ast$    & $    {\rm H}_3^+   +      {\rm Mg}                 \rightleftharpoons        {\rm H}    +    {\rm H}_2    +   {\rm Mg}^+ $     &
       R86    & $             {\rm H}_3{\rm O}^+   +    {\rm gr}^{--}   \rightarrow       {\rm H}_2     +   {\rm OH}    +   {\rm gr}^- $     \\
 
       R21   & $      {\rm H}_2   +    {\rm OH}^+             \rightleftharpoons                  {\rm H}  +    {\rm H}_2{\rm O}^+ $     &
       R57$^\ast$   & $    {\rm C}^+    +     {\rm Mg}                     \rightleftharpoons                  {\rm C}    +    {\rm Mg}^+ $     &
       R87      & $            {\rm Li}^+   +    {\rm gr}^{--}                    \rightarrow                 {\rm Li}   +    {\rm gr}^- $     \\
 
       R22   & $      {\rm H}_2   +   {\rm H}_2{\rm O}^+    \rightleftharpoons                  {\rm H}   +   {\rm H}_3{\rm O}^+ $     &
       R58$^\ast$   & $  {\rm HCO}^+    +     {\rm Mg}                       \rightleftharpoons                {\rm HCO}    +    {\rm Mg}^+ $     &
       R88      & $            {\rm Mg}^+   +    {\rm gr}^{--}                      \rightarrow                 {\rm Mg}   +    {\rm gr}^- $     \\
 
       R23   & $     {\rm H}_2{\rm O}    +   {\rm HCO}^+  \rightleftharpoons                 {\rm CO}   +    {\rm H}_3{\rm O}^+ $     &
       R59$^\ast$  & $    {\rm H}   +    {\rm HCO}              \rightleftharpoons                 {\rm H}_2   +     {\rm CO} $     &
       R89    & $              {\rm gr}^+   +     {\rm gr}^-                    \rightarrow                 {\rm gr}    +    {\rm gr} $     \\
 
       R24  & $     e    +    {\rm OH}^+                              \rightleftharpoons                  {\rm H}    +     {\rm O} $     &
       R60$^\dagger$ & $         {\rm H}    +    {\rm CH}  \rightleftharpoons               {\rm H}_2      + {\rm C}                 $     &    
       R90    & $              {\rm gr}^+    +   {\rm gr}^{--}                      \rightarrow                 {\rm gr}  +     {\rm gr}^- $     \\
 
       R25  & $    e    +   {\rm H}_2{\rm O}^+                 \rightleftharpoons                  {\rm H}   +     {\rm OH} $    &
       R61$^\dagger$ & $         {\rm H}    +    {\rm CH}_2  \rightleftharpoons                {\rm H}_2     + {\rm CH}               $     &
       R91$^\ast$    & $                          {\rm O}  \rightleftharpoons                         {\rm O(p)} $     \\
 
       R26  & $    e  +  {\rm H}_2{\rm O}^+                     \rightleftharpoons                 {\rm H}_2      +   {\rm O} $     &
       R62$^\dagger$  & $     {\rm H}   + {\rm C}  \rightleftharpoons     {\rm CH}   + \gamma     $     &
      R92$^\ast$   & $                           {\rm C}  \rightleftharpoons                         {\rm C(p)} $     \\
 
       R27   & $          e    +   {\rm H}_2{\rm O}^+              \rightleftharpoons        2{\rm H} +        {\rm O} $     &
       R63$^\dagger$  & $     {\rm H}_2   + {\rm C}  \rightleftharpoons     {\rm CH}_2   + \gamma     $     &
       R93$^\ast$    & $                         {\rm OH} \rightleftharpoons                        {\rm OH(p)} $     \\
 
       R28  & $     e    +   {\rm H}_3{\rm O}^+                \rightleftharpoons                  {\rm H}   +    {\rm H}_2{\rm O} $     &
       R64$^\dagger$ & $  {\rm CH}   +  {\rm O}  \rightleftharpoons     {\rm H}  +        {\rm CO}           $     &
      R94$^\ast$   & $                          {\rm CO}  \rightleftharpoons                       {\rm CO(p)} $     \\
 
       R29   & $          e    +   {\rm H}_3{\rm O}^+              \rightleftharpoons                 {\rm H}_2   +     {\rm OH} $     &
       R65$^\dagger$ & $  {\rm CH}   +  {\rm O}  \rightleftharpoons      e   +    {\rm HCO}^+        $     &
      R95$^\ast$   & $                         {\rm H}_2{\rm O} \rightleftharpoons                       {\rm H}_2{\rm O(p)} $     \\
 
       R30  & $      e   +    {\rm H}_3{\rm O}^+             \rightleftharpoons        2{\rm H} +      {\rm OH} $     &
       R66$^\dagger$ & $    {\rm C}  +      {\rm H}_3{\rm O}^+  \rightleftharpoons        {\rm H}_2  + {\rm HCO}^+   $     &
       CR1$^\ast$   & $      {\rm H}    +    {\rm CR}                   \rightarrow                  e      +   {\rm H}^+ $     \\

       R31  & $    e    +   {\rm HCO}^+                           \rightleftharpoons                  {\rm H}   +     {\rm CO} $     &
                  &                                                                                                                                                                        &
       CR2$^\ast$   & $     {\rm He}    +    {\rm CR}                    \rightarrow                  e      +  {\rm He}^+ $     \\
 
       R32   & $       e    +    {\rm Mg}^+                            \rightleftharpoons                 {\rm Mg}    +    \gamma $     &
                  &                                                                                                                                                                        &
       CR3$^\ast$  & $      {\rm H}_2    +    {\rm CR}               \rightarrow        {\rm H}    +     e     +    {\rm H}^+ $     \\
 
       R33   & $      {\rm H}^+    +    {\rm Mg}                      \rightleftharpoons                  {\rm H}    +   {\rm Mg}^+ $     &
                  &                                                                                                                                                                        &
       CR4$^\ast$   & $       {\rm H}_2    +    {\rm CR}              \rightarrow                  e   +    {\rm H}_2^+ $     \\
 
       R34   & $      e    +    {\rm Li}^+                                    \rightleftharpoons                 {\rm Li}    +    \gamma $     &
                  &                                                                                                                                                                        &
       CR5$^\ast$   & $      {\rm H}   +     {\rm CR ph}                      \rightarrow                e      +   {\rm H}^+ $     \\
 
       R35   & $     {\rm H}^+   +     {\rm Li}                           \rightleftharpoons        {\rm H}   +    {\rm Li}^+    +     \gamma $     &
                  &                                                                                                                                                                        &
       CR6$^\ast$   & $      {\rm He}   +     {\rm CR ph}                   \rightarrow                  e     +   {\rm He}^+ $\\
 
       R36   & $        {\rm H}_2    +    {\rm Li}                             \rightleftharpoons       {\rm H}_2    +     e    +    {\rm Li}^+ $     &
                  &                                                                                                                                                                        &
       CR7$^\ast$   & $     {\rm C}     +    {\rm CR ph}                     \rightarrow                  e     +    {\rm C}^+ $      \\
 
                  &                                                                                                                                                                        &
                  &                                                                                                                                                                        &
       CR8$^\ast$   & $     {\rm O}_2    +     {\rm CR ph}                 \rightarrow                2{\rm O} $          \\
\hline
\end{tabular}}
\end{center}
{\bf Notes.} Reactions with asterisks~(or daggers) are needed only in the presence~(or absence, respectively) of ionization sources.
\label{tab:chem_react}
\end{table*}


\begin{figure*}
\begin{center}
\begin{tabular}{lll}
{\includegraphics[scale=0.7]{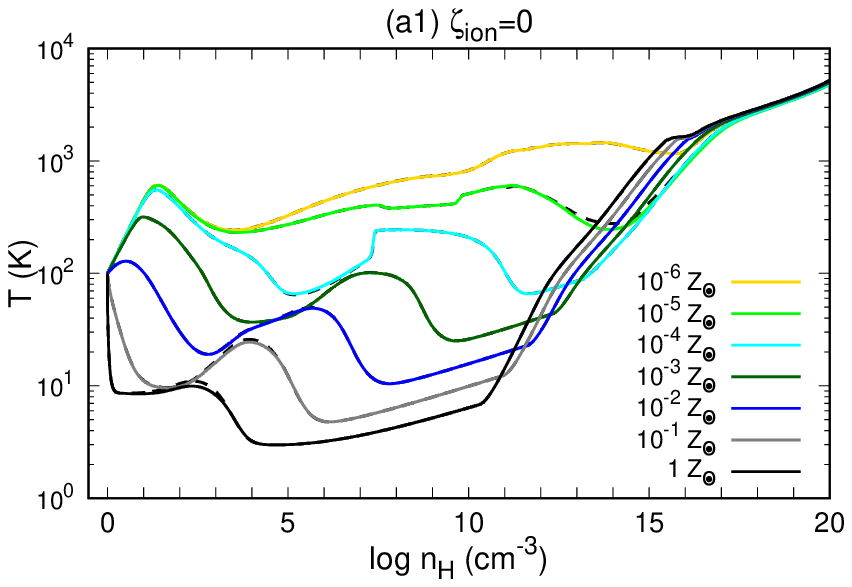}}
{\includegraphics[scale=0.7]{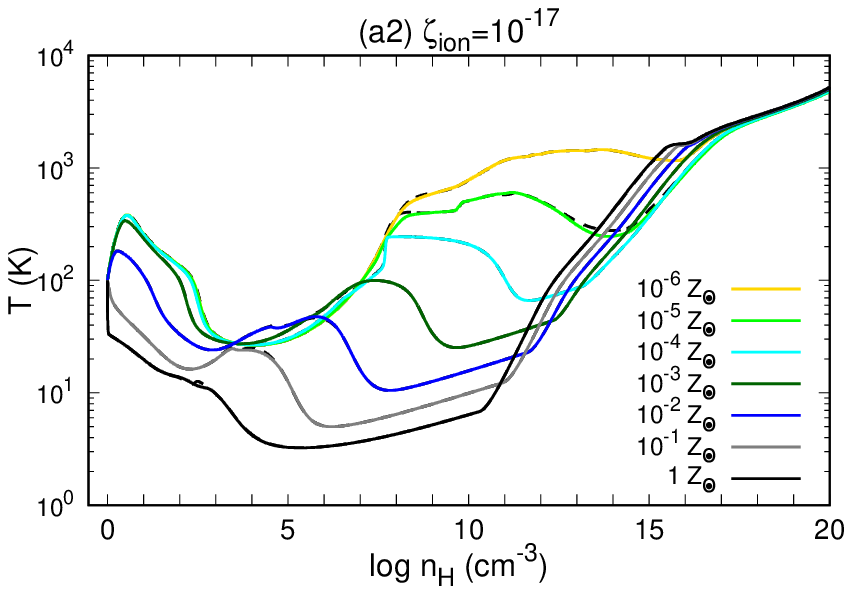}}
{\includegraphics[scale=0.7]{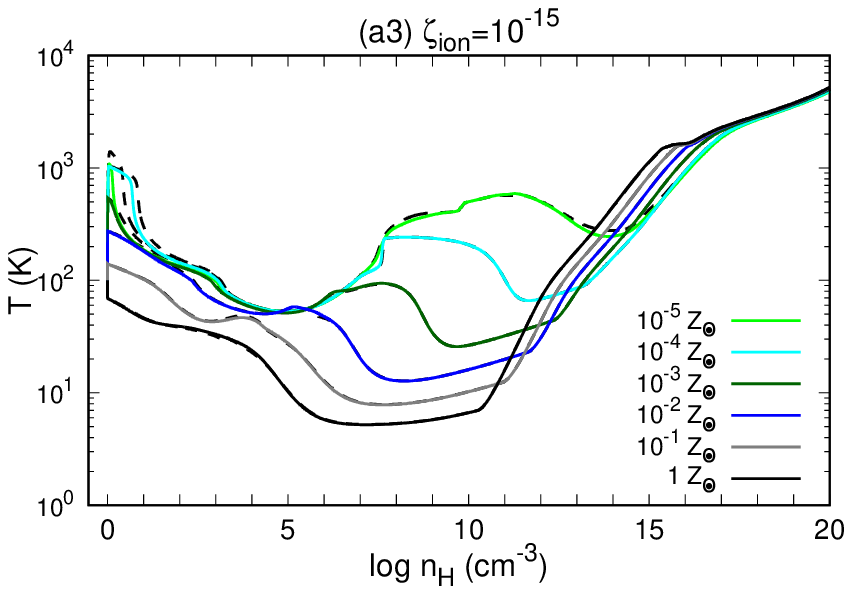}}\\
{\includegraphics[scale=0.7]{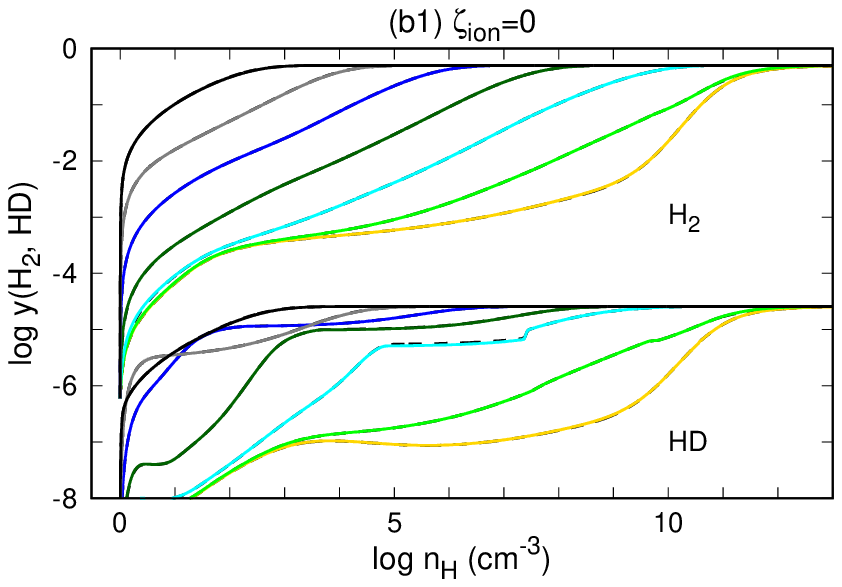}}
{\includegraphics[scale=0.7]{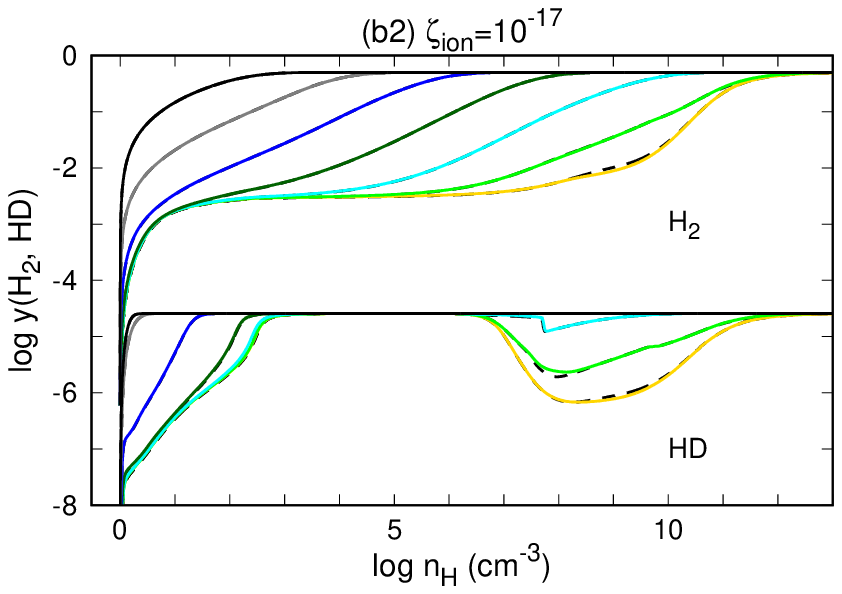}}
{\includegraphics[scale=0.7]{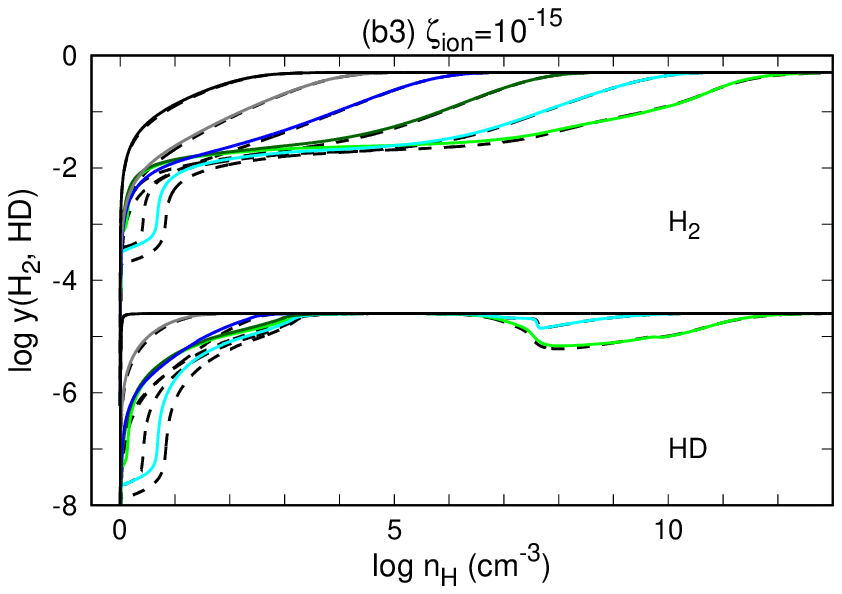}}\\
{\includegraphics[scale=0.7]{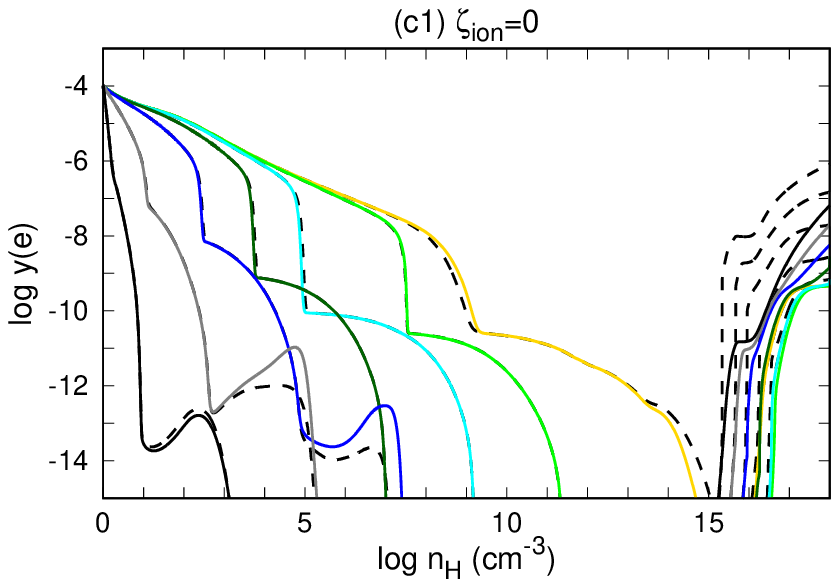}}
{\includegraphics[scale=0.7]{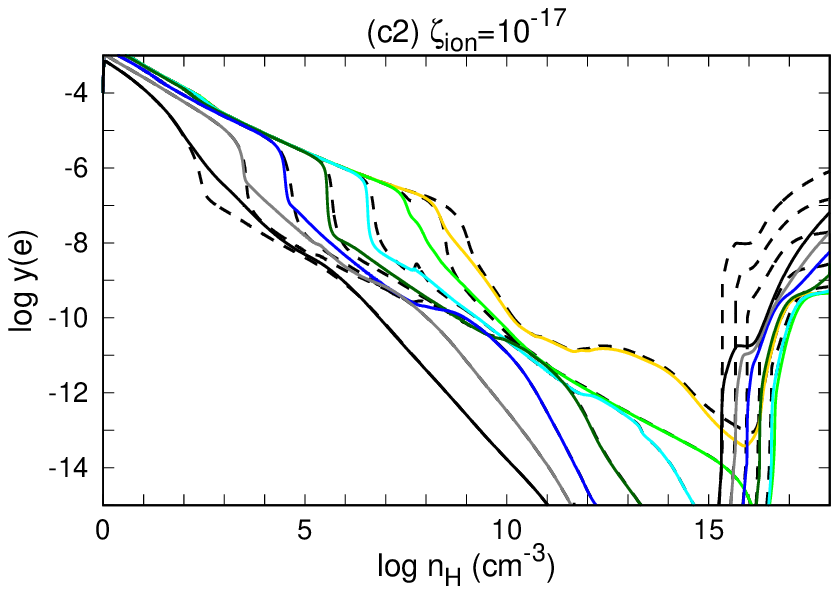}}
{\includegraphics[scale=0.7]{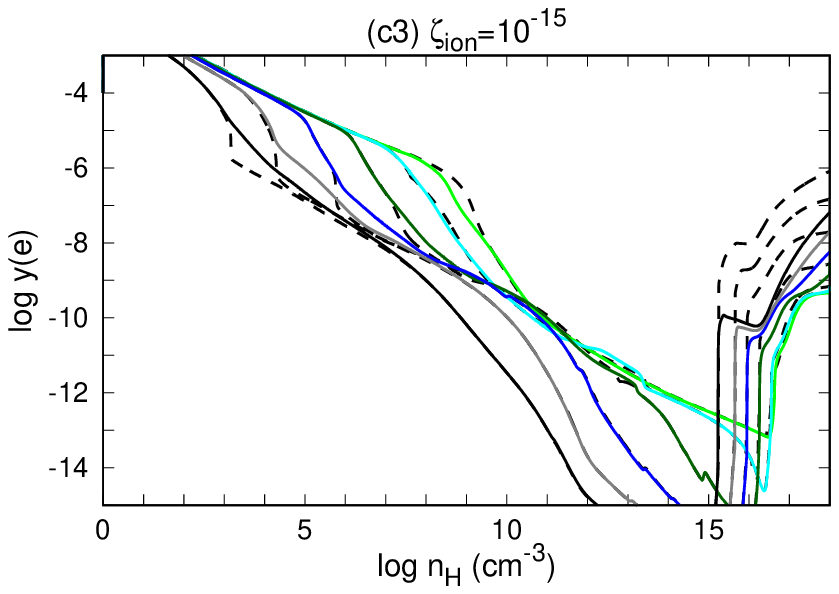}}\\
{\includegraphics[scale=0.7]{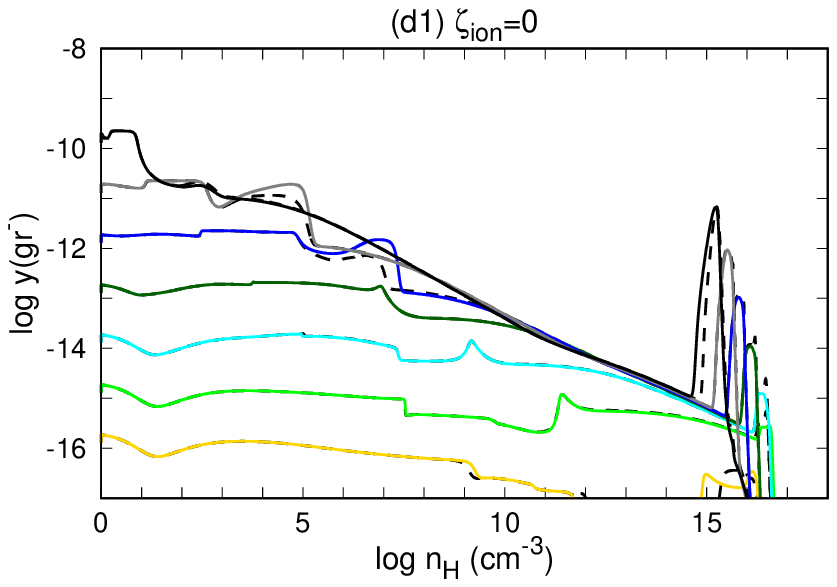}}
{\includegraphics[scale=0.7]{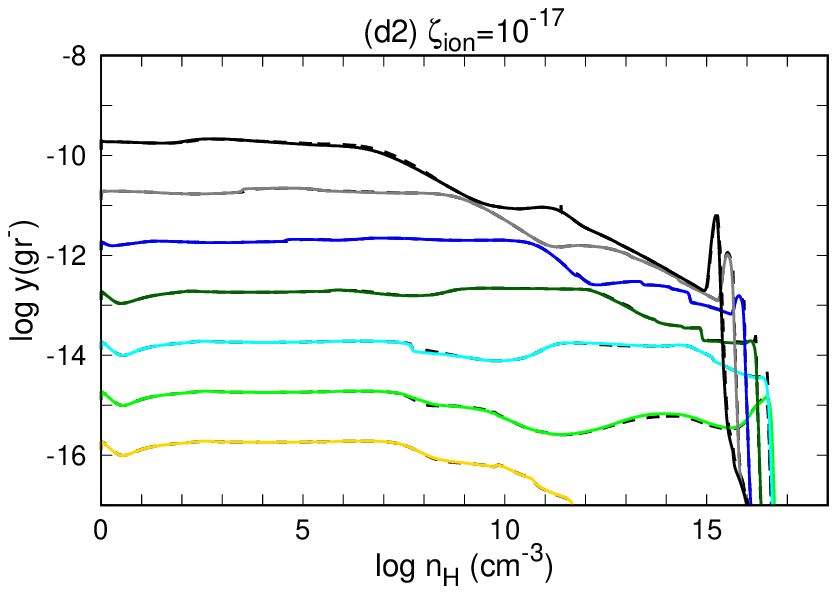}}
{\includegraphics[scale=0.7]{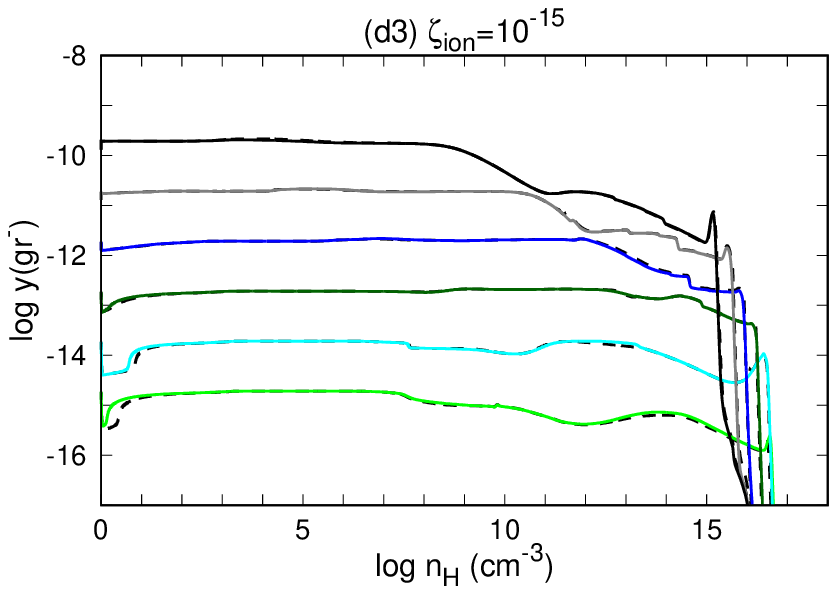}}\\
\end{tabular}
\caption{Comparison of the (a) temperature evolution and the abundances of (b) H$_2$ and HD, (c) $e$, and (d) gr$^-$, which are calculated by the reduced~(solid) and full~(dashed curves) networks, for the cases with \crzero~(left panels), \crfid~(middle panels), and \crmax~(right panels).}
\label{fig:redchem}
\end{center}
\end{figure*}

The full chemical network developed above involves a large number of species~(89) and reactions~(1482).
It requires great computational resources to implement the full network into multi-D (M)HD calculations.
To save the computational time, we develop a reduced network to reproduce the fractional abundances of the main coolants and charged species, by identifying the important processes associated with these species.
The list of the reactions included in the reduced network is shown in Table \ref{tab:chem_react}.
H$_2$ and HD formation on grain surfaces~(reactions R6 and R11) is implemented by using the simple formulae of Eqs. \eqref{eq:k_gr_H2} and \eqref{eq:k_gr_HD}.
In Figure \ref{fig:redchem}, we compare the (a) temperature evolution and the abundances of (b) H$_2$ and HD, (c) $e$, and (d) gr$^-$, which are calculated by using the reduced~(solid) and full~(dashed curves) chemical networks, for the cases with \crzero~(left panels), \crfid~(middle panels), and \crmax~(right panels).

Without ionization sources~(\crzero), the reduced network consists of 104 reactions~(reactions with no symbols and daggers in Table \ref{tab:chem_react}) among the following 28 species:
\noindent H, H$_2$, $e$, H$^+$, H$^-$, D,  D$^+$, HD, C, CH, CH$_2$,
O, OH, CO, H$_2$O, O$^+$, OH$^+$, H$_2$O$^+$, HCO$^+$, H$_3$O$^+$,
Li, Li$^+$, Mg, Mg$^{+}$, gr$^0$, gr$^+$, gr$^-$, gr$^{--}$.
The following six species, H$^-$, D$^+$, O$^+$, OH$^+$, H$_2$O$^+$, and HCO$^+$, appear as the intermediate products of the major chemical species, and their abundances are replaced by the chemical equilibrium values.

According to the left panels in Figure \ref{fig:redchem}, the temperature evolution and the abundances of H$_2$ and HD are reproduced almost completely by the reduced network.
The abundances of $e$ and gr$^-$ in the reduced model follows those in the full model over most of the range of densities except at $\nH \sim 10^3\mbox{-}10^5\pcc$~(or $10^5\mbox{-}10^7\pcc$) for $Z/\zsun = 10^{-1}$~(or $10^{-2}$, respectively), where $y(e)$ and $y({\rm gr}^-)$ are overestimated by up to an order of magnitude and by a factor of a few, respectively.
These deviations can be reduced, if the freeze-out of C and O on grain surfaces~(reactions R91 and R92) is taken into account.
However, these deviations are observed only at a limited range of density and metallicity, so that the freeze-out of C and O is omitted in the reduced model with \crzero.
At $\nH > 10^{15}\pcc$, the electron fraction is underestimated in the reduced model by up to three orders of magnitude, owing to the lack of the collisional ionization of K and Na~(Eq. \ref{eq:therm_ionization}).
Even in this case, the ionization degree is high enough to couple magnetic fields coherent over the cloud size to the gas, so that these reactions are negligible, as long as focusing on ordered magnetic fields.

With ionization sources, the reduced network consists of 161 reactions~(reactions with no symbols and asterisks in Table \ref{tab:chem_react}) among the following 38 species:
\noindent H, H$_2$, $e$, H$^+$, H$_2^+$, H$_3^+$, H$^-$, He, He$^+$
D,  D$^+$, HD, C, C$^+$, O, O$_2$, OH, CO, H$_2$O, HCO, O$^+$, OH$^+$, H$_2$O$^+$, HCO$^+$, H$_3$O$^+$,
Li, Li$^+$, Mg, Mg$^{+}$, gr$^0$, gr$^+$, gr$^-$, gr$^{--}$, C(p), O(p), OH(p), CO(p), H$_2$O(p).
For the following eight species, H$_2^+$, H$^-$, D$^+$, HCO, O$^+$, OH$^+$, H$_2$O$^+$, and HCO$^+$, the abundances can be given by the equilibrium values.

According to the middle and right panels in Figure \ref{fig:redchem}, the temperature evolution as well as the abundances of H$_2$, HD, and gr$^-$ are reproduced almost completely by the reduced model for both cases with \crfid and \crmax.
The electron fraction is also reproduced over most of the range of densities.
Deviations from the full model appear only a limited range of densities and are regulated below an order of magnitude, in each metallicity case.
With ionization sources, molecular ions, H$_3$O$^+$ and HCO$^+$, which are formed via the reactions R44 and R45, respectively, can be the dominant cations at some densities~(Figure \ref{fig:charge_w_cr}).
The abundances of H$_3$O$^+$ and HCO$^+$ depend on those of H$_2$O and CO, which also depend on those of O, C, and OH via the reactions R13-R16.
Most of these neutral atoms and molecules are depleted on the grain surface at high densities. 
Therefore, the freeze-out of O, C, OH, CO, and H$_2$O~(reaction R91-R95) is included in the reduced model with ionization sources.

\vspace{10mm}
\section{Magnetic field dissipation in star-forming clouds}
\label{sec:magnetic}

\begin{figure*}
\begin{center}
\begin{tabular}{lll}
{\includegraphics[scale=0.8]{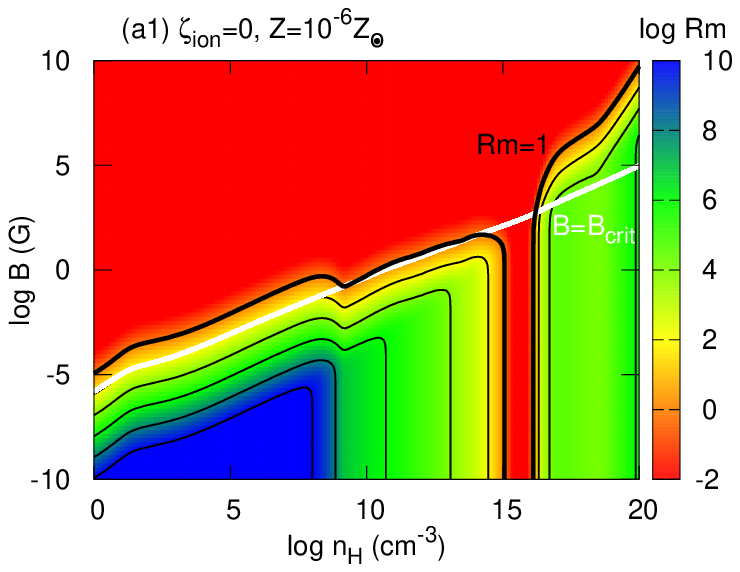}}
{\includegraphics[scale=0.8]{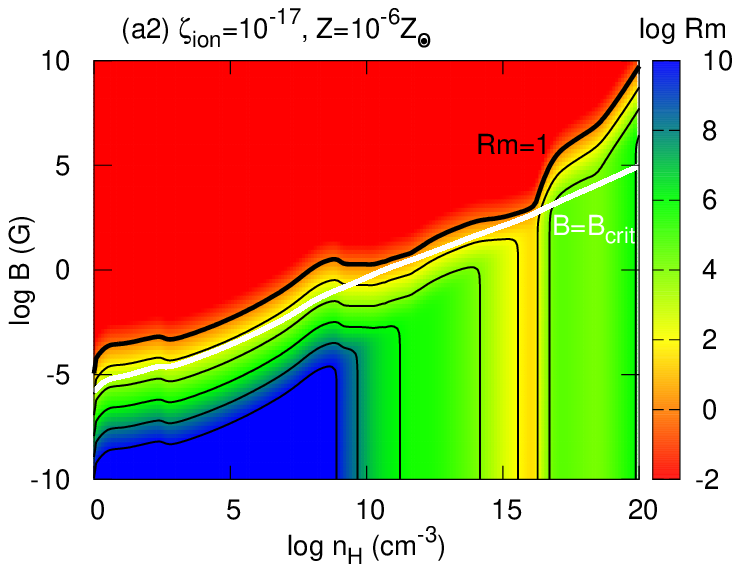}}
{\includegraphics[scale=0.8]{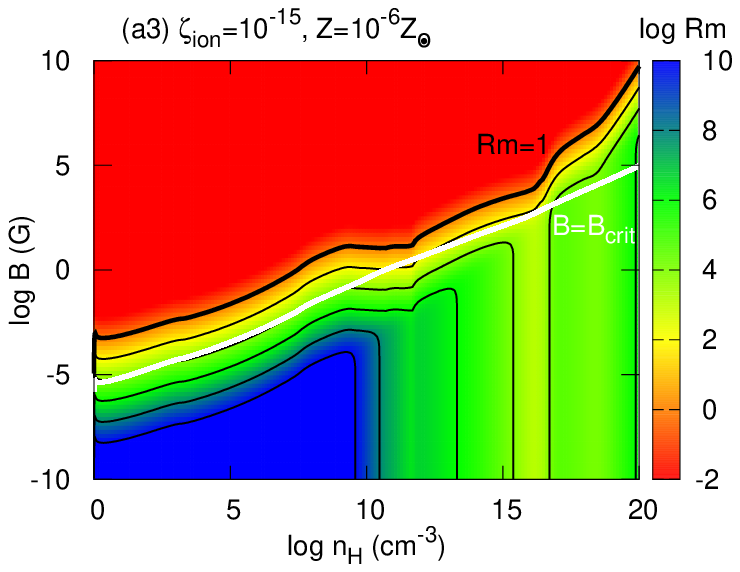}}\\
{\includegraphics[scale=0.8]{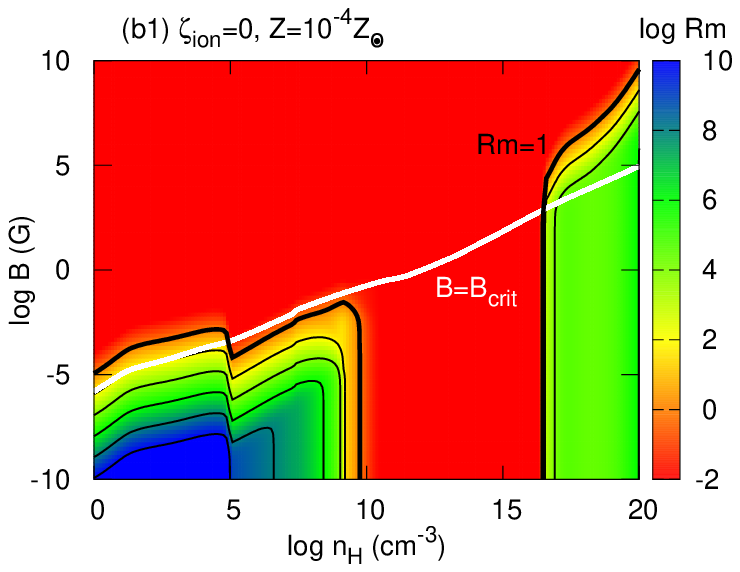}}
{\includegraphics[scale=0.8]{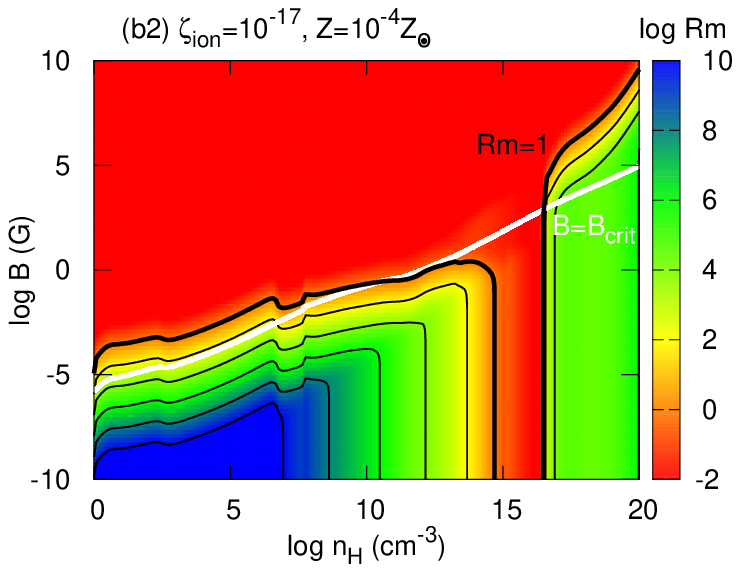}}
{\includegraphics[scale=0.8]{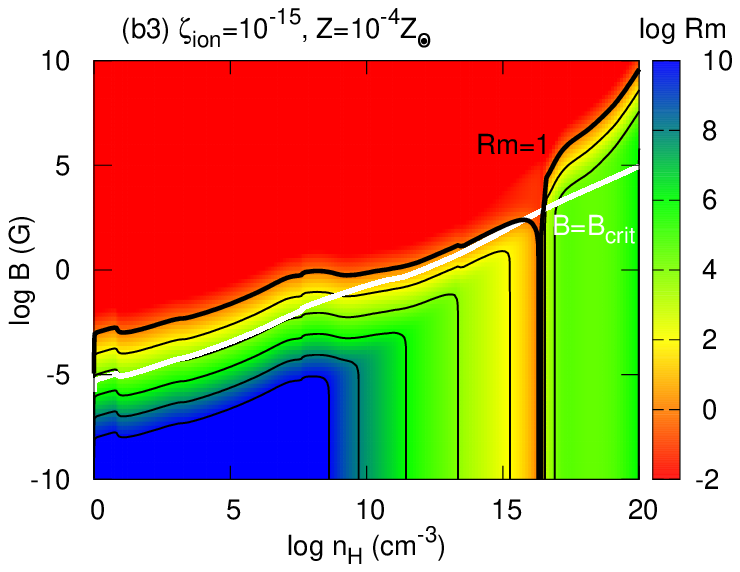}}\\
{\includegraphics[scale=0.8]{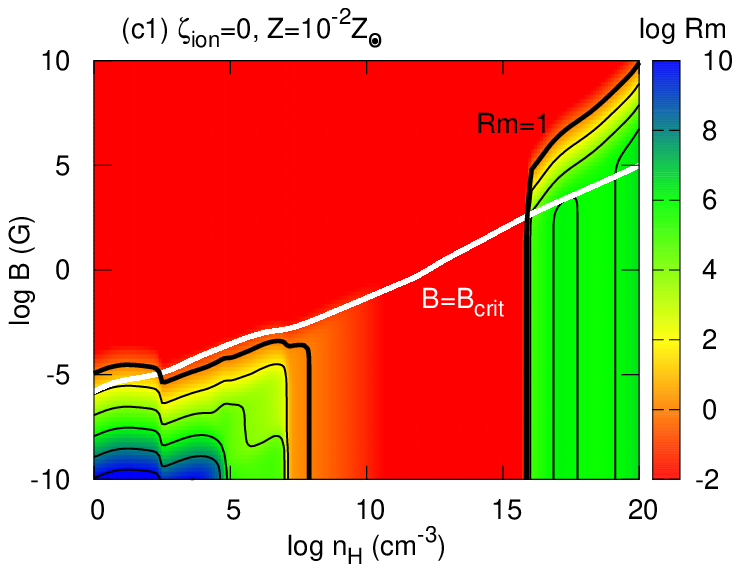}}
{\includegraphics[scale=0.8]{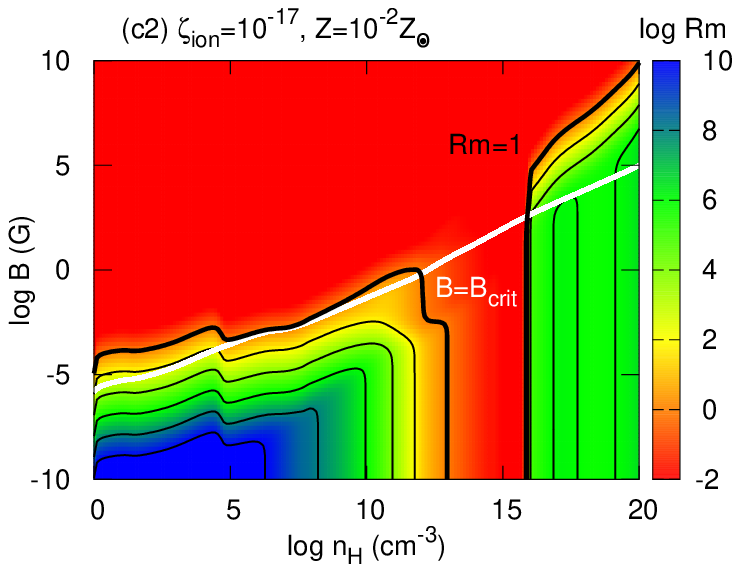}}
{\includegraphics[scale=0.8]{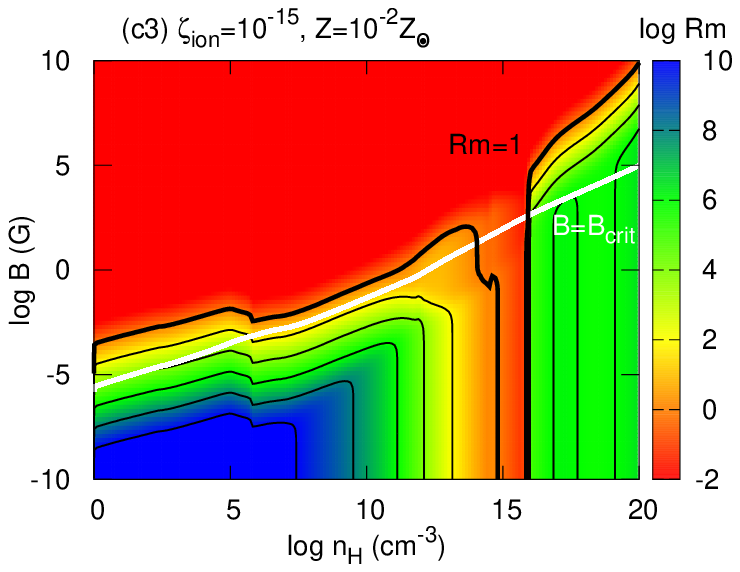}}\\
{\includegraphics[scale=0.8]{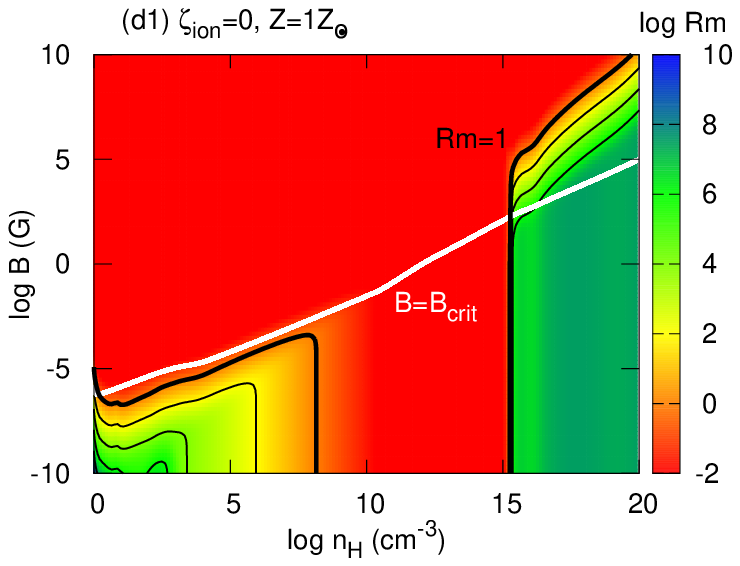}}
{\includegraphics[scale=0.8]{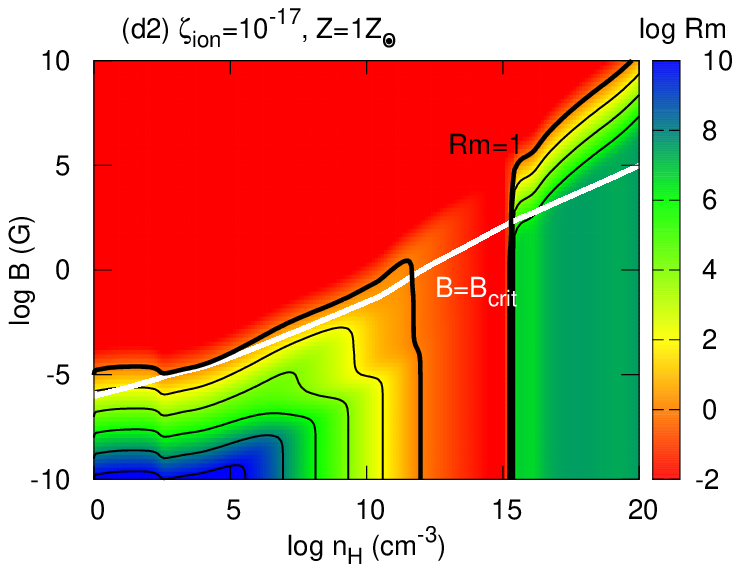}}
{\includegraphics[scale=0.8]{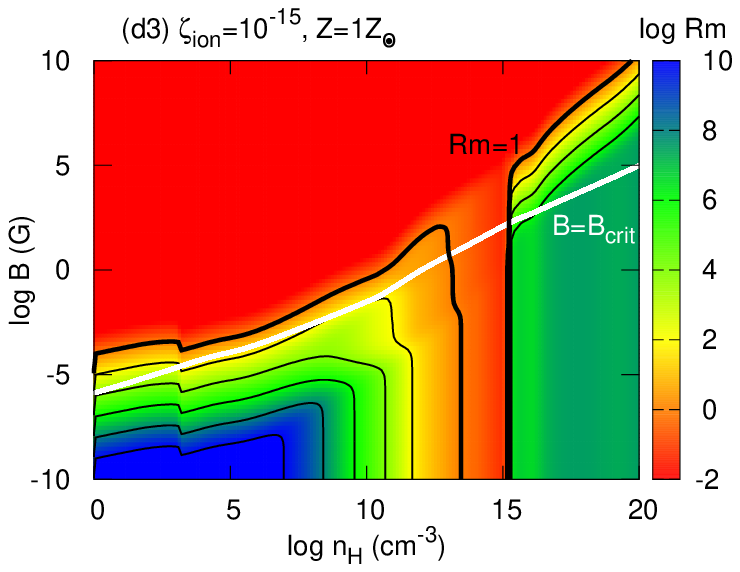}}\\
\end{tabular}
\caption{The contour map of the magnetic Reynolds number, for the cases with \crzero~(left panels), \crfid~(middle panels), and \crmax~(right panels).
The thick and thin black curves represent the contours of $\log \Rm = 0$ and $\log \Rm = 2, 4, 6, 8$, and $10$.
In the red-shaded regions, $\Rm < 1$ and the magnetic field dissipates.
The white curve corresponds to the critical field strength $\Bcr$ above which magnetic pressure suppresses the gravitational contraction of a star-forming cloud.}
\label{fig:Rm}
\end{center}
\end{figure*}

As we have discussed in the previous section, the ionization degree in a star-forming cloud decreases monotonically with increasing density until dust grains evaporate at $\nH \sim 10^{15}\mbox{-}10^{17}\pcc$.
In a cloud threaded by magnetic fields, the ionization degree can be low enough to weaken the coupling between the gas and magnetic fields via such resistive MHD effects as ambipolar diffusion and Ohmic loss.
In this section, we discuss the conditions of magnetic dissipation in a star-forming cloud with various metallicities.

The diffusion coefficients for ambipolar dissipation and Ohmic loss, $\eta_{\rm ambi}$ and $\eta_{\rm Ohm}$, are calculated from the following equations~\citep[e.g.,][]{Wardle1999}:
\begin{align}
\eta_{\rm ambi} &= \frac{c^2}{4 \pi} \frac{\sigma_{\rm P}}{\sigma_{\rm P}^2+ \sigma_{\rm H}^2} - \eta_{\rm Ohm},\nonumber\\
\eta_{\rm Ohm} &= \frac{c^2}{4 \pi \sigma_{\rm O}},
\label{eq:resistivity}
\end{align}
where $\sigma_{\rm P}, \sigma_{\rm H}$, and $\sigma_{\rm O}$ are the Pedersen, Hall, and Ohmic conductivities, respectively:
\begin{align}
\sigma_{\rm P} &= \left(\frac{c}{B}\right)^2 \sum_\nu \frac{\rho_\nu \tau_\nu \omega_\nu^2}{1 + \tau_\nu^2 \omega_\nu^2} \nonumber\\
\sigma_{\rm H} &= \left(\frac{c}{B}\right)^2 \sum_\nu \frac{q_\nu}{|q_\nu|} \frac{ \rho_\nu \omega_\nu}{1 + \tau_\nu^2 \omega_\nu^2}, \nonumber\\
\sigma_{\rm O} &= \left(\frac{c}{B}\right)^2 \sum_\nu \rho_\nu \tau_\nu \omega_\nu^2.
\label{eq:conductivity}
\end{align}
Here the subscript `$\nu$' indicates a charged species with electric charge $q_\nu$, mass density $\rho_\nu = m_\nu y(\nu) n_{\rm H}$, and cyclotron frequency $\omega_\nu = e |q_\nu| B / m_\nu c$.
$\tau_\nu$ is the collision timescale between the charged and neutral particles~\citep{Nakano1986}:
\begin{equation}
\tau_\nu^{-1} = \sum_{\rm n} \tau_{\nu, {\rm n}}^{-1} = \sum_{\rm n} \frac{\mu_{\nu, {\rm n}} y(\nu) y({\rm n}) n_{\rm H}^2 \langle \sigma {\rm v} \rangle_{\nu, {\rm n}}}{\rho_\nu},
\label{eq:tau_nu}
\end{equation}
where the subscript `n' stands for the major neutral species~(H, H$_2$, and He), $\mu_{\nu, {\rm n}}$ the reduced mass, and $\langle \sigma {\rm v} \rangle_{\nu, {\rm n}}$ the collision rate coefficient.
For the collisions between ($e$, H$^+$, gr$^\pm$, and gr$^{2\pm}$)-(H, H$_2$, and He), H$_3^+$-H$_2$, C$^+$-H, and HCO$^+$-H$_2$, the values for $\langle \sigma {\rm v} \rangle_{\nu, {\rm n}}$ are taken from \cite{Pinto2008}, and for the other cases from \cite{Osterbrock1961}.

The dependence of $\eta_{\rm ambi}$ and $\eta_{\rm Ohm}$ on the chemical compostion and magnetic field strength can be seen from Eq. \eqref{eq:resistivity}, by extracting the dominant terms in the summation of Eq. \eqref{eq:conductivity}:
\begin{align}
\eta_{\rm ambi} &\simeq \frac{1}{4 \pi} \left(\frac{B}{n_{\rm H}}\right)^2 \left(\mu_{\nu, {\rm n}} \langle \sigma {\rm v} \rangle_{\nu, {\rm n}} y(\nu) y({\rm n})\right)^{-1}, \nonumber\\ 
\eta_{\rm Ohm} &\simeq \frac{c^2}{4 \pi e^2} \mu_{\nu, {\rm n}} \langle \sigma {\rm v} \rangle_{\nu, {\rm n}} y({\rm n}) y(\nu)^{-1}.
\label{eq:resistivity_app}
\end{align}
Eq. \eqref{eq:resistivity_app} indicates that $\eta_{\rm ambi}$ scales quadratically with the magnetic field strength, whereas $\eta_{\rm Ohm}$ does not depend on the field strength.
Moreover, both coefficients are inversely proportional to the ionization degree.

The condition of magnetic dissipation is discussed by calculating the magnetic Reynolds number Rm defined below:
\begin{equation}
\Rm(L_{\rm B}) \equiv \frac{{\rm v}_{\rm n} L_{\rm B}}{\eta_{\rm ambi} + \eta_{\rm Ohm}},
\label{eq:Rm}
\end{equation}
where ${\rm v}_{\rm n}$ is the fluid velocity, $L_{\rm B}$ the coherent length of the field.
Magnetic Reynolds number is expressed as a ratio of the magnetic dissipation timescale $t_{\rm dis}(L_{\rm B}) = L_{\rm B}^2/(\eta_{\rm ambi} + \eta_{\rm Ohm})$ to the fluid dynamical timescale $t_{\rm dyn}(L_{\rm B}) = L_{\rm B}/{\rm v}_{\rm n}$:
\begin{equation}
\Rm(L_{\rm B}) = \frac{t_{\rm dis}(L_{\rm B})}{t_{\rm dyn}(L_{\rm B})}.
\label{eq:Rm_2}
\end{equation}
When inequality $\Rm(L_{\rm B}) < 1$ is satisfied, magnetic fields decouples from the fluid motion and dissipate.

First, we consider ordered magnetic field lines which are dragged with cloud contraction and coherent over the cloud size.
Without turbulent motions in the cloud, magnetic field fluctuations at smaller scales are neglected.
In this case, the magnetic Reynolds number is calculated by substituting the fluid velocity ${\rm v}_{\rm n} \sim {\rm v}_{\rm ff} \equiv \lambda_{\rm J} / (3t_{\rm ff})$ and coherent length $L_{\rm B} \sim \lambda_{\rm J}$ into Eq. \eqref{eq:Rm}.
In Figure \ref{fig:Rm}, we show the contour map of the magnetic Reynolds number $\Rm(\lambda_{\rm J})$, for the cases with \crzero~(left panels), \crfid~(middle panels), and \crmax~(right panels).
The thick and thin black curves represent the contours of $\log \Rm = 0$ and $\log \Rm = 2, 4, 6, 8$, and $10$.
In the red-shaded regions, $\Rm(\lambda_{\rm J}) < 1$ and the magnetic field dissipates in the cloud.
The white curve corresponds to the critical field strength $\Bcr$
\begin{equation}
\Bcr = \left(\frac{4 \pi G M_{\rm J} \rho}{\lambda_{\rm J}}\right)^{1/2},
\label{eq:B_crit}
\end{equation}
above which magnetic pressure suppresses the cloud contraction.
Now that a star-forming cloud is concerned, we consider magnetic fields weaker than $\Bcr$~(corresponding to the region below the white curves in Figure \ref{fig:Rm}).

Without ionization sources~($\zetaion = 0$; Figure \ref{fig:Rm} left panels), inequality $\Rm(\lambda_{\rm J}) > 1$ holds in the beginning of collapse and magnetic dissipation is negligible.
In the presence of such a strong field as $B \sim \Bcr$, magnetic field lines slip away from the contracting cloud via ambipolar diffusion at $\nH \gtrsim 10^5, 10^3$, and $10^1\pcc$ for $Z/\zsun = 10^{-4}, 10^{-2}$, and $1$.
When charged dust grains become the dominant charge carrier, the resistivity is elevated because the ionization degree is decreased to an extremely low value and dust grains have large inertia and collision cross section.
Then the magnetic field is dissipated irrespective of its strength via Ohmic loss at $\nH \gtrsim 10^{15}, 10^{10}, 10^8$, and $10^8\pcc$ for $Z/\zsun = 10^{-6}, 10^{-4}, 10^{-2}$, and $1$~(red-shaded regions).
This continues until dust grains evaporate at $\nH \sim 10^{15}\mbox{-}10^{17}\pcc$, where the ionization degree jumps up via the thermionic emission and thermal ionization of alkali metals.
We find that the magnetic field recovers strong coupling with the cloud at much earlier stages~(by a few orders of magnitude in density), compared to the previous work
which showed by neglecting the above ionization processes that the recoupling occurs at $\nH \sim 10^{17}\mbox{-}10^{18}\pcc$~(see Figure 7 of \citealt{Susa2015}).

With increasing ionization rate, the ionization degree at a given density becomes higher, elevating the value of $\Rm(\lambda_{\rm J})$~(c.f., Figure \ref{fig:Rm} middle and right panels).
As a result, the resistive MHD effects are weakened, and the regions of magnetic dissipation~(red-shaded regions) shrink, compared to the cases without ionization sources.
Especially, in such metal-poor clouds as $Z/\zsun \lesssim 10^{-5}$~(or $10^{-4}$) for \crfid~(or \crmax, respectively), the magnetic field remains coupled with cloud contraction throughout the evolution.
In higher metallicity cases, the magnetic field restores strong coupling with the cloud at $\nH \sim 10^{15}\mbox{-}10^{17}\pcc$, which is the same as in the case with \crzero.
This is because grain evaporation and thermal ionization of alkali metals proceed much more rapidly than radioactive ionization.
Compared to the previous work, the recoupling occurs at lower densities by a few orders of magnitude~(c.f. Figure 7 of \citealt{Susa2015}).


\begin{figure*}
\begin{center}
\begin{tabular}{lll}
{\includegraphics[scale=0.8]{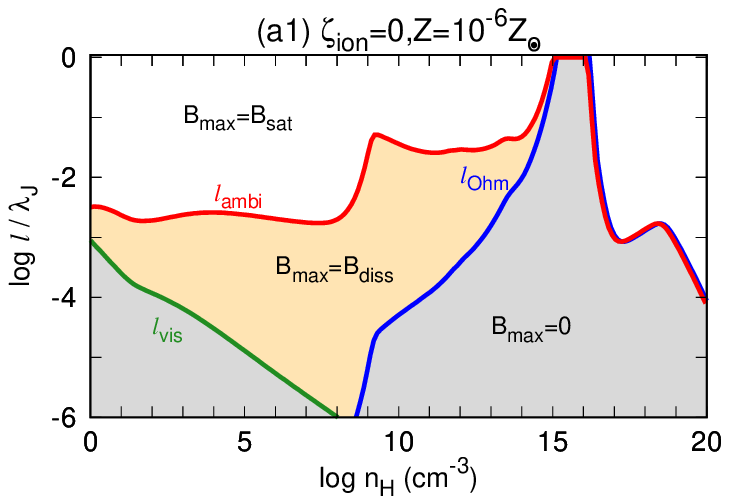}}
{\includegraphics[scale=0.8]{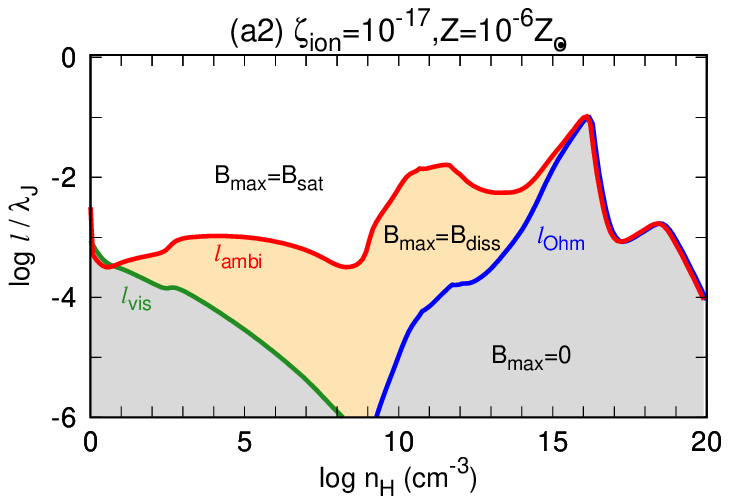}}
{\includegraphics[scale=0.8]{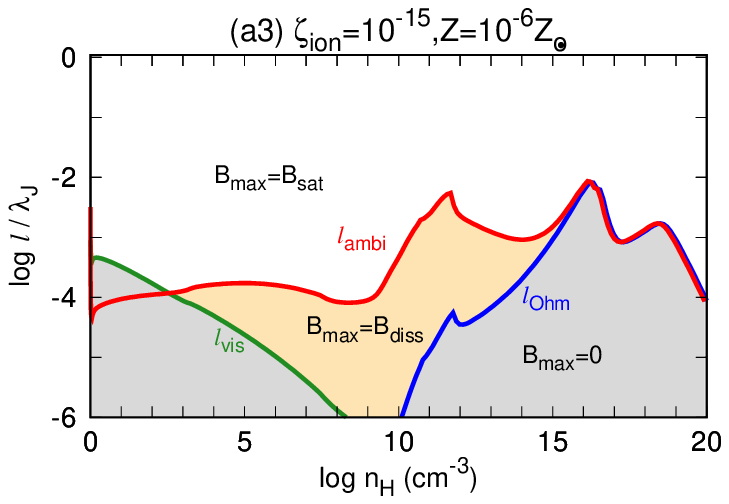}}\\
{\includegraphics[scale=0.8]{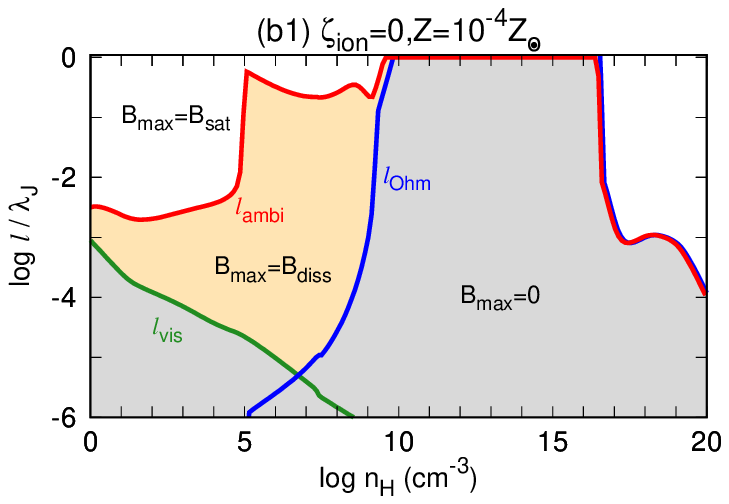}}
{\includegraphics[scale=0.8]{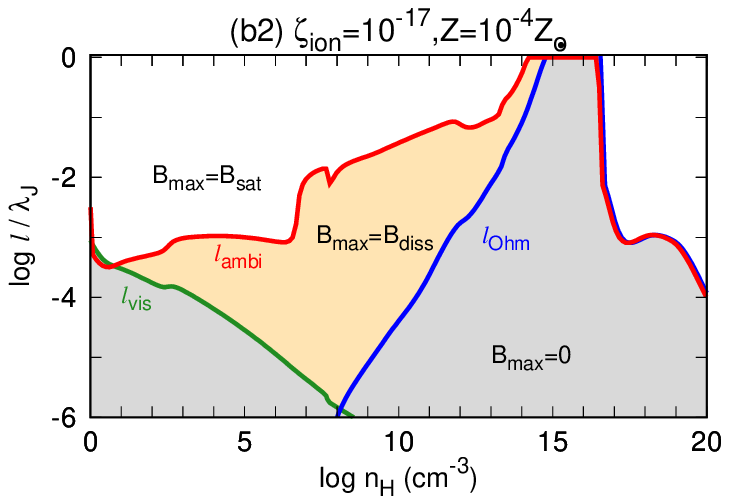}}
{\includegraphics[scale=0.8]{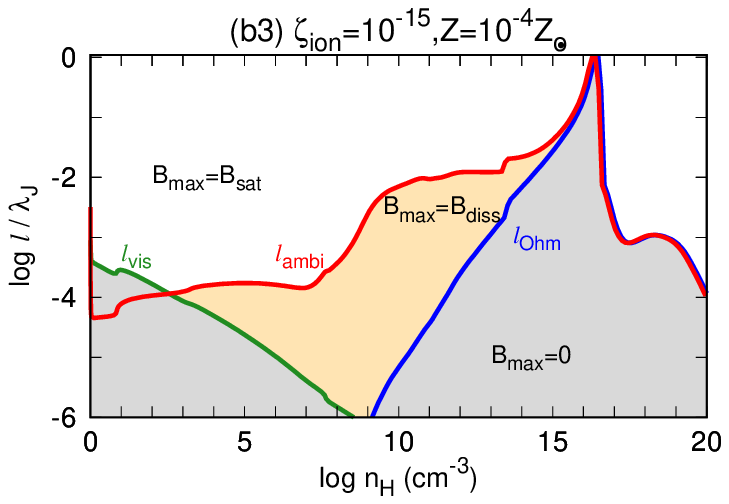}}\\
{\includegraphics[scale=0.8]{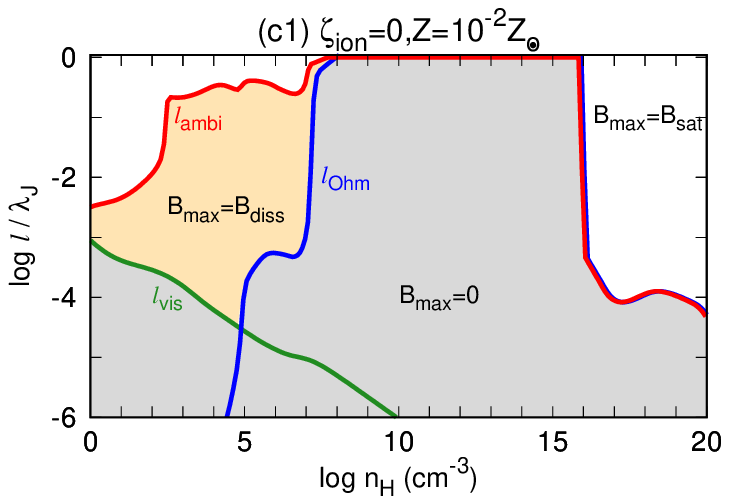}}
{\includegraphics[scale=0.8]{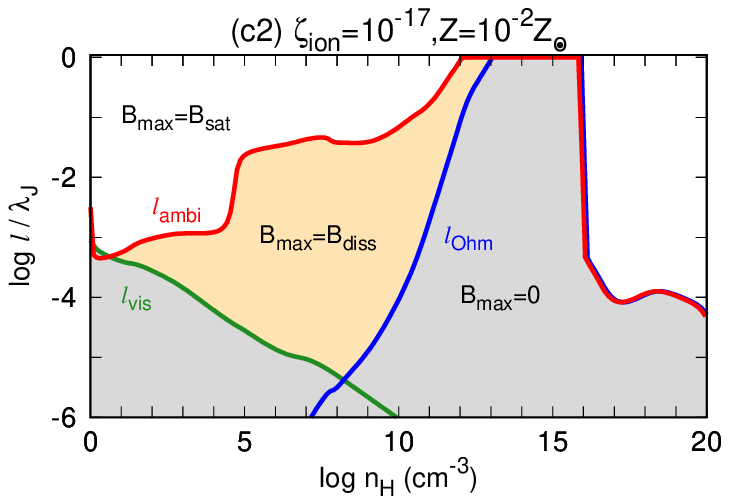}}
{\includegraphics[scale=0.8]{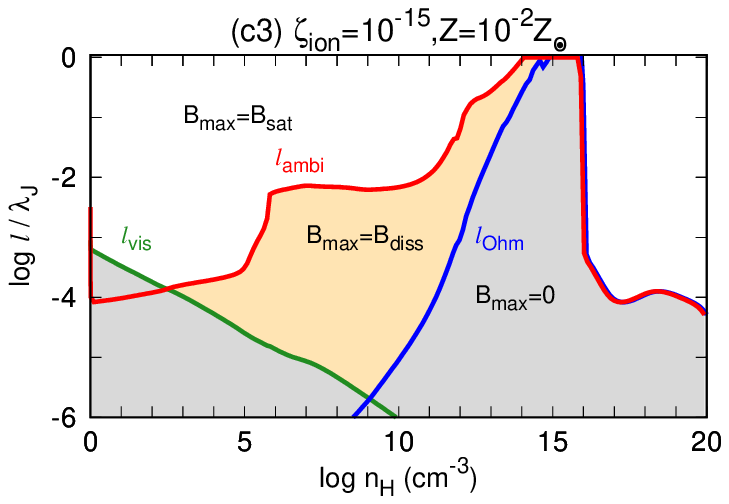}}\\
{\includegraphics[scale=0.8]{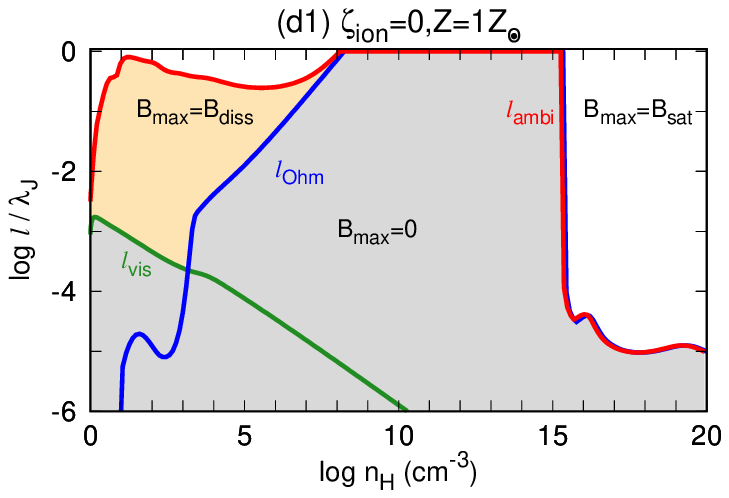}}
{\includegraphics[scale=0.8]{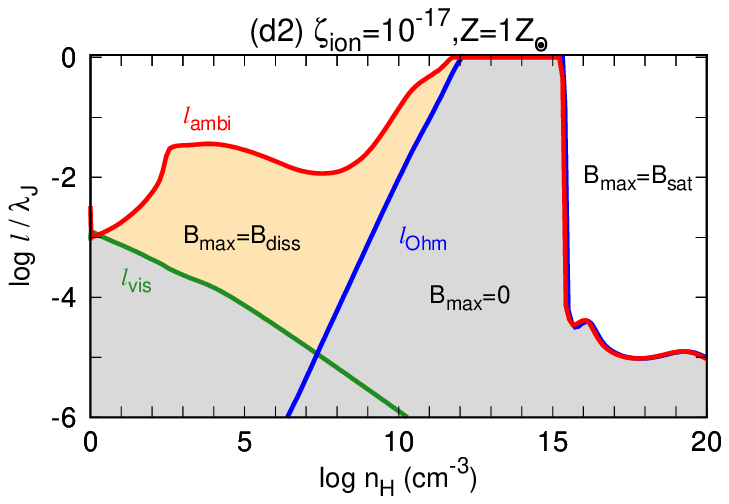}}
{\includegraphics[scale=0.8]{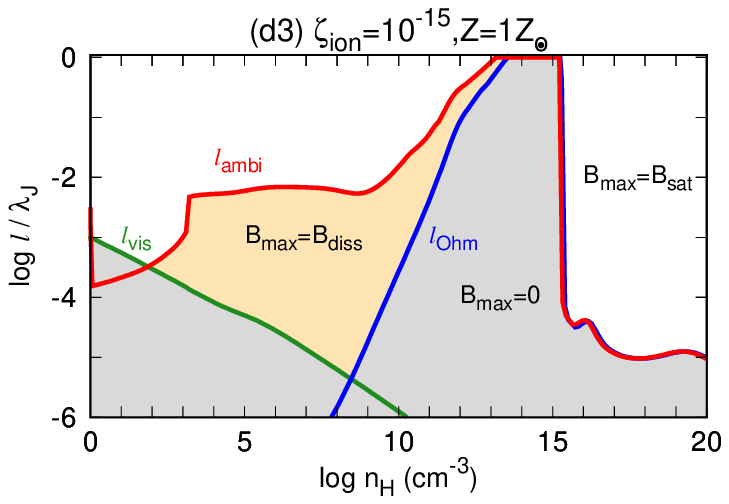}}\\
\end{tabular}
\caption{The maximum magnetic field strength achievable by the small scale dynamo action at each coherent length as a function of $\nH$, for the cases with \crzero~(left panels), \crfid~(middle panels), and \crmax~(right panels).
Here, we set $\vartheta = 1/2$ and $f_{\rm sat} = 10^{-2}$.
The colored curves represent the scales of viscous~($l_{\rm vis}$; green), Ohmic~($l_{\rm Ohm}$; blue), and ambipolar~($l_{\rm ambi}$; red) dissipation.
In the scales smaller than $l_{\rm vis}$ and $l_{\rm Ohm}$, the turbulent field is damped completely~($B_{\rm max}=0$; gray-shaded).
In the scales smaller than $l_{\rm ambi}$, $B_{\rm max}$ is limited to $B_{\rm diss}$ by the ambipolar dissipation~($B_{\rm max}=B_{\rm diss}$; yellow-shaded).
In the scales larger than $l_{\rm ambi}$, the amplification up to the saturation level is permitted~($B_{\rm max}=B_{\rm sat}$; non-filled).}
\label{fig:dynamo}
\end{center}
\end{figure*}


Next, we consider that turbulent motions in a cloud twist and stretch magnetic field lines, generating random fields fluctuating at $L_{\rm B} = l < \lambda_{\rm J}$~\citep{Brandenburg2005}.
Such small scale fields are more subject to dissipation.
The dissipation condition is judged at each scale $l$ by the magnetic Reynolds number
\begin{equation}
{\rm Rm}(l) = \frac{{\rm v}_{\rm eddy}(l)\ l}{\eta_{\rm ambi} + \eta_{\rm Ohm}}.
\label{eq:Rm_l}
\end{equation}
In Eq. \eqref{eq:Rm_l}, the velocity spectrum of the turbulent motion ${\rm v}_{\rm eddy}(l)$ is assumed to follow a power law
\begin{equation}
{\rm v}_{\rm eddy}(l) = {\rm v}_{\rm ff} \left(\frac{l}{\lambda_{\rm J}}\right)^{\vartheta} \;\;\; ( l_{\rm vis} \leq l \leq  \lambda_{\rm J} ),
\label{eq:v_eddy}
\end{equation}
where the viscous scale $l_{\rm vis}$ is calculated by using the kinematic viscosity $\nu_{\rm vis} = c_s/(n_{\rm H} \sigma_{\rm nn})$ and collision cross-section among neutral particles $\sigma_{\rm nn}$ as~\citep{Subramanian1998}
\begin{equation}
\frac{l_{\rm vis}}{\lambda_{\rm J}} = \left(\frac{\nu_{\rm vis}}{{\rm v}_{\rm ff}\lambda_{\rm J}}\right)^{1/(\vartheta + 1)}.
\label{eq:l_vis}
\end{equation}
The power-law index takes $\vartheta = 1/3$ for subsonic incompressive turbulence~\citep{Kolmogorov1941} and $\vartheta = 1/2$ for supersonic compressive turbulence~\citep{Burgers1948}.
The Burgers law is found to be more typical in the observations of Galactic molecular clouds~\citep{Heyer2004}, and is also confirmed in the 3D numerical simulations of supersonic turbulence~\citep{Federrath2013}.
The dissipation condition~(${\rm Rm}(l) < 1$) can be rewritten in terms of the field strength by substituting the approximate form of $\eta_{\rm ambi}$~(Eq. \ref{eq:resistivity_app}) into Eq. \eqref{eq:Rm_l} as:
\begin{equation}
\begin{split}
B > B_{\rm diss}(l) &\sim \left(4 \pi \nH^2 \mu_{\nu, {\rm n}} \langle \sigma {\rm v} \rangle_{\nu, {\rm n}} y({\rm n}) y(\nu) {\rm v}_{\rm ff} \lambda_{\rm J} \right)^{1/2} \\
&\times \left[\left(\frac{l}{\lambda_{\rm J}}\right)^{(\vartheta + 1)} - \left(\frac{l_{\rm Ohm}}{\lambda_{\rm J}}\right)^{(\vartheta + 1)}\right]^{1/2},
\end{split}
\label{eq:B_ambi}
\end{equation}
where $l_{\rm Ohm}$ is the Ohmic dissipation scale defined by
\begin{equation}
l_{\rm Ohm} \equiv \lambda_{\rm J} \left(\frac{\eta_{\rm Ohm}}{u_{\rm ff}\lambda_{\rm J}}\right)^{1/(\vartheta + 1)}.
\label{eq:l_Ohm}
\end{equation}
Fluctuations at scales smaller than $l_{\rm Ohm}$ is damped, whereas at scales larger than $l_{\rm Ohm}$, the fluctuating field strength is limited below $B_{\rm diss}(l)$ via ambipolar diffusion.

On the basis of the above results, we discuss the maximum field strength achieved if the small-scale dynamo action continues long enough time.
Amplification proceeds faster at smaller scales in the eddy timescale of $t_{\rm eddy}(l) = l/{\rm v}_{\rm eddy}(l) = t_{\rm ff}(l/\lambda_{\rm J})^{1-\vartheta}$.
Without dissipation, the amplification continues until the magnetic energy reaches a fraction $f_{\rm sat}$ of  the total turbulent kinetic energy,~i.e., $B_{\rm sat}^2/(8\pi \rho) \sim f_{\rm sat} {\rm v}_{\rm eddy}(\lambda_{\rm J})^2/2$,
\begin{equation}
B_{\rm sat} \sim f_{\rm sat}^{1/2} B_{\rm crit}.
\label{eq:B_eq}
\end{equation}
According to the numerical simulations, the saturation level of the turbulent dynamo changes in the range of $10^{-3} \lesssim f_{\rm sat} \lesssim 1$, depending strongly on the properties of the turbulent flow, in particular on the sonic Mach number and on the driving mode of the turbulence~\citep{Federrath2011}, and on the magnetic Prandtl number, defined as the ratio of kinematic viscosity to magnetic diffusivity~\citep{Federrath2014b,Federrath2016}.
Amplification above the saturation level is prohibited by the backreaction from the field.
With dissipation considered, any small-scale fluctuation is damped at $l < l_{\rm Ohm}$ by Ohmic loss, whereas at $l \geq l_{\rm Ohm}$, the field strength is limited below $B_{\rm diss}(l)$ via ambipolar dissipation.
Therefore, magnetic field reaches the saturation level only at scales where the inequality $B_{\rm diss}(l) \geq B_{\rm sat}$ holds.
In summary, the maximum field strength achieved by dynamo action is:
\begin{equation}
B_{\rm max}(l) = \begin{cases}
0 & ( l \leq {\rm max}[l_{\rm vis}, l_{\rm Ohm}] ), \\
B_{\rm diss}(l) &  ({\rm max}[l_{\rm vis}, l_{\rm Ohm}] < l \leq l_{\rm ambi}), \\
B_{\rm sat}   &  (l_{\rm ambi} <  l),
\end{cases} 
\label{eq:B_max}
\end{equation}
where $l_{\rm ambi}$ corresponds to the scale where the equality $B_{\rm diss}(l_{\rm ambi}) = B_{\rm sat}$ holds.

In Figure \ref{fig:dynamo}, we show the maximum magnetic field strength achievable by the small scale dynamo action at each coherent length as a function of $\nH$, for the cases with \crzero~(left panels), \crfid~(middle panels), and \crmax~(right panels).
Referring to the numerical simulations of supersonic turbulence, we adopt $\vartheta = 1/2$ and $f_{\rm sat} = 10^{-2}$ as fiducial values~\citep{Federrath2011,Federrath2014b,Federrath2016}.
The colored curves represent the scales of viscous~($l_{\rm vis}$; green), Ohmic~($l_{\rm Ohm}$; blue), and ambipolar~($l_{\rm ambi}$; red) dissipation.
In the scales smaller than $l_{\rm vis}$ and $l_{\rm Ohm}$, the turbulent field is damped completely~($B_{\rm max}=0$; gray-shaded).
In the scales smaller than $l_{\rm ambi}$, $B_{\rm max}$ is limited to $B_{\rm diss}$ by the ambipolar dissipation~($B_{\rm max}=B_{\rm diss}$; yellow-shaded).
In the scales larger than $l_{\rm ambi}$, the amplification up to the saturation level is allowed~($B_{\rm max}=B_{\rm sat}$; non-filled).

Without ionization sources~($\zetaion=0$; left panels), dissipation is negligible in the beginning of collapse at scales larger than $l/\lambda_{\rm J} \sim 10^{-3}\mbox{-}10^{-2}$~(non-filled regions), where magnetic fields are amplified to the saturation level successively from the smaller scales.
With increasing density and decreasing ionization fraction, ambipolar dissipation becomes effective over a wider range of scales~(yellow-shaded regions), where the amplification is limited below $B_{\rm diss}(l)$.
Dissipation works from lower densities with increasing metallicity.
As charged grains become more abundant by capturing electrons and ions, Ohmic loss works at larger and larger scales, finally damping turbulent magnetic fields at all scales~(gray-shaded regions).
This is the case until grain evaporation at $\nH \gtrsim 10^{15}\mbox{-}10^{17}\pcc$, where both ambipolar and Ohmic dissipation scales drop rapidly.
Afterwards, amplification to the saturation level becomes possible from very small scales of $l/\lambda_{\rm J} > 10^{-5}\mbox{-}10^{-4}$ to the largest scales of $l/\lambda_{\rm J} \sim 1$, irrespective of metallicity.

With increasing ionization rate and ionization fraction, turbulent magnetic fields couple to the gas more strongly at each density, and both ambipolar dissipation and Ohmic loss works only at smaller and smaller scales~(c.f. Figure \ref{fig:dynamo} middle and right panels).
At the largest scales of $l/\lambda_{\rm J} \sim 1$, magnetic fields can be amplified to the saturation level in the timescale of $t_{\rm ff}$, over a wider range of densities, compared to the case without ionization sources.
In the very metal-poor cases of $Z/\zsun \sim 10^{-5}$~(or $10^{-4}$) for \crfid~(or \crmax, respectively), such efficient amplification continues throughout the evolution.

\vspace{10mm}
\section{Summary and Discussion}
\label{sec:summary}

In this paper, we have calculated the temperature and ionization-degree evolution in a star-forming cloud for various metallicities $Z/\zsun = 10^{-6}, 10^{-5}, 10^{-4}, 10^{-3}, 10^{-2}, 10^{-1}$, and $1$ by using updated chemical network reversing all the gas-phase processes and accounting for grain-surface chemistry, including grain evaporation, thermal ionization of alkali metals, and thermionic emission.
We have also discussed the dissipation conditions of magnetic fields that are ordered over the cloud scale, as well as that fluctuating at smaller scales.
Below, we briefly summarize the results of our work:

\begin{itemize}

\item At low densities, the ionization degree decreases as the major ions, such as H$^+$, H$_3^+$, H$_3$O$^+$, and HCO$^+$, recombine with electrons~(Figures \ref{fig:charge_CR00} and \ref{fig:charge_w_cr}).
With increasing density, dust grains capture electrons and ions with their large recombination cross section, and become the major charge carriers.
Charged grains also neutralize each other via collision.
When the temperature exceeds $T \sim 1000\K$ at $\nH \sim 10^{15}\mbox{-}10^{17}\pcc$, dust grains evaporate and the ionization degree turns to rise via thermionic emission, and thermal ionization of alkali metals.
The ionization degree is elevated continuously by hydrogen ionization afterwards.

\item With increasing ionization rate, the ionization fraction becomes higher at a given density~(Figure \ref{fig:charge_w_cr}).
Ions and electrons remain as major charge carriers over a wider range of densities compared to the case without ionization sources.
In very low-metallicity cases of $Z/\zsun \lesssim 10^{-5}$~($10^{-4}$) for \crfid~(\crmax, respectively), charged grains never become dominant populations.

\item The ionization degree at $\nH \sim 10^{15}\mbox{-}10^{19}\pcc$ became up to eight orders of magnitude higher than that obtained in the previous model~(Figure \ref{fig:CR00_ion_comp}).
This is due to the thermionic emission and thermal ionization of vaporized alkali metals, which are not included so far.
As a result, magnetic fields recover strong coupling to the gas at much earlier stages~(by a few orders of magnitude in density), compared to the previous work.

\item We have developed a reduced chemical network that reproduces the chemical abundances of the major coolants and charged species~(Figure \ref{fig:redchem}).
The reduced network consists of 104~(or 161) reactions among 28~(38) species in the absence~(presence, respectively) of ionization sources~(Table \ref{tab:chem_react}).
The reduced model includes H$_2$ and HD formation on grain surfaces by using simple formulae obtained in Appendix \ref{sec:dust_chem}.
With ionization sources, the reduced model includes the depletion of O, C, OH, CO, and H$_2$O on grain surfaces to reproduce the abundances of molecular ions, H$_3$O$^+$ and HCO$^+$, which become dominant cations at some density ranges.

\item The coupling of ordered magnetic fields to the gas becomes weak as the ionization degree decreases with increasing density~(Figure \ref{fig:Rm}).
If magnetic fields are so strong as to suppress the contraction, magnetic field lines gradually drift out of the cloud via ambipolar diffusion.
Once charged grains become the dominant population, the ionization degree drops to a value as low as to dissipate magnetic fields irrespective of their strength via Ohmic loss.
Magnetic fields remain decoupled until the ionization degree jumps up at $\nH \sim 10^{15}\mbox{-}10^{17}\pcc$ due to grain evaporation.

\item Magnetic fields couple with the gas more strongly for increasing ionization rate.
Parameter space of magnetic dissipation shrinks, compared to the cases without ionization sources~(Figure \ref{fig:Rm}).
In the cases of $Z/\zsun \lesssim 10^{-5}$~(or $10^{-4}$) for \crfid~(or \crmax, respectively), magnetic fields remain coupled with the gas throughout the evolution.

\item Dissipation of turbulent magnetic fields is negligible in the beginning of collapse at scales larger than $l/\lambda_{\rm J} \sim 10^{-3}\mbox{-}10^{-2}$~(Figure \ref{fig:dynamo}).
Dissipation is also negligible at $\nH > 10^{15}\mbox{-}10^{17}\pcc$ at very small scales of $l/\lambda_{\rm J} \sim 10^{-5}\mbox{-}10^{-4}$.
In these cases, magnetic fields can be amplified by dynamo action until their energy becomes a fraction of the turbulent kinetic energy.
At intermediate densities, with increasing density and decreasing ionization fraction, ambipolar diffusion or Ohmic loss operates over a wider range of scales.
Ambipolar diffusion regulates the magnetic field amplification below $B_{\rm diss}(l)$~(Eq. \ref{eq:B_ambi}), whereas Ohmic loss damps turbulent magnetic fields almost completely.

\end{itemize}

So far, we have assumed that the grain-size distribution follows the MRN-type power law~\citep{Mathis1977} throughout the evolution.
Dust grains grow in their size by accreting gas-phase species and by colliding with other grains~\citep[e.g.,][]{Flower2005}.
This occurs more frequently with increasing density, so that the size distribution would deviate from the MRN.
If large grains become more abundant, gas-phase ions and electrons could recombine on grain surfaces more frequently.
Then, charged dust grains may become the dominant charge carrier from lower densities, extending the domain of Ohmic dissipation.
We have also assumed that dust grains evaporate immediately maintaining the MRN distribution above the vaporization temperature, whereas grains evaporate by reducing their size~\citep[e.g.,][]{Lenzuni1995}.
This assumption remains valid, since grain evaporation proceeds with strong temperature dependence.

Our calculation is based on the simple one-zone model by assuming spherical symmetry and by neglecting the backreaction of magnetic fields on cloud contraction.
Multi-dimensional effects, such as rotation and turbulence, are prevalent in a star-forming cloud, so that multi-dimensional resistive MHD calculations are required for discussing magnetic field effects.
\cite{Higuchi2018,Higuchi2019} performed 3D resistive MHD calculations of low-metallicity clouds by using the lookup tables of the barotropic EOS and resistivity coefficients calculated by \cite{Susa2015}.
We have found that the magnetic fields recover the coupling with the cloud at earlier stages~(by a few orders of magnitude in density), compared to \cite{Susa2015}.
If resistive MHD calculations were performed by using our resistivity coefficients, such magnetic effects as magnetically-driven outflows and magnetic braking would work at earlier stages, compared to \cite{Higuchi2018,Higuchi2019}.
Since these effects lower the star-formation efficiency and suppress the formation of massive accretion disc around a protostar, the formation efficiency of binary and multiple stellar systems could be lowered.

\cite{Higuchi2018,Higuchi2019} also assumed that the temperature evolves following the barotropic EOS, without solving the energy equation consistently with the MHD equations.
In the presence of strong magnetic fields, the magnetic pressure support delays the cloud contraction compared to the free-fall rate, weakening the compressional heating.
In addition, the magnetic energy dissipated by ambipolar diffusion and Ohmic loss can heat the gas.
From these effects, temperature evolution could deviate from the track given by the barotropic EOS.
In the primordial clouds, these effects are found to be small both by analytical~(\paperI) and 3D numerical~(Sadanari et al. in prep.) calculations.
In metal-enriched clouds, where resistive MHD effects work more strongly, the heating via magnetic energy dissipation could be more important.
To clarify this point, multi-dimensional resistive MHD calculations by consistently solving the energy equation with the MHD equations are needed.

\section*{Acknowledgments}
The authors wish to express their cordial thanks to Prof. Toyoharu Umebayashi for his continual interests and kind suggestions. 
We also thank Drs. Motomichi Tashiro, Attila G. Cs\'asz\'ar, and Kenji Furuya for fruitful discussions.
Numerical calculations are performed by the computer cluster, {\tt Draco}, supported by the Frontier Research Institute for Interdisciplinary Sciences in Tohoku University.
This work is supported in part by the Grant-in-Aid from the Ministry of Education, Culture, Sports, Science and Technology (MEXT)
of Japan (DN:17H06360, 16J02951, KO:17H06360, 17H01102, HS:17H02869, 17H01101).

\section*{Data Availability}
The data underlying this article will be shared on reasonable request to the corresponding author.




\bibliographystyle{mnras}



\appendix
\section{Partition function and heat of reaction}
\label{sec:part_fnc}

The partition functions are referred from \cite{Popovas2016} for H$_2$, from \cite{Vidler2000} for H$_2$O, from the ExoMol database~\citep{Tennyson2012}\footnote{\url{http://exomol.com/data/molecules/}} for HD$^+$, H$_3^+$, and LiH$^+$, from the HITRAN database~\citep{Gamache2017}\footnote{\url{http://hitran.org/docs/iso-meta/}} for HD, CH$_3$, CH$_4$, CO$_2$, H$_2$O$_2$, O$_2$H, and H$_2$CO, and from \cite{Barklem2016} for the remaining atoms and diatomic molecules.
The remaining polyatomic molecules are regarded as rigid rotators, and their partition functions are calculated by considering the rotation degree of freedom alone~\citep[e.g.,][]{Irwin1988}.
The values of the rotational constants are cited from the Cologne Database for Molecular Spectroscopy\footnote{\url{https://www.astro.uni-koeln.de/cdms}} for CH$_2$, HCO, HCO$^+$, H$_2$O$^+$, H$_3$O$^+$, HCO$_2^+$, and H$_3$CO$^+$, from the NIST Computational Chemistry Comparison and Benchmark Database\footnote{\url{http://cccbdb.nist.gov/}} for H$_2$CO$^+$, O$_2$H$^+$, CH$_2^+$, CH$_3^+$, and CH$_4^+$, and from \cite{Thompson2005} for CH$_5^+$.

The heat of reaction $\Delta E$ is calculated from the ionization~(or dissociation) energy of atoms~(or molecules) involved in the reaction.
The values of ionization~(or dissociation) energy are referred from the KIDA database~\citep{Wakelam2012}\footnote{\url{http://kida.obs.u-bordeaux1.fr/}} for the atoms and molecules composed of H, He, C, O, Na, Mg, and K nuclei, from the Active Thermochemical Table\footnote{\url{https://atct.anl.gov/Thermochemical Data/}} for those containing D nuclei, and from the NIST-JANAF Thermochemical Table~\citep{Chase1998}\footnote{\url{https://janaf.nist.gov/}} for those containing Li nuclei, except for ${\rm LiH}^+$, whose value is taken from \cite{Stancil1996}.

\section{Grain surface chemistry}
\label{sec:dust_chem}
The grain-surface chemistry is considered by calculating kinetic rate equations in a similar form to the gas-phase chemistry~\citep[e.g.,][]{Hasegawa1992}.
The grain-surface chemistry is divided into three categories: (i) the adsorption of a gas-phase species onto the grain surface, (ii) the desorption of a grain-surface species into the gas-phase, and (iii) molecule formation.
For the grain-surface chemistry, 148 reactions are considered and are listed in Table \textcolor{blue}{B1}.
Dust grains also exchange an electric charge with gas-phase ions, electrons, and other grains.
For grain charging, 150 reactions are considered and they are summarized in Table \ref{tab:charge_transfer}.
Below, we derive the rate coefficient of each reaction in the unit of $\pcc \ps$, following \cite{Draine1987}, \cite{Hocuk2016}, and \cite{Esplugues2016,Esplugues2019}.

\subsection{Molecule formation on dust grains}

\subsubsection{Adsorption~(reactions 1-23)}
Atoms and molecules stick on grain surfaces via two types of interactions: they are bound weakly via van der Waals interaction and strongly via covalent bond.
The former~(so-called physisorption) is considered for all the neutral species, whereas the latter~(so-called chemisorption) is considered only for H and D, following \cite{Hocuk2016}.
For a gas-phase species X, its physisorbed and chemisorbed counterparts are denoted as X(p) and X(c), respectively.

A gas-phase species X sticks at the physisorption site on a grain surface at the rate of
\begin{equation}
k_{\rm ads}({\rm X}) = y_{\rm gr} \langle \sigma_{\rm gr} \rangle {\rm v}({\rm X}) S(T, T_{\rm gr}), 
\label{eq:adsorption}
\end{equation}
where $y_{\rm gr}=n_{\rm gr}/n_{\rm H}$ is the total grain fraction relative to hydrogen nuclei, $\langle \sigma_{\rm gr} \rangle$ the geometrical cross section of a dust grain averaged over the MRN distribution, ${\rm v}({\rm X})=(8k_{\rm B}T/\pi m_{\rm X})^{1/2}$ the thermal velocity of X, and
\begin{equation}
S(T, T_{\rm gr}) =  \left(1+0.4 \left(\frac{T+T_{\rm gr}}{100\ {\rm K}}\right)^{0.5}+0.2 \frac{T}{100\ {\rm K}}+0.08 \left(\frac{T}{100\ {\rm K}}\right)^{2}\right)^{-1}, 
\label{eq:sticking_probability}
\end{equation}
the sticking probability of X derived by \cite{Hollenbach1979}.

A gas-phase H~(or D) sticks directly at a chemisorption site, if it can cross the energy barrier $E_{\rm act}$ between a physisorption and chemisorption site.
The reaction barrier is overcome via thermal hopping or quantum tunneling, so that the transmission probability is given by~\citep{Esplugues2016}
\begin{equation}
T_{\rm chem} = \exp{\left(-\frac{E_{\rm act}}{T_{\rm gr}}\right)} + \exp{\left(-2 \Delta \sqrt{\frac{2m_{\rm red}k_{\rm B}E_{\rm act}}{\hbar^2}}\right)},
\label{eq:T_chem}
\end{equation}
where $\Delta \sim 3$ \AA~is the width of the barrier.
The direct chemisorption into an empty site occurs in the rate of
\begin{equation}
k_{\rm gc}({\rm X}) = k_{\rm ads}({\rm X}) T_{\rm chem} (1-f_{\rm chem}). 
\label{eq:k_gc}
\end{equation}
In Eq. \eqref{eq:k_gc}, $f_{\rm chem}$ is the fraction of the filled chemisorption sites represented by
\begin{equation}
f_{\rm chem} =  \frac{n[{\rm H(c)}]+n[{\rm D(c)}]}{n_{\rm gr} N_{\rm site}},
\label{eq:f_chem}
\end{equation}
where $N_{\rm site}$ is the total number of adsorption sites on a grain calculated as $N_{\rm site}=4 \langle \sigma_{\rm gr} \rangle /d_{\rm pp}^2$, by assuming that adsorption sites are located at the average distance of $d_{\rm pp} \sim 3$ \AA~on a dust grain.

A physisorbed H~(or D) also moves to an empty chemisorption site via thermal hopping and quantum tunneling, at the rate of \citep{Cazaux2010,Esplugues2016}
\begin{equation}
k_{\rm pc}({\rm H}) = \alpha_{\rm pc}({\rm H}) (1-f_{\rm chem}). 
\label{eq:k_pc}
\end{equation}
In Eq. \eqref{eq:k_pc}, the mobility $\alpha_{\rm pc}({\rm H})$ is calculated from
\begin{align}
\alpha_{\rm pc}({\rm X}) =&  8 \sqrt{\pi T_{\rm gr}} \nu_0 \frac{\sqrt{E_{\rm bind}({\rm H}_{\rm c})-E_{\rm bind}({\rm H}_{\rm p})}}{E_{\rm bind}({\rm H}_{\rm c})-E_{\rm s}} \nonumber \\
\times& \exp{\left(-2 \Delta \frac{\sqrt{2m_{\rm H}k_{\rm B}(E_{\rm bind}({\rm H}_{\rm p})-E_{\rm s})}}{\hbar}\right)} \nonumber \\
+& 4 \nu_0 \sqrt{\frac{E_{\rm bind}({\rm H}_{\rm p})-E_{\rm s}}{E_{\rm bind}({\rm H}_{\rm c})-E_{\rm s}} }\exp{\left(-\frac{E_{\rm bind}({\rm H}_{\rm p})-E_{\rm s}}{T_{\rm gr}}\right)},
\label{eq:alpha_pc}
\end{align}
where $\nu_0=10^{12}\ {\rm s}^{-1}$ is the oscillation frequency, $E_{\rm s}=200\K$, and $E_{\rm bind}({\rm H}_{\rm p})$ and $E_{\rm bind}({\rm H}_{\rm c})$ the binding energies of H(p) and H(c), whose values are referred from \cite{Cazaux2010}.

\subsubsection{Thermal desorption and CR-induced desorption~(reactions 24-44)}
When the grain is heated, grain-surface species X obtains enough thermal energy to escape the binding, and X is released into the gas phase.
The thermal desorption rate depends on the binding energy of X, which changes depending on whether the adsorption site is covered by H$_2$O ice or not.
The fraction of the adsorption sites covered by H$_2$O ice is
\begin{equation}
f_{\rm ice} =  {\rm min}\left\{\frac{n[{\rm H}_2{\rm O(p)}]}{n_{\rm gr} N_{\rm site}},\ 1\right\},
\label{eq:f_ice}
\end{equation}
whereas that of the bare sites is $f_{\rm bare} = 1- f_{\rm ice}$.
Then the rate coefficient for thermal desorption is calculated from
\begin{equation}
k_{\rm des}({\rm X}) =  \nu_0 \left[f_{\rm bare} \exp\left(-\frac{E_{\rm bare, X}}{T_{\rm gr}}\right) + f_{\rm ice} \exp\left(-\frac{E_{\rm ice, X}}{T_{\rm gr}}\right) \right],
\label{eq:desorption}
\end{equation}
where $\nu_0=10^{12}\ {\rm s}^{-1}$ is the oscillation frequency, and $E_{\rm bare, X}$~(and $E_{\rm ice, X}$) the binding energy of X on bare~(and icy, respectively) surfaces, whose values are referred from \cite{Esplugues2019} Table A3.
When the grain temperature is so low as $<10\K$, despite its low binding energy, H$_2$ is depleted on grain surface and forms H$_2$-iced layers.
Once the grain surface is covered by H$_2$ ice, the binding on grain surface is weakened, and the binding energy is set to $E_{{{\rm H}_2}-{\rm H}_2} = 100\K$~\citep{Nakano1971}.

CR particles directly hit and heat the grain surfaces, leading to the desorption of the grain-surface species.
CR particles also excite the gas-phase H$_2$, and the UV photons emitted with H$_2$ de-excitation enable the desorption of surface species.
These two effects are considered following \cite{Hasegawa1993} and \cite{Hocuk2016}, respectively.

\subsubsection{Molecule formation~(reactions 45-127)}
Grain-surface species move around the grain surface via thermal diffusion and meet other species at one adsorption site, producing a new molecule.
The rate coefficients of the two-body reactions are calculated following \cite{Esplugues2016}.

On the bare substrate, two physisorbed species X and Y meet via thermal diffusion in the frequency of
\begin{equation}
\mathcal{R}_{\rm bare} = \nu_0 P_{\rm bare},
\label{eq:R_bare}
\end{equation}
where
\begin{equation}
P_{\rm bare} = f_{\rm bare} \left[\exp\left(-\frac{2}{3}\frac{E_{\rm bare, X}}{T_{\rm gr}}\right) + \exp\left(-\frac{2}{3}\frac{E_{\rm bare, Y}}{T_{\rm gr}}\right)\right].
\label{eq:P_bare}
\end{equation}
On the icy substrate, the frequency $\mathcal{R}_{\rm ice}$ is calculated by changing the subscript  `bare' into `ice'.
The energy barrier between two adjacent sites of physisorption is assumed to be $2/3$ of the binding energy following \cite{Esplugues2016}.

Reactants produce a new molecule immediately in their encounter, if the reaction barrier is negligibly small.
In this case, the rate coefficient for molecule-forming reaction is calculated from
\begin{equation}
k_{\rm 2body}({\rm X}, {\rm Y}) = (\mathcal{R}_{\rm bare}\delta_{\rm bare} + \mathcal{R}_{\rm ice} \delta_{\rm ice}) / (n_{\rm gr}N_{\rm site}),
\label{eq:two_body_reaction}
\end{equation}
where $\delta_{\rm bare}$~(and $\delta_{\rm ice}$) indicates the desorption probability of the products on the bare~(and icy, respectively) substrate.
The values of $\delta_{\rm bare}$ and $\delta_{\rm ice}$ are referred from Table A4 in \cite{Esplugues2019}.

Inversely, for reactions with a high activation barrier~($E_{\rm act}$), their rates are determined by the competition between the probability of reactants' encounter~($P_{\rm bare}$ or $P_{\rm ice}$) and that of overcoming the reaction barrier via thermal crossing or quantum tunneling~($P_{\rm cross}$).
Therefore, the rate coefficient is decreased from Eq. \eqref{eq:two_body_reaction} by the reaction probability given by~\citep{Garrod2011,Esplugues2016}:
\begin{equation}
P_{\rm react} = \frac{P_{\rm cross}}{P_{\rm cross} + P_{\rm bare} + P_{\rm ice}}.
\label{eq:p_react}
\end{equation}
The reactants overcome the reaction barrier via thermal crossing with the probability of
\begin{equation}
P_{\rm therm} = \exp{\left(-\frac{E_{\rm act}}{T_{\rm gr}}\right)},
\label{eq:p_diffusion}
\end{equation}
and via quantum tunneling with
\begin{equation}
P_{\rm tunnel} = \exp{\left(-2 \Delta \sqrt{\frac{2m_{\rm red}k_{\rm B}E_{\rm act}}{\hbar^2}}\right)},
\label{eq:p_tunnel}
\end{equation}
where $\Delta \sim 1$ \AA~ the width of the barrier, and $m_{\rm red}$ the reduced mass of the reactants.
Then the crossing probability is represented by the maximum of the two as
\begin{equation}
P_{\rm cross} = {\rm max} \left\{P_{\rm therm}, P_{\rm tunnel}\right\}.
\label{eq:p_max}
\end{equation}
The value of $E_{\rm act}$ is also referred from Table A4 in \cite{Esplugues2019}.

There are three additional pathways of H$_2$~(or HD) formation on grain surfaces.
When a gas-phase H hits a physisorbed and chemisorbed H directly~(reactions 119-124), H$_2$ formation proceeds at the rate of
\begin{equation}
k_{\rm H_2, gp} = k_{\rm ads}({\rm H}) / (n_{\rm gr}N_{\rm site}),
\label{eq:k_H2_gp}
\end{equation}
and
\begin{equation}
k_{\rm H_2, gc} = k_{\rm ads}({\rm H}) T_{\rm chem} / (n_{\rm gr}N_{\rm site}),
\label{eq:k_H2_gc}
\end{equation}
respectively.
When a physisorbed H moves to an adjacent chemisorption site filled by another H~(reactions 125-127), H$_2$ is produced at the rate of
\begin{equation}
k_{\rm H_2, pc} = \alpha_{\rm pc}({\rm H}) / (n_{\rm gr}N_{\rm site}).
\label{eq:k_H2_pc}
\end{equation}

\begin{table}
\caption{Grain surface chemistry.}
\begin{center}
{\begin{tabular}{llc}
\hline
Number & Reaction & Reference \\
\hline
   1 &  H(g) $\rightarrow$  H(p)                       & 1 \\
   2 &  H$_2$(g) $\rightarrow$  H$_2$(p)       & 1 \\
   3 &  D(g) $\rightarrow$  D(p)                       & 2 \\
   4 &  HD(g) $\rightarrow$  HD(p)                  & 2 \\
   5 &  O(g) $\rightarrow$  O(p)                       & 1 \\
   6 &  O$_2$(g) $\rightarrow$  O$_2$(p)       & 1 \\
   7 &  OH(g) $\rightarrow$  OH(p)                  & 1 \\
   8 & CO(g) $\rightarrow$  CO(p)                   & 1 \\
   9 &  CO$_2$(g) $\rightarrow$  CO$_2$(p)                           & 1 \\
 10 &  H$_2$O(g) $\rightarrow$  H$_2$O(p)                          & 1 \\
 11 &  HO$_2$(g) $\rightarrow$  HO$_2$(p)                           & 1 \\
 12 &  H$_2$O$_2$(g) $\rightarrow$  H$_2$O$_2$(p)           & 1 \\
 13 &  HCO(g) $\rightarrow$  HCO(p)                          & 1 \\
 14 &  H$_2$CO(g) $\rightarrow$  H$_2$CO(p)          & 1 \\
 15 &  C(g) $\rightarrow$  C(p)                                     & 1 \\
 16 &  CH(g) $\rightarrow$  CH(p)                                & 1 \\
 17 &  CH$_2$(g) $\rightarrow$  CH$_2$(p)                & 1 \\
 18 &  CH$_3$(g) $\rightarrow$  CH$_3$(p)                & 1 \\
 19 &  CH$_4$(g) $\rightarrow$  CH$_4$(p)                & 1 \\
 20 &  H(g) $\rightarrow$  H(c)                       & 1 \\
 21 &  D(g) $\rightarrow$  D(c)                      & 2 \\
 22 &  H(p) $\rightarrow$ H(c)                          & 3 \\
 23 &  D(p) $\rightarrow$ D(c)                          & 2 \\
 \hline
 24 &  H(p) $\rightarrow$  H(g)                         & 1 \\
 25 &  H$_2$(p) $\rightarrow$  H$_2$(g)         & 1 \\
 26 &  D(p) $\rightarrow$  D(g)                         & 2 \\
 27 &  HD(p) $\rightarrow$  HD(g)                    & 2 \\
 28 &  O(p) $\rightarrow$  O(g)                         & 1 \\
 29 &  O$_2$(p) $\rightarrow$  O$_2$(g)         & 1 \\
 30 &  OH(p) $\rightarrow$  OH(g)                    & 1 \\
 31 &  CO(p) $\rightarrow$  CO(g)                    & 1 \\
 32 &  CO$_2$(p) $\rightarrow$  CO$_2$(g)    & 1 \\
 33 &  H$_2$O(p) $\rightarrow$  H$_2$O(g)    & 1 \\
 34 &  HO$_2$(p) $\rightarrow$  O(g) $+$ OH(g)                     & 1 \\
 35 &  H$_2$O$_2$(p) $\rightarrow$  H$_2$O$_2$(g)            & 1 \\
 36 &  HCO(p) $\rightarrow$  HCO(g)                           & 1 \\
 37 &  H$_2$CO(p) $\rightarrow$  H$_2$CO(g)           & 1 \\
 38 &  C(p) $\rightarrow$  C(g)                                      & 1 \\
 39 &  CH(p) $\rightarrow$  CH(g)                                 & 1 \\
 40 &  CH$_2$(p) $\rightarrow$  CH$_2$(g)                 & 1 \\
 41 &  CH$_3$(p) $\rightarrow$  CH$_3$(g)                 & 1 \\
 42 &  CH$_4$(p) $\rightarrow$  CH$_4$(g)                 & 1 \\
 43 &  H(c) $\rightarrow$  H(g)                         & 1 \\
 44 &  D(c) $\rightarrow$  D(g)                         & 2 \\
 \hline
 45 &  H(p) $+$ H(p) $\rightarrow$              H$_2$(p)                 & 1 \\
 46 &  H(p) $+$ D(p) $\rightarrow$              HD(p)                      & 2 \\
 47 &  H(p) $+$ O(p) $\rightarrow$              OH(p)                      & 1 \\
 48 &  H(p) $+$ OH(p) $\rightarrow$            H$_2$O(p)              & 1 \\
 49 &  H(p) $+$ O$_2$(p) $\rightarrow$           HO$_2$(p)         & 1 \\
 50 &  H(p) $+$ CO(p) $\rightarrow$        HCO(p)                       & 1 \\
 51 &  H(p) $+$ HO$_2$(p) $\rightarrow$        H$_2$O$_2$(p) & 1 \\
 52 &  H(p) $+$ HCO(p) $\rightarrow$        H$_2$CO(p)           & 1 \\
 53 &  H(p) $+$ C(p) $\rightarrow$        CH(p)                           & 1 \\
 54 &  H(p) $+$ CH(p) $\rightarrow$        CH$_2$(p)                & 1 \\
 55 &  H(p) $+$ CH$_2$(p) $\rightarrow$        CH$_3$(p)        & 1 \\
 56 &  H(p) $+$ CH$_3$(p) $\rightarrow$        CH$_4$(p)       & 1 \\
 57 &  O(p) $+$ O(p) $\rightarrow$        O$_2$(p)                   & 1 \\
 58 &  O(p) $+$ C(p) $\rightarrow$        CO(p)                         & 1 \\
 59 &  O(p) $+$ CO(p) $\rightarrow$        CO$_2$(p)              & 1 \\
 60 &  OH(p) $+$ OH(p) $\rightarrow$     H$_2$O$_2$(p)      & 1 \\
 \hline
\end{tabular}}
\end{center}
\label{tab:dust_molecule}
\end{table}

\begin{table}
\contcaption{}
\begin{center}
{\begin{tabular}{llc}
\hline
Number & Reaction & Reference \\
\hline
 61 &  H(p) $+$ H(p) $\rightarrow$              H$_2$(g)                                              & 1 \\
 62 &  H(p) $+$ D(p) $\rightarrow$              HD(g)                                                   & 2 \\
 63 &  H(p) $+$ O(p) $\rightarrow$              OH(g)                                                  & 1 \\
 64 &  H(p) $+$ OH(p) $\rightarrow$            H$_2$O(g)                                         & 1 \\
 65 &  H(p) $+$ O$_2$(p) $\rightarrow$           HO$_2$(g)                                     & 1 \\
 66 &  H(p) $+$ CO(p) $\rightarrow$        HCO(g)                                                   & 1 \\
 67 &  H(p) $+$ HO$_2$(p) $\rightarrow$        H$_2$O$_2$(g)                              & 1 \\
 68 &  H(p) $+$ HCO(p) $\rightarrow$        H$_2$CO(g)                                        & 1 \\
 69 &  H(p) $+$ C(p) $\rightarrow$        CH(g)                                                        & 1 \\
 70 &  H(p) $+$ CH(p) $\rightarrow$        CH$_2$(g)                                             & 1 \\
 71 &  H(p) $+$ CH$_2$(p) $\rightarrow$        CH$_3$(g)                                     & 1 \\
 72 &  H(p) $+$ CH$_3$(p) $\rightarrow$        CH$_4$(g)                                     & 1 \\
 73 &  O(p) $+$ O(p) $\rightarrow$        O$_2$(g)                                                 & 1 \\
 74 &  O(p) $+$ C(p) $\rightarrow$        CO(g)                                                       & 1 \\
 75 &  O(p) $+$ CO(p) $\rightarrow$        CO$_2$(g)                                            & 1 \\
 76 &  OH(p) $+$ OH(p) $\rightarrow$     H$_2$O$_2$(g)                                    & 1 \\
 77 &  H(p) $+$ H$_2$O(p) $\rightarrow$        H$_2$(p) $+$ OH(p)                    & 1 \\
 78 &  H(p) $+$ HO$_2$(p) $\rightarrow$        OH(p) $+$ OH(p)                         & 1 \\
 79 &  H(p) $+$ HO$_2$(p) $\rightarrow$        OH(g) $+$ OH(g)                         & 1 \\
 80 &  H(p) $+$ H$_2$O$_2$(p) $\rightarrow$     OH(p) $+$  H$_2$O(p)           & 1 \\
 81 &  H(p) $+$ H$_2$O$_2$(p) $\rightarrow$     OH(g) $+$  H$_2$O(p)           & 1 \\
 82 &  H(p) $+$ H$_2$O$_2$(p) $\rightarrow$     OH(g) $+$  H$_2$O(g)           & 1 \\ 
 83 &  H(p) $+$ HCO(p) $\rightarrow$    H$_2$(p) $+$ CO(p)                             & 1 \\
 84 &  H(p) $+$ HCO(p) $\rightarrow$    H$_2$(g) $+$ CO(p)                             & 1 \\
 85 &  H(p) $+$ HCO(p) $\rightarrow$    H$_2$(g) $+$ CO(g)                             & 1 \\
 86 &  H(p) $+$ H$_2$CO(p) $\rightarrow$  H$_2$(p) $+$ HCO(p)                    & 1 \\
 87 &  H(p) $+$ H$_2$CO(p) $\rightarrow$  H$_2$(g) $+$ HCO(p)                    & 1 \\
 88 &  H(p) $+$ H$_2$CO(p) $\rightarrow$  H$_2$(g) $+$ HCO(g)                    & 1 \\
 89 &  H(p) $+$ CO$_2$(p) $\rightarrow$   OH(p) $+$ CO(p)                             & 1 \\
 90 &  H(p) $+$ CH(p) $\rightarrow$   H$_2$(p)  $+$   C(p)                                & 1 \\
 91 &  H(p) $+$ CH(p) $\rightarrow$   H$_2$(g)  $+$   C(p)                                & 1 \\
 92 &  H(p) $+$ CH(p) $\rightarrow$   H$_2$(g)  $+$   C(g)                                & 1 \\
 93 &  H(p) $+$ CH$_2$(p) $\rightarrow$   H$_2$(p)  $+$   CH(p)                     & 1 \\
 94 &  H(p) $+$ CH$_2$(p) $\rightarrow$   H$_2$(g)  $+$   CH(p)                     & 1 \\
 95 &  H(p) $+$ CH$_2$(p) $\rightarrow$   H$_2$(g)  $+$   CH(g)                     & 1 \\
 96 &  H(p) $+$ CH$_3$(p) $\rightarrow$   H$_2$(p)  $+$   CH$_2$(p)             & 1 \\
 97 &  H(p) $+$ CH$_4$(p) $\rightarrow$   H$_2$(p)  $+$   CH$_3$(p)             & 1 \\
 98 &  O(p) $+$ OH(p) $\rightarrow$        H(p) $+$ O$_2$(p)                             & 1 \\
 99 &  O(p) $+$ OH(p) $\rightarrow$        H(g) $+$ O$_2$(p)                             & 1 \\
 100 &  O(p) $+$ OH(p) $\rightarrow$        H(g) $+$ O$_2$(g)                             & 1 \\
 101 &  O(p) $+$ HO$_2$(p) $\rightarrow$        O$_2$(p) $+$ OH(p)                 & 1 \\
102 &  O(p) $+$ HO$_2$(p) $\rightarrow$        O$_2$(g) $+$ OH(p)                 & 1 \\
103 &  O(p) $+$ HO$_2$(p) $\rightarrow$        O$_2$(g) $+$ OH(g)                 & 1 \\
104 &  O(p) $+$ HCO(p) $\rightarrow$     H(p) $+$ CO$_2$(p)                          & 1 \\
105 &  O(p) $+$ HCO(p) $\rightarrow$     H(g) $+$ CO$_2$(p)                          & 1 \\
106 &  O(p) $+$ HCO(p) $\rightarrow$     H(g) $+$ CO$_2$(g)                          & 1 \\
107 &  O(p) $+$ H$_2$CO(p) $\rightarrow$   H$_2$(p) $+$ CO$_2$(p)            & 1 \\
108 &  O(p) $+$ H$_2$CO(p) $\rightarrow$   H$_2$(g) $+$ CO$_2$(p)            & 1 \\
109 &  O(p) $+$ H$_2$CO(p) $\rightarrow$   H$_2$(g) $+$ CO$_2$(g)            & 1 \\
110 &  H$_2$(p) $+$ OH(p) $\rightarrow$     H(p) $+$ H$_2$O(p)                     & 1 \\
111 &  H$_2$(p) $+$ OH(p) $\rightarrow$     H(g) $+$ H$_2$O(p)                     & 1 \\
112 &  OH(p) $+$ CO(p) $\rightarrow$     H(p) $+$ CO$_2$(p)                          & 1 \\
113 &  OH(p) $+$ CO(p) $\rightarrow$     H(g) $+$ CO$_2$(p)                          & 1 \\
114 &  OH(p) $+$ CO(p) $\rightarrow$     H(g) $+$ CO$_2$(g)                          & 1 \\
115 &  OH(p) $+$ HCO(p) $\rightarrow$     H$_2$(p) $+$ CO$_2$(p)               & 1 \\
116 &  OH(p) $+$ HCO(p) $\rightarrow$     H$_2$(g) $+$ CO$_2$(p)               & 1 \\
117 &  OH(p) $+$ HCO(p) $\rightarrow$     H$_2$(g) $+$ CO$_2$(g)               & 1 \\
118 &  H$_2$(p) $+$ HO$_2$(p) $\rightarrow$     H(p) $+$ H$_2$O$_2$(p)    & 1 \\
 \hline
\end{tabular}}
\end{center}
\end{table}

\begin{table}
\contcaption{}
\begin{center}
{\begin{tabular}{llc}
\hline
Number & Reaction & Reference \\
\hline
119 &  H(g) $+$ H(p) $\rightarrow$ H$_2$(g)   & 2 \\
120 &  H(g) $+$ D(p) $\rightarrow$ HD(g)        & 2 \\
121 &  D(g) $+$ H(p) $\rightarrow$ HD(g)        & 2 \\
122 &  H(g) $+$ H(c) $\rightarrow$ H$_2$(g)   & 3 \\
123 &  H(g) $+$ D(c) $\rightarrow$ HD(g)        & 2 \\
124 &  D(g) $+$ H(c) $\rightarrow$ HD(g)        & 2 \\
125 &  H(p) $+$ H(c) $\rightarrow$ H$_2$(g)   & 3 \\
126 &  H(p) $+$ D(c) $\rightarrow$ HD(g)        & 2 \\
127 &  D(p) $+$ H(c) $\rightarrow$ HD(g)        & 2 \\
\hline
\end{tabular}}
\end{center}
{\bf References.} 1) \cite{Esplugues2019}. 2) \cite{Thi2018}. 3) \cite{Hocuk2016}.
\end{table}

\subsection{Simple formulae for the rate coefficients of H$_2$ and HD formation on dust grains}

Here, we derive simple formulae for the rate coefficients of H$_2$ and HD formation via grain-surface reactions.
Without relying on the abundances of grain-surface H and D, these formulae are represented as the fraction of H~(and D) atoms that are adsorbed on the grain surface and return into the gas phase as H$_2$~(and HD, respectively).

First, H$_2$ is produced dominantly via the reaction between physisorbed and chemisorbed H atoms~(reaction 125), so that the abundances of H(p) and H(c) are relevant to  the formation efficiency.
The H(p) abundance is determined by the balance among the reactions 1, 22, 24, and 125 as
\begin{equation}
y[{\rm H}_{\rm p}] = \frac{k_{\rm ads}({\rm H}) y({\rm H}) n_{\rm H}}{\alpha_{\rm pc}({\rm H}) + k_{\rm des}({\rm H}_{\rm p})},
\label{eq:y_Hp}
\end{equation}
whereas the H(c) abundance is calculated by the balance between the reactions 22 and 125 as
\begin{equation}
y[{\rm H}_{\rm c}] = \frac{1}{2}\frac{N_{\rm site}n_{\rm gr}}{n_{\rm H}}.
\label{eq:y_Hc}
\end{equation}
Eq. \eqref{eq:y_Hc} indicates that half the chemisorption sites are occupied by H atoms in the steady state.
By using Eqs. \eqref{eq:y_Hp} and \eqref{eq:y_Hc}, the H$_2$ formation rate can be summarized as
\begin{equation}
\frac{dy({\rm H}_2)}{dt} = k_{\rm H2, pc} y[{\rm H}_{\rm p}] y[{\rm H}_{\rm c}] \nH = k_{\rm gr}({\rm H}_2) y({\rm H}) n_{\rm H} 
\label{eq:dyH2_dt}
\end{equation}
where $k_{\rm gr}({\rm H}_2)$ is the rate coefficient given by
\begin{equation}
k_{\rm gr}({\rm H}_2) = \frac{1}{2} k_{\rm ads}({\rm H}) f_{\rm gr}({\rm H}_2)
\label{eq:k_gr_H2}
\end{equation}
and
\begin{equation}
f_{\rm gr}({\rm H}_2) = \left(1+\frac{k_{\rm des}({\rm H}_{\rm p})}{\alpha_{\rm pc}({\rm H})} \right)^{-1}.
\label{eq:f_gr_H2}
\end{equation}
$f_{\rm gr}({\rm H}_2)$ represents the fraction of H atoms that stick to grain surface and return into the gas phase as H$_2$.
This rate coefficient Eq. \eqref{eq:k_gr_H2} is consistent with that in \cite{Cazaux2008}.

Next, on the grain surface, HD is produced efficiently via the reactions between physisorbed H~(or D) and chemisorbed D~(or H, respectively; reactions 126 and 127), so that the abundances of D(p) and D(c), in addition to those of H(p) and H(c), are relevant to the HD formation rate.
The balance among the reactions 3, 23, 26, and 127 determines the D(p) abundance as
\begin{equation}
y[{\rm D}_{\rm p}] = \frac{k_{\rm ads}({\rm D}) y({\rm D}) n_{\rm H}}{\alpha_{\rm pc}({\rm D}) + k_{\rm des}({\rm D}_{\rm p})},
\label{eq:y_Dp}
\end{equation}
whereas that between the reactions 23 and 126 determines the D(c) abundance as
\begin{equation}
y[{\rm D}_{\rm c}] = \frac{\alpha_{\rm pc}({\rm D})}{2\alpha_{\rm pc}({\rm H})}\frac{y[{\rm D}_{\rm p}]}{y[{\rm H}_{\rm p}]}\frac{N_{\rm site}n_{\rm gr}}{n_{\rm H}}.
\label{eq:y_Dc}
\end{equation}
By using Eqs. \eqref{eq:y_Hp}, \eqref{eq:y_Hc}, \eqref{eq:y_Dp} and \eqref{eq:y_Dc}, the HD formation rate can be summarized as
\begin{equation}
\frac{d y({\rm HD})}{dt} = k_{\rm gr}({\rm HD}) y({\rm D}) n_{\rm H} 
\label{eq:dyHD_dt}
\end{equation}
where
\begin{equation}
k_{\rm gr}({\rm HD}) = k_{\rm ads}({\rm D}) f_{\rm gr}({\rm HD})
\label{eq:k_gr_HD}
\end{equation}
and
\begin{equation}
f_{\rm gr}({\rm HD}) = \left(1+\frac{k_{\rm des}({\rm D}_{\rm p})}{\alpha_{\rm pc}({\rm D})} \right)^{-1}.
\label{eq:f_gr_HD}
\end{equation}
These formulae are also consistent with those in \cite{Cazaux2009}.

\subsection{Collisional charging of dust grains}

Dust grains obtain an electric charge, when gas-phase ions and electrons recombine with grain-surface species.
Charged grains transfer its electric charge to other grains via collision with each other.
The list of the reactions for collisional charging of dust grains is summarized in Table \ref{tab:charge_transfer}.
Dust grains are assumed to have five charge states: gr$^0$, gr$^{\pm}$, and gr$^{2\pm}$, since the abundances of more than triply charged grains are negligibly small~\citep[e.g.,][]{Nakano2002}.

The rate coefficients for the grain charging via gas-grain and grain-grain collisions are calculated following \cite{Draine1987}.
A gas-phase ion~(or an electron) with an electric charge $q_{\rm x}e$ hits a charged grain with a charge $q_{\rm gr}e$~(reactions G1-G11) in the frequency of 
\begin{equation}
k(q_{\rm x}, q_{\rm gr}) =  S(T, T_{\rm gr}) {\rm v}({\rm X}) \pi a_{\rm gr}^2 \tilde{J}(\tau, \nu), 
\label{eq:collisional_charging_gas_grain}
\end{equation}
where $\tau = \frac{a_{\rm gr} k_{\rm B} T}{(q_{\rm x} e)^2},\ \nu = \frac{q_{\rm gr}}{q_{\rm x}}$,
\begin{align}
\tilde{J}(\tau, \nu = 0) =& 1 + \left(\frac{\pi}{2 \tau}\right)^{1/2},\nonumber\\ 
\tilde{J}(\tau, \nu < 0) =& \left[1-\frac{\nu}{\tau}\right] \left[1 + \left(\frac{2}{\tau-2\nu}\right)^{1/2}\right],\\
\tilde{J}(\tau, \nu > 0) =& \left[1 + (4 \tau + 3 \nu)^{-1/2}\right] \exp\left(-\frac{\theta_\nu}{\tau}\right),\nonumber
\label{eq:J_tilde}
\end{align}
and $\theta_\nu = \frac{\nu}{1+\nu^{-1/2}}$.
The mutual neutralization via grain-grain collisions~(reactions G12-G15) occurs at the rate of
\begin{equation}
k(q_1, q_2) = \pi (a_1 + a_2)^2 {\rm v}({\rm gr}) (\tilde{J}(\tau_1, \nu_1) + \tilde{J}(\tau_2, \nu_2))/2,
\label{eq:collisional_charging_grain_grain}
\end{equation}
between two grains with electric charges $q_{1}$ and $q_{2}$, and radii of $a_{1}$ and $a_{2}$.

When a grain is heated above the temperature of $\sim 500\K$, an electron bound on the grain surface obtains enough thermal energy to escape into the gas phase~(thermionic emission; reactions G16-G19).
The rate coefficient of thermionic emission is given by the Richardson law~\citep{Desch2015}
\begin{equation}
k_{\rm TE}(q_{\rm gr}) = 4 \pi a_{\rm gr}^2 \lambda_{\rm R} \frac{4 \pi m_e (k_{\rm B} T_{\rm gr})^2}{h^3} \exp\left(-\frac{W+q_{\rm gr}e^2/a_{\rm gr}}{k_{\rm B} T_{\rm gr}}\right),
\label{eq:k_TE}
\end{equation}
where $W=5$ eV is the work function of the grain material, and $\lambda_{\rm R}=1/2$.

\begin{table}
\caption{Collisional charging of dust grains.}
\begin{center}
{\begin{tabular}{clc}
\hline
Number & Reaction & Reference \\
\hline
G1 & $ {\rm X}^+  +  {\rm gr}  \rightarrow  {\rm gr}^{+}  +  {\rm X} $    &  1  \\
G2 & $ {\rm X}^{2+}  +  {\rm gr}  \rightarrow  {\rm gr}^{2+}  +  {\rm X} $    & 1   \\
G3 & $ {\rm X}^-  +  {\rm gr}  \rightarrow  {\rm gr}^{-}  +  {\rm X} $    &  1  \\
G4 & $ {\rm X}^+      +  {\rm gr}^{+}   \rightarrow  {\rm gr}^{2+} +  {\rm X} $  & 1   \\
G5 & $ {\rm X}^-       +  {\rm gr}^{+}   \rightarrow  {\rm gr}  +  {\rm X}       $  &  1  \\
G6 & $ {\rm X}^-       +  {\rm gr}^{2+} \rightarrow  {\rm gr}^{+}  +  {\rm X} $ &  1  \\
G7 & $ {\rm X}^{+}    +  {\rm gr}^{-}    \rightarrow   {\rm gr}  +  {\rm X}     $  &  1  \\
G8 & $ {\rm X}^{2+}  +  {\rm gr}^{-}    \rightarrow   {\rm gr}^+  +  {\rm X}  $  &  1  \\
G9 & $ {\rm X}^-       +  {\rm gr}^{-}    \rightarrow  {\rm gr}^{2-} +  {\rm X} $ &  1  \\
G10 & $ {\rm X}^+      +  {\rm gr}^{2-}  \rightarrow   {\rm gr}^-  +  {\rm X} $    &  1  \\
G11 & $ {\rm X}^{2+}  +  {\rm gr}^{2-}  \rightarrow   {\rm gr}  +  {\rm X} $    &  1  \\
\hline
G12 & $ {\rm gr}^+  +  {\rm gr}^-  \rightarrow  {\rm gr}  +  {\rm gr} $    & 1   \\
G13 & $ {\rm gr}^+  +  {\rm gr}^{2-}  \rightarrow  {\rm gr}  +  {\rm gr}^- $    & 1   \\
G14 & $ {\rm gr}^{2+}  +  {\rm gr}^{2-}  \rightarrow  {\rm gr}  +  {\rm gr} $  & 1     \\
G15 & $ {\rm gr}^{2+}  +  {\rm gr}^-  \rightarrow  {\rm gr}  +  {\rm gr}^+ $  & 1 \\
\hline
G16 & $ {\rm gr}^{2-}  \rightarrow  {\rm gr}^{-} + e^- $    &  2  \\
G17 & $ {\rm gr}^{-}  \rightarrow  {\rm gr} + e^- $    & 2   \\
G18 & $ {\rm gr}  \rightarrow  {\rm gr}^+ + e^- $    &  2  \\
G19 & $ {\rm gr}^{+}  \rightarrow  {\rm gr}^{2+} + e^- $    &  2  \\
\hline
\end{tabular}}
\end{center}
{\bf References.} 1) \cite{Draine1987}. 2) \cite{Desch2015}.
\label{tab:charge_transfer}
\end{table}




\label{lastpage}
\end{document}